\documentclass[a4paper,twoside,12pt,openright]{book}
\usepackage{cite}
\usepackage{amsmath}
\usepackage{amsthm}
\usepackage{amssymb}
\usepackage{amsfonts,amsxtra}
\usepackage[latin1]{inputenc}
\usepackage{psfrag,graphicx,float}
\usepackage[nottoc]{tocbibind}
\usepackage{fancyhdr}
 \clearpage{\pagestyle{empty}\cleardoublepage}
\makeatletter
\def\cleardoublepage{\clearpage\if@twoside \ifodd\c@page\else
    \hbox{}
    \thispagestyle{plain}
    \newpage
    \if@twocolumn\hbox{}\newpage\fi\fi\fi}
\makeatother \clearpage{\pagestyle{plain}\cleardoublepage}
\newlength\encuadernacion \newlength\longB 
\theoremstyle{theorem}
\newtheorem{theorem}{Theorem}

\newtheorem{proposition}[theorem]{Proposition}

\newtheorem{corollary}[theorem]{Corollary}

\newtheorem{lemma}[theorem]{Lemma}

\newtheorem{conjecture}[theorem]{Conjecture}

\newtheorem{claim}[theorem]{Claim}

\theoremstyle{definition}

\newtheorem{example}[theorem]{Example}

\theoremstyle{remark}
\newtheorem{remark}[theorem]{Remark}
\newtheorem{gloss}[theorem]{Gloss}

\include{header}
\numberwithin{theorem}{chapter}
\numberwithin{equation}{chapter}
\numberwithin{teorema}{chapter}

\newcommand{\al}{\alpha}
\newcommand{\be}{\beta}
\newcommand{\de}{\delta}
\newcommand{\ep}{\epsilon}
\newcommand{\vep}{\varepsilon}
\newcommand{\ga}{\gamma}
\newcommand{\ka}{\kappa}
\newcommand{\la}{\lambda}
\newcommand{\om}{\omega}
\newcommand{\si}{\sigma}

\newcommand{\vp}{\varphi}
\newcommand{\ze}{\zeta}
\newcommand{\La}{\Lambda}
\newcommand{\Si}{\Sigma}
\newcommand{\bx}{\mathbf{x}}

\newcommand{\bA}{\mathbf{A}}
\newcommand{\bsi}{\boldsymbol{\si}}
\newcommand{\btau}{\boldsymbol{\tau}}
\newcommand{\bxi}{{\boldsymbol{\xi}}}
\newcommand{\bs}{\mathbf{s}}
\newcommand{\bS}{\mathbf{S}}

\newcommand{\bv}{\mathbf{v}}
\newcommand{\bk}{\mathbf{k}}
\newcommand{\bn}{\mathbf{n}}
\newcommand{\bm}{\mathbf{m}}
\newcommand{\bz}{\mathbf{z}}

\newcommand{\bal}{\boldsymbol{\alpha}}
\newcommand{\bee}{\mathbf{e}}
\newcommand{\bB}{\mathbf{B}}
\newcommand{\bp}{\mathbf{p}}
\newcommand{\bJ}{\mathbf{J}}
\newcommand{\bde}{{\boldsymbol{\delta}}}
\newcommand\grad{\boldsymbol{\nabla}}
\newcommand{\tPsi}{\widetilde{\Psi}}
\newcommand{\Th}{\tilde{h}}

\newcommand{\tcH}{\widetilde{\cH}}
\newcommand{\tK}{\tilde{K}}
\newcommand{\tS}{\tilde{S}}

\newcommand{\tih}{\tilde{h}}

\newcommand{\tbs}{\tilde{\bs}}
\newcommand{\tPhi}{\widetilde{\Phi}}

\newcommand{\sse}{\mathsf{e}}
\newcommand{\ssE}{\mathsf{E}}
\newcommand{\ssh}{\mathsf{h}}

\newcommand{\ssH}{\mathsf{H}}

\newcommand{\ssZ}{\mathsf{Z}}
\newcommand{\hq}{\hat{q}}

\newcommand{\hPhi}{\widehat{\Phi}}
\newcommand{\hPsi}{\widehat{\Psi}}

\newcommand{\hcH}{\widehat{\mathcal H}}

\def\CC{\mathbb{C}}
\def\NN{\mathbb{N}}
\def\RR{\mathbb{R}}
\def\ZZ{\mathbb{Z}}

\def\TT{\mathbb{T}}
\def\HH{\mathbb{H}}
\newcommand{\cB}{{\mathcal B}}

\newcommand{\cD}{{\mathcal D}}

\newcommand{\cF}{{\mathcal F}}
\newcommand{\cH}{{\mathcal H}}

\newcommand{\cJ}{{\mathcal J}}
\newcommand{\cL}{{\mathcal L}}
\newcommand{\cM}{{\mathcal M}}
\newcommand{\cN}{{\mathcal N}}
\newcommand{\cO}{{\mathcal O}}
\newcommand{\cP}{{\mathcal P}}

\newcommand{\cS}{{\mathcal S}}
\newcommand{\cT}{{\mathcal T}}
\newcommand{\cX}{{\mathcal X}}
\newcommand{\cV}{{\mathcal V}}

\newcommand{\fD}{{\mathfrak D}}
\newcommand{\fB}{{\mathfrak B}}

\newcommand{\fK}{{\mathfrak K}}
\newcommand{\fP}{{\mathfrak P}}
\newcommand{\fS}{{\mathfrak S}}
\newcommand{\fW}{{\mathfrak W}}
\newcommand{\fZ}{{\mathfrak Z}}
\newcommand{\su}{{\mathfrak{su}}}

\newcommand{\fU}{{\mathfrak{U}}}
\newcommand{\fg}{{\mathfrak{g}}}
\def\Bx{\bar x}
\def\Bmu{\hat\mu}
\def\Bbe{\overline\beta}

\def\BH{\,\overline{H}{}}
\def\BV{\,\overline{V}{}}
\newcommand{\BcH}{\overline{\cH}{}}

\def\BP{\,\overline{\!P}{}}
\def\BQ{\,\overline{\!Q}{}}
\def\BR{\,\overline{\!R}{}}

\def\BcB{\overline{\cB}{}}

\newcommand{\pa}{\partial}
\newcommand{\pd}{\partial}
\newcommand\minus\backslash

\def\ket#1{|#1\rangle}

\def\BStrut{\vrule height12pt depth6pt width0pt}
\let\ds\displaystyle
\newcommand{\ms}{\mspace{1mu}}
\newcommand\lan\langle
\newcommand\ran\rangle

\newcommand{\spec}{\operatorname{spec}}
\newcommand{\tr}{\operatorname{tr}}

\newcommand{\card}{\operatorname{card}}
\newcommand{\erf}{\operatorname{erf}}
\newcommand{\Ad}{\operatorname{Ad}}

\newcommand{\iu}{{\mathrm i}}
\newcommand{\I}{{\mathrm i}}
\newcommand{\e}{{\mathrm e}}
\newcommand{\dd}{{\mathrm d}}
\newcommand{\super}[1]{^{\mathrm{#1}}}
\newcommand\sca{_{\mathrm{sc}}}
\newcommand\scal{^{\mathrm{sc}}}
\newcommand\Hsc{H\scal}
\DeclareMathOperator\SU{SU} \DeclareMathOperator\End{End}
\DeclareMathOperator\SL{SL} 
\newcommand\Emph[1]{\textbf{\emph{{#1}}}}
\newcommand{\Eps}{{\ep\ep'}}
\newcommand\mg{^{\mathrm{mag}}}

\renewcommand\leq\leqslant
\renewcommand\geq\geqslant

\begin{document}
\frontmatter

\thispagestyle{empty}
\begin{center}
  \vspace*{1cm}
  {\Large \bf Spin models of Calogero--Sutherland type and associated
    spin chains}

  \vspace*{2cm}
  {\large\bf Alberto Enciso Carrasco}

  \vfill

  {\normalsize Memoria de Tesis Doctoral\\[1mm]
  presentada al Departamento de Física Teórica II para optar al grado de\\
  [3mm] Doctor en Física\\[3mm]
4/5/2007}
  \vspace*{0.9cm}

  % Put your university logo here if you wish.
   \begin{center}
   \includegraphics{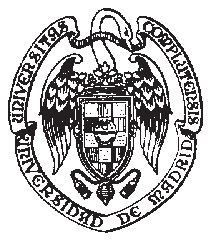}
   \end{center}

  {\normalsize Dirigida por:\\[3mm]
  Federico Finkel Morgenstern\\
  Artemio González López\\
  Miguel Ángel Rodríguez González\\[3mm]
  Departamento de Física Teórica II\\
  Universidad Complutense de Madrid}

\end{center}

%%%%%%%%%%%%%%%%%%%%%%%%%%%%%%%%%%%%%%%%%%%%%%%%%%%%%%%%%%%%%%%%%%%%%%%%%%%
%%%%%%%%%%%%%%%% The dedication page, of you have one  %%%%%%%%%%%%%%%%%%%%
%%%%%%%%%%%%%%%%%%%%%%%%%%%%%%%%%%%%%%%%%%%%%%%%%%%%%%%%%%%%%%%%%%%%%%%%%%%
\newpage
\thispagestyle{empty}\phantom{sdgg}
\newpage
\thispagestyle{empty}

%{\large\em\hfill A Pilar.
%
%\hfill A mis padres y a mi hermana.}

\begin{center}
 \vspace*{2cm}
  \hfill\begin{minipage}{6.6cm}\textit{\large\hfill A Pilar.\\[2mm]
A mis padres y a mi hermana.}
  \end{minipage}
\vfill

\hfill\begin{minipage}{9cm}
\begin{verse}
Per correr miglior acque alza le vele\\
omai la navicella del mio ingegno,\\
che lascia dietro a sé mar sì crudele;\\
e canterò di quel secondo regno\\
dove l'umano spirito si purga\\
e di salire al ciel diventa degno.\\[3mm]
Dante Alighieri, \textit{Divina Commedia}
\end{verse}
\end{minipage}
\end{center}
\newpage
\thispagestyle{empty}\phantom{sdgg}

\mainmatter

\tableofcontents

\chapter*{Preface}
\addcontentsline{toc}{chapter}{Preface}

This dissertation intends to present a unified view of my research in the area
of integrable systems, which has been carried out over the last four years
under the insightful supervision of Profs.\ F.\ Finkel, A.\ González-López and
M.~A.\ Rodríguez. Particular emphasis has been laid on the connection between
(partially) solvable spin models of Calogero--Sutherland (CS) type and
Haldane--Shastry (HS) chains, which can be rightly regarded as the leitmotif of
this thesis.

The dissertation is divided into five chapters. In the first one we summarize
some aspects of the theory of CS and HS models that are needed in later
chapters and introduce some notation. Related original results are introduced
when appropriate. In Chapter~\ref{Ch:BCN} we rigorously analyze the CS/HS
connection, which enables us to compute the partition function of the $BC_N$
HS chain in closed form. In Chapter~\ref{Ch:NN} we turn our attention to
quasi-exactly solvable (QES) generalizations of the standard CS models
presenting only near-neighbors interactions. A systematic explanation of their
quasi-exact solvability is given in terms of known invariant flags. In
Chapter~\ref{Ch:NNchain} we make use of the results in the two preceding
chapters to analyze the algebraic spectrum of a novel spin chain with
nearest-neighbors interactions which is naturally related to a QES Schrödinger
operator of CS type. This work ends with a short summary and a
concluding note on future projects.

The most remarkable contributions presented in this thesis are the following:
\begin{enumerate}
\item We complete the first mathematically rigorous treatment of the CS/HS connection.
Our approach hinges on several convergence results that make
Polychronakos's ``freezing trick'' precise.

\item We provide the first systematic analysis of spin QES models of CS type presenting
near-neighbors interactions, thereby obtaining several families of totally
explicit eigenfunctions.

\item We define an HS chain with nearest-neighbors interactions
that presents some of the features of a QES system. The study of
its algebraic states relies on a nontrivial refinement of the
freezing trick and on the explicit results obtained in
Chapter~\ref{Ch:NN}.

\item We introduce a natural Hamiltonian system admitting the maximum
number of independent first integrals, just as the original Calogero and
hyperbolic Sutherland models. This is the first example of a maximally
superintegrable system in an $N$-dimensional space of nonconstant curvature.
\end{enumerate}

A significant part of the material presented in this memoir has been adapted
from
Refs.~\cite{BEHR07,EP06b,EP06,EFGR06,EFGR07b,EFGR07,EFGR05c,EFGR05,EFGR05b}.
The presentation of the main results in Chapters~\ref{Ch:BCN}
and~\ref{Ch:NNchain}, however, greatly differs from that of the original
articles. In particular, most of the rigorous proofs provided in these chapters
appear in printed form here for the first time.

\chapter{Calogero--Sutherland models}
\label{Ch:CS}

\section{Introduction}

The discovery of Calogero--Sutherland (CS) models~\cite{Ca71,Su71,Su72} is a
landmark in the theory of integrable systems which has exerted a far-reaching
influence on many different areas of Mathematics and Physics. The intrinsic
mathematical interest of these models and their classical
analogs~\cite{Mo75,OP81} lies in their manifold connections with such disparate
topics as representation theory~\cite{BEG03,FGP05,SV04,EG02}, the theory of
special functions and orthogonal polynomials~\cite{BF97,Du98,Ta03,Ta05},
harmonic analysis~\cite{HO97,CV93,Ro98}, and complex and algebraic
geometry~\cite{Hi87,HN05,Ne96,Wi98,Et06}. On the other hand, CS models have
found a wide variety of applications to diverse areas of Physics, including
quantum field and string theory~\cite{MP93,GN94,FGNR00,DP98,Ma99,AJ06}, the
quantum Hall effect~\cite{AI94,Po01,KN06}, soliton
theory~\cite{Ka95,AW05,BH96,KBBT95}, fractionary
statistics~\cite{Po89,Ha94,CL99} and random matrix theory~\cite{Sh01,CM04}.

In this chapter we summarize the basic properties of CS models that shall be
needed in the sequel. We prefer not to follow a historical approach to this
subject, but rather to develop a modern bird's-eye presentation of the
essentials of the theory. Thus the most advanced machinery becomes available
from the very beginning, while the interest of the reader is not suffocated by
unnecessary details. Our survey of CS models does not intend to be exhaustive.
In particular, we have omitted several aspects of the theory (e.g., spin CM
models~\cite{Wo85,LX00,IS01b,Li04}, Ruijsenaars--Schneider systems and
$q$-Dunkl operators~\cite{Li06,RS86,AF98,Di95,BF97b,Ch95},
supersymmetry~\cite{GT04,DLM01,DG01} and connections with random matrix
theory~\cite{Sh01,CM04,CU96}), which could fit perfectly within the mainstream
of this dissertation and will indeed have some bearing on the content of
Chapter~\ref{Ch:Final}.

The core of this section is the brief survey of spin CS models and Dunkl
operators developed in Sections~\ref{S:DunklCh1} and~\ref{S:SpinCS}. While we
introduce a wealth of relevant concepts, we do not include a single rigorous
proof or statement: our presentation of these topics should be regarded as a
large dry bone that will be fleshed in Chapters~\ref{Ch:BCN} and~\ref{Ch:NN}.

We occasionally present some original results as we go along; these are mainly
taken from Refs.~\cite{EP06b,BEHR07,EP06}.

\section{Scalar CS models}
\label{S:scalarCS}

Let $R\subset\RR^N$ be a (crystallographic, possibly nonreduced) root system,
which we assume to be irreducible without loss of generality, and let
$R_+\subset R$ a subsystem of positive roots~\cite{Hu90}. Given $\bal\in R$, we
represent by
\[
\si_{\bal}(\bx)=\bx-2\frac{\bal\cdot\bx}{\|\bal\|^2}\bal
\]
the reflection across the hyperplane orthogonal to $\bal$. The Weyl
group corresponding to $R$ is denoted by
$\fW(R)=\lan\si_{\bal}:\bal\in R_+\ran$.

Consider a function $v\in C^\infty(\RR\minus\{0\})$ of one of the
following forms:
\begin{equation}\label{v}
v(x)=\begin{cases}x^{-2}\qquad&\text{(rational)}\\
\sinh^{-2}x &\text{(hyperbolic)}\\
\sin^{-2}x &\text{(trigonometric)}.
\end{cases}
\end{equation}
The (scalar) $N$-body \Emph{CS model} of type $R$ is defined by
the Hamiltonian
\begin{equation}\label{HCS}
H\sca=-\Delta+\sum_{\bal\in R_+}g_{\bal}^2\,v(\bal\cdot\bx)\,,
\end{equation}
where
\[
\Delta=\sum_{i=1}^N\pd_{x_i}^2
\]
and the coupling constants $g_{\bal}^2$ are assumed to take the same value for
roots of the same length. In the rational case we shall frequently include a
confining harmonic term, i.e.,
\[
H\sca=-\Delta+\sum_{\bal\in R_+}\frac{g_{\bal}^2}{(\bal\cdot\bx)^2}+\om^2r^2\,,
\]
with $r=\|\bx\|$. If we define a unitary action $K$ of $\fW(R)$ on $L^2(\RR^N)$
by
\begin{equation}\label{Kbal}
K(\si_{\bal})\equiv K_{\bal}:f\mapsto f\circ\sigma_{\bal}\,,
\end{equation}
it is clear that $\fW(R)$ is a symmetry group of the
Hamiltonian~\eqref{HCS}.

\begin{remark}
More generally~\cite{OP83}, the function $v$ in Eq.~\eqref{v} can be taken to
be a Weierstrass $\wp$ function with suitable periods. Despite the relevance of
this case and the abundant related literature~(cf.\ e.g.\ \cite{La00,La04} and
references therein), we have preferred to omit elliptic potentials in our
presentation of CS models.  On the one hand, they do not appear anywhere in the
sequel; on the other, their solvability and integrability properties are rather
different from those of their limiting cases listed in Eq.~\eqref{v} (with
$\om$ set to zero). The conditions on the root system can be also relaxed
somewhat~\cite{OP83,BCS98}.
\end{remark}

The double pole that $v$ possesses at 0 imposes that any function $\psi$ in the
domain of $H\sca$ must vanish at the hyperplanes $\bx\cdot\bal=0$ with $\bal\in R_+$,
which means that the particles cannot overtake one another. Hence one can
assume that the configuration space for $H\sca$ is the Weyl chamber
\begin{equation}\label{Wchamber}
C=\big\{\bx\in\RR^N:\bal\cdot\bx>0,\;\forall\bal\in R_+\big\}
\end{equation}
in the rational and hyperbolic cases, and the Weyl alcove
\begin{equation}\label{Walcove}
C=\big\{\bx\in\RR^N:\bal\cdot\bx>0,\;\forall\bal\in
R_+;\;\bal_{\max}\cdot\bx<\pi\big\}
\end{equation}
in the trigonometric one. Here we denote by $\bal_{\max}$ the highest root of
the system $R$.

\begin{remark}\label{R:Wchamber}
The symmetry of $H\sca$ under $\fW(R)$ guarantees the equivalence of the
eigenvalue problems for $H\sca$ in $\RR^N$ and in $C$. Namely, for any
eigenfunction $\vp\in L^2(\RR^N)$ of $H\sca$ there exists an eigenfunction
$\psi\in L^2(C)$ of $H\sca$ with the same eigenvalue and a one-dimensional
representation $\chi:\fW(R)\to\{±1\}$ of the Weyl group such that for any
$w\in\fW(R)$ satisfying $w\bx\in C$ one has
\[
\vp(\bx)=\chi(w)\,\psi(w\bx)\,.
\]
Clearly, the inclusion map defines a one-to-one correspondence
between the eigenfunctions of $H\sca$ on $L^2(C)$ and its
eigenfunctions on $L^2(\RR^N)$ with a given symmetry.
\end{remark}

We shall be mainly interested in the root systems $A_{N-1}$ and $BC_N$, which
are respectively given by
\begin{gather*}
R=\big\{\bee_i-\bee_j:1\leq i\neq j\leq N\big\}
\end{gather*}
and
\begin{align*}
R&=\big\{±\bee_i,±2\ms\bee_i,±\ms\ms\bee_i±\bee_j:1\leq i<j\leq
N\big\}\,,
\end{align*}
$\{\bee_i:1\leq i\leq N\}$ being the canonical basis of $\RR^N$. An appropriate
choice of the coupling constants of the CS models associated with these systems
suffices to exhaust all the possibilities arising from the non-exceptional Lie
algebras. Physically, these systems are important because they are the only
ones which allow for an arbitrary number $N$ of particles, given by the rank of
the algebra. For the sake of completeness and future reference, let us mention
that set of positive roots $R_+$ and the highest root $\bal_{\max}$ for the
$A_{N-1}$ and $BC_N$ systems can be taken as
\begin{gather*}
R_+=\big\{\bee_i-\bee_j:1\leq i<j\leq N\big\}\,,\qquad
\bal_{\max}=\bee_1-\bee_N\,,
\end{gather*}
and
\begin{gather*}
R_+=\big\{\bee_i±\bee_j,\bee_i,2\ms\bee_i:1\leq i<j\leq N\big\}\,,\qquad
\bal_{\max}=2\ms\bee_1\,.
\end{gather*}
The $A_{N-1}$ and $BC_N$ scalar CS models~\eqref{HCS} with
rational and trigonometric potentials can therefore be written as
\begin{equation}\label{Hsc}
H\sca=-\Delta+V\sca\,,
\end{equation}
where the potential is in each case given by
\begin{subequations}\label{Vsc}
\begin{align}
V\sca&=\sum_{i\neq j}\frac{a(a-1)}{(x_i-x_j)^2}+\om^2r^2\,,\label{HscAr}\\
V\sca&=\sum_{i\neq j}\frac{a(a-1)}{\sin^2(x_i-x_j)}\,,\label{HscAt}\\
V\sca&=a(a-1)\sum_{i\neq
j}\bigg[\frac{1}{(x_i-x_j)^2}+\frac{1}{(x_i+x_j)^2}\bigg]+\sum_i\frac{b(b-1)}{x_i^2}+\om^2r^2\,,\label{HscBr}\\
V\sca&=a(a-1)\sum_{i\neq
  j}\bigg[\frac{1}{\sin^2(x_i-x_j)}+\frac{1}{\sin^2(x_i+x_j)}\bigg]\notag\\
&\hphantom{{}=a(a-1)\sum_{i\neq
  j}\bigg[\frac{1}{\sin^2(x_i-x_j)}}{}+\sum_i\frac{b(b-1)}{\sin^2x_i}+\sum_i
\frac{b'(b'-1)}{\cos^2x_i}\,.\label{HscBt}
\end{align}
\end{subequations}
Here and in what follows the indices in all sums and products range from 1 to
$N$ unless otherwise stated.

\section{Integrability and solvability}
\label{S:IntSolv}

CS models are the paradigm of quantum integrable $N$-body problems. The first
satisfactory explanation of their remarkable algebraic properties was given by
Olshanetsky and Perelomov~\cite{OP83}, who constructed $N$ commuting first
integrals for the Hamiltonian~\eqref{HCS} associated with any non-exceptional
root system. A new proof of the integrability of these systems (which applies
even to the case of non-crystallographic root systems) has been recently
developed by Bordner, Manton and Sasaki~\cite{BMS00} using universal quantum
Lax pairs, and previously by Heckman and Opdam~\cite{He94,He97} (in the context
of degenerate Hecke algebras) and by Oshima and Sekiguchi \cite{OS95}. For
simplicity, we shall henceforth assume that $R$ is a non-exceptional root
system, typically $A_{N-1}$ or $BC_N$.

For these root systems, the models~\eqref{HCS} are \Emph{exactly solvable} in
the sense that all their eigenvalues and eigenfunctions can be computed in
closed form. (The solvability properties of CS models associated with exceptional
Lie algebras are still uncertain, even though there have been some recent
developments in this area~\cite{BTV05}.) The eigenfunctions $\psi_{\bn}$
are labeled by integer multiindices $\bn=(n_1,\dots,n_N)$ and factorize as
\begin{equation}\label{factor}
\psi_{\bn}(\bx)=\mu(\bx)\,P_{\bn}(\bz)\,,
\end{equation}
where $\mu\equiv\psi_{\mathbf0}$ is the ground state function,
$\bz=(z_1,\dots,z_N)$ with
\[
z_i=\begin{cases}x_i\qquad&\text{(rational)}\\
\e^{2x_i} &\text{(hyperbolic)}\\
\e^{2\I x_i} &\text{(trigonometric)},
\end{cases}
\]
and $P_\bn$ is a polynomial of degree
\begin{equation}\label{bn}
|\bn|=\sum_i |n_i|\,.
\end{equation}
Physically, the multiindex $\bn$ can be thought of as the
quasimomentum of the particles~\cite{Po99b,Po06}, and the energy of
the corresponding eigenfunction can be expressed neatly in terms of
this quantity and the parameters appearing in the Hamiltonian.

\begin{example}\label{Ex:Calogero}
The original \Emph{Calogero model}~\eqref{HscAr} is obtained by setting
$R=A_{N-1}$ and choosing the rational potential with a harmonic term. Its
eigenfunctions are labeled by a multiindex $\bn\in\NN_0^N$ and can be written as
$\psi_\bn=\mu\ms P_\bn$, where
\[
\mu(\bx)=\e^{-\frac\om2r^2}\prod_{i<j}|x_i-x_j|^a
\]
is the ground state function and $P_\bn$ is a homogeneous polynomial of degree
$|\bn|$). These polynomials are known as {generalized Hermite
  polynomials}~\cite{La91,UW96}, since when $N=1$ they reduce to Hermite
polynomials. Explicit formulas can be consulted, e.g., in
Refs.~\cite{BHV92,Ka96,HL06}. The energy of the eigenfunction $\psi_\bn$ is
given by
\[
E_\bn=2\om|\bn|+E_0\,,
\]
with $E_0=\om N(a(N-1)+1)$.
\end{example}

\begin{example}\label{Ex:Sutherland}
If we keep the $A_{N-1}$ root system and choose the trigonometric potential, we
obtain the original \Emph{Sutherland model}. Its eigenfunctions can be written
as $\psi_\bn(\bx)=\mu(\bx)\ms P_\bn(\bz)$, where
\[
\mu(\bx)=\prod_{i<j}|\sin(x_i-x_j)|^a\,,
\]
$\bn\in\ZZ^N$, and $P_\bn$ is a polynomial of degree $|\bn|$ in $\bz=(\e^{2\I
  x_1},\dots,\e^{2\I x_N})$.  These polynomials are called {Jack
  polynomials}~\cite{Ja70}, and play a fundamental role in Heckman and Opdam's
theory of hypergeometric functions for root
systems~\cite{Op88,Op88b,HO87,He87}.

A hyperbolic version of the trigonometric Sutherland system can be readily
obtained. Its spectrum, however, is not discrete, and in fact this model only
has a finite number of eigenvalues.
\end{example}

\section{Orthogonal polynomials}
\label{S:OrthPol}

Whereas the theory of orthogonal polynomials in one variable is a classical
branch of Mathematics, significant results on multivariate polynomials were
not obtained until the second half of the XX century~\cite{DX01}. Since then,
orthogonal polynomials of several variables have become a major field of
research in modern analysis and
combinatorics~\cite{Ma03,Di99,Be91,St89b,Ch95}.  As can be seen from
Eq.~\eqref{factor}, multivariate orthogonal polynomials appear as
eigenfunctions of CS models, and the role that these models play in the latter
theory is analogous to that of the Laguerre, Hermite or Jacobi equations in
the case of classical orthogonal polynomials in one variable.

As a matter of fact, the analogy between the eigenfunctions of CS models and
classical orthogonal polynomials is even closer. We showed in
Example~\ref{Ex:Calogero} that the eigenfunctions of the rational $A_{N-1}$ CS
model (the Hi-Jack polynomials) reduce to the Hermite polynomials in the case
$N=1$ and indeed coincide with their multivariate generalization~\cite{La91}.
It is also well known~\cite{BF97} that the multivariate generalizations of the
Laguerre~\cite{La91b} and Jacobi~\cite{La91c} polynomials are similarly
recovered from the $BC_N$ rational and trigonometric CS models, respectively.
Generally, CS models give rise to generalized hypergeometric functions
associated with root systems, as extensively studied by Heckman and
Opdam~\cite{He87,HO87,Op88,Op88b}.

A recent development is some kind of converse statement to the latter comment.
Hallnäs and Langmann have constructed an algorithm that allows one to go from
a family of orthogonal polynomials in one variable to an $N$-body Hamiltonian
whose eigenfunctions are a symmetric generalization of this family to $N$
variables. The $N$-body models associated with the classical families of
orthogonal polynomials are in fact previously known CS Hamiltonians.

The fruitful connection between CS models and the theory of orthogonal
polynomials has lead to a plethora of closed formulas for the eigenfunctions
of these Hamiltonians (cf.~\cite{DX01,Ma03,HL06} and references therein).
Furthermore, an exciting new ingredient of this theory are supersymmetric
(spin) polynomials~\cite{DLM03,DLM04}, related directly to the spin CS models
which will be the main subject of the remaining chapters of this memoir.

\section{A classical detour}
\label{S:CM}

It can be argued that the conceptually simplest approach to the exceptional
solvability properties of CS models is through their classical counterparts,
the Calogero--Moser (CM) models. Hence, we deem it important to present a
brief account of the theory of CM models, which, apart from its academic
interest, is essential in order to understand some recent developments on
classical Haldane--Shastry spin chains~\cite{IS01}.

\Emph{CM models}~\cite{OP81} are the Hamiltonian systems on $\RR^N$ whose
Hamiltonian function is obtained from Eq.~\eqref{HCS} by substitution of the
quantum kinetic energy (the Laplacian $-\Delta$) for the classical one (the
squared momentum $\|\bp\|^2$). The mathematical intricacies that lurk behind
quantum integrability~\cite{EP06,We92,Ku85,CP90} are absent in the classical
setting, the Liouville integrability of CM models being related to the
existence of an invariant foliation by Lagrangian cylinders in phase
space~\cite{AM78}. The Liouville integrability of CM models for any
(generalized) root system stems from Bordner, Corrigan and Sasaki's
construction~\cite{BCS99} of a universal Lax representation for these systems.
Nonetheless, their integrability was also explicitly obtained~\cite{Op88b} as a
corollary in a series of papers by Heckman and Opdam ten years earlier by means
of a detailed study of the corresponding CS models and a suitable theorem on
the classical limit of their quantum first integrals. An advantage of the first
authors' approach is that the equations of motion of the system can be
explicitly integrated for any root system (resp.\ for root systems associated
with classical Lie algebras) when the potential is rational (resp.\
trigonometric or hyperbolic.)

Most remarkably, the $A_N$ CM Hamiltonians with rational and hyperbolic
potentials are also~\Emph{maximally superintegrable} for any value of $N$,
i.e., they admit the maximum number $2N-1$ of functionally independent global
first integrals. This property is quite exceptional: apart from these two
models, the only known (natural) Hamiltonian systems which are maximally
superintegrable in arbitrary dimension are the geodesic flow on the simply
connected space forms, the (generalized) Kepler problem~\cite{Ev90}, the
Smorodinsky--Winternitz system~\cite{FMSUW65,Ev90b}, their generalizations to
the simply connected spaces of constant sectional curvature (cf.~\cite{BHSS03}
and references therein), the nonisotropic oscillator with rational frequencies,
the nonperiodic $A_N$ Toda lattice~\cite{To67,ADS06}, and a very recent example
living on a manifold of nonconstant curvature~\cite{BEHR07}, namely
\[
H(\bx,\bp)=\frac{\|\bp\|^2+\om^2r^2+\sum_jb_jx_j^{-2}}{1+r^2}\,.
\]
This system can be understood as the intrinsic Smorodinsky--Winternitz model on
an $N$-dimensional generalization of the Darboux space of type III~\cite{Ko72}.

The exceptional properties of the CM models can be understood by means of
(singular) Hamiltonian reduction~\cite{OR04}. This technique, independently
developed by Olshanetsky and Perelomov~\cite{OP76,OP76b} and by Kazhdan,
Konstant and Sternberg~\cite{KKS78}, was recently clarified and placed in a
broader context by Fehér and Pusztai~\cite{FP06,FP06b,FP06c}. The latter works
provide some kind of symplectic analog of Harish-Chandra's theory of
$c$-functions~\cite{He00,EG02}.

We sketch the gist of the method for the simple case of the $A_{N-1}$
trigonometric Sutherland model\footnote{I am indebted to Prof.\ L.\ Fehér for
valuable discussions on the state of the art of the projection method in CM
models.} (cf.\ Example~\ref{Ex:Sutherland}), as developed by Kazhdan, Konstant
and Sternberg~\cite{KKS78}. Let us consider the geodesic motion on $K=\SU(N)$,
which is given by the free Hamiltonian
\begin{equation*}
H_0(k,\la)=\|\la\|^2
\end{equation*}
on $T^*K$, where we trivialize $T^*K\cong K×\mathfrak k^*$ (with $\mathfrak
k=\mathrm{Lie}(K)$) by right translations and identify $\mathfrak
k^*\cong\mathfrak k$ by means of the inner product. Here the invariant
symplectic form on $T^*K$ is $\om=\dd\lan\la,(\dd k)k^{-1}\ran$. The
Hamiltonian action
\begin{equation}\label{conjugationaction}
(k,\la)\mapsto (hkh^{-1},h\la h^{-1})\,,\qquad h\in K
\end{equation}
is a symmetry of $H_0$ and gives rise to a momentum map $\psi: T^*K\to\mathfrak
k$ defined by
\[
\psi(k,\la)=\la-k^{-1}\la k\,.
\]
Now let us look at the (singular) Marsden--Weinstein reduced phase spaces
\[
M_\mu=\psi^{-1}(\mu)/K_\mu\,,
\]
where $K_\mu$ denotes the isotropy group of $\mu\in\mathfrak k$. (Generally
speaking, these symplectic spaces are not manifolds, but stratified sets whose
strata are differentiable manifolds.) The easiest nontrivial choice for $\mu$
is
\[
\mu=\I\big(uu^\dagger-N^{-1}g^2\big)\,,
\]
$u\in\CC^N$ being an arbitrary vector of norm $g$. In this case, the reduced
phase space actually turns out to be $M_\mu=T^*C\cong C×\mathrm{Cartan}(K)\cong
C×\RR^N$, $C$ being a Weyl alcove, and the reduced Hamiltonian is given by the
trigonometric Sutherland model, with $(\bx,\bp)\in C×\RR^N$. The integrability
and solvability properties of the latter system now follow from those of the
geodesic flow on $K$. As a matter of fact~\cite{FP06b}, these ideas can be also
applied to obtain a Lax representation. These ideas apply to any compact,
connected simple Lie group $K$, but actually it can be proved that a scalar CM
model is only obtained in the case that we have reviewed, whereas for a general
$K$ a spin CM model arises.

The above construction can be reinterpreted as follows. Let $G=K× K$ and
consider its involution $\Theta(k_1,k_2)=(k_2,k_1)$. If $G_+=\{(k,k):k\in K\}$
and $G_+\super R$ (resp.\ $G_+\super L$) denotes its right (resp.\ left) action
on $G$, it is clear that geodesic motion on $G$ enjoys a $G_+\super R×
G_+\super L$ symmetry. If we reduce $T^*G$ using the $G_+\super R$ symmetry at
the zero value of the momentum map, we arrive precisely at the symplectic
system $(T^*K,\om,H)$ analyzed in the preceding paragraph. Moreover, the
$G_+\super L$ symmetry still survives, and its action on $K\simeq G/G_+\super
R$ in fact coincides with the conjugation action~\eqref{conjugationaction} of
$K$ on itself. In this framework, hyperbolic CS models (possibly with spin) are
obtained when considering the complexification $G$ of $K$ given by
$\fg=\mathfrak k+\I \mathfrak k$, $\Theta$ being a Cartan involution of $G$.
The hyperbolic Sutherland model is obtained when $K=\SU(N)\simeq G_+$, so that
$G=\SL(N,\CC)$, $G/G_+\super R\simeq \exp(\I\mathfrak k)$ and $M_\mu=T^*C$ for
the above prescription for $\mu$, $C$ now being a Weyl chamber.

The above results also hold true, mutatis mutandis, if we take $G$ to be any
noncompact simple Lie group with finite center, $\Theta$ a Cartan involution,
and $G_+=\{g\in G:\Theta(g)=g\}$, so that it is a maximal compact subgroup. Any
element $g\in G$ can be uniquely decomposed as $g=g_+g_-$, where $g_±\in
G_±$ and $G_-=\{g\in G:\Theta(g)=g^{-1}\}$ is diffeomorphic to $\fg_-$ via
the exponential map. In this framework, one usually obtains spin CM models on
$T^*C$, and further generalizations have been also considered. Moreover, the
quantum analogs of these results are being actively pursued by the same
authors, and connections with flat symmetric spaces have also been studied
recently~\cite{AKLM03,Ho04}.

The related topic of the reduction of CS models by discrete symmetries has
also been considered in the Physics literature~\cite{Po99,Po02}, both in the
spin and spinless cases. A wide panoply of CS models can be obtained from the
original Calogero model through this procedure, partially explaining why this
system (with or without the harmonic confining potential) is sometimes
regarded as the mother of all CM systems~\cite{Po06}.

Interestingly, some purely quantum mechanical techniques in CS models were
originally suggested by a careful analysis of their classical analogs. For
instance, Polychronakos's rediscovery of Dunkl operators~\cite{Po92} was
motivated by a careful derivation of the scattering Calogero model by reduction
of free motion. Moreover, a short calculation allows one to prove that the only
asymptotic effect of the interaction with the potential~\eqref{HscAr} is an
effective reshuffling of the particles. This property carries over to the
quantum case, where it acquires an even greater relevance through the concept
of identical particles. It is therefore amusing that some ``physical'' effects
of the CS models were first observed at a classical level but only become
really fundamental in the quantum realm. For instance, the ideas that we have
just sketched lie at the heart of the applications of CS models to generalized
statistics in one dimension.

Another fruitful connection between CM and CS models, which actually holds for
a wide variety of one-dimensional quantum systems, appears when considering
the exact \Emph{bosonization} of one such model. This technique essentially
consists~\cite{Bl33,To50,Wi84} in describing a fermionic system by means of a
bosonic field theory. More precisely, we require that the algebra of bosonic
operators be irreducible on the fermionic Hilbert space and that the bosonic
field theory reproduce the exact eigenvalues of the system up to
non-perturbative effects. It has been recently discovered~\cite{EP06b} that an
exact one-dimensional bosonization can be constructed via Weyl quantization of
the classical picture, thus obtaining a new realization of the $W_\infty$ algebra.
This result is expected to shed new light on higher dimensional bosonization,
but the difficulties encountered in this case are considerable.

\section{Dunkl operators}
\label{S:DunklCh1}

The first attempt to provide a conceptual explanation of the exact solvability
of CS models was made by Olshanetsky and Perelomov~\cite{OP83}. In the rational
(with $\om=0$) and trigonometric cases, they showed that there exist some
particular values of the coupling constants for which the
Hamiltonian~\eqref{HCS} is equivalent to the radial part of the Laplacian on a
symmetric space associated with the root system $R$. The radial parts of the
Laplacians on these spaces have been thoroughly studied~\cite{He00}, and quite
explicit formulas are available. The general rational ($\om\neq0$) and
hyperbolic cases could also be dealt with using appropriate modifications. The
main drawback of Olshanetsky and Perelomov's approach was that for the $BC_N$
case and arbitrary coupling constants one did not obtain the radial Laplacian.
This, however, has been recently amended by Fehér and Pusztai (although so far
only the classical case has been carried out in detail~\cite{FP06}).

Concerning the connection between CS models and the theory of orthogonal
polynomials, a major breakthrough was the introduction of \emph{Dunkl {\em (or}
exchange{\em)} operators}~\cite{Du89}
\begin{equation}\label{Dunkl.rat.Ch1}
J\super{rat}_i=\pd_{x_i}-\sum_{\bal\in
R_+}a_{\bal}\frac{\bal\cdot\bee_i}{\bal\cdot\bx}\,(1-K_{\bal})\,,
\end{equation}
to study spherical harmonics associated with measures invariant under the Weyl
group $\fW(R)$. Dunkl operators have become an essential tool in harmonic
analysis, combinatorics and CS models~\cite{He97,Ro03}. The
differential-difference operators~\eqref{Dunkl.rat.Ch1} have two key
properties: they are homogeneous of degree $-1$, i.e., their action on a
polynomial lowers its degree by one, and are mutually commutative. In this
sense, one can think of the Dunkl operators as deformations of the usual
partial derivatives that preserve most of their analytic properties. In fact,
one can construct a Dunkl transform~\cite{Je93,Je06} using the spectral
resolution of the anti-self-adjoint operator $\bJ\super{rat}$, and the
resulting properties are fairly similar to those of the usual Fourier
transform.

In the Physics community, Dunkl operators were rediscovered by
Polychronakos~\cite{Po92} within the context of CS models. To explain the
essentials of the connection between these objects, let us restrict ourselves
to the case $v(x)=x^{-2}$ and define
\[
\mu(\bx)=\prod_{\bal\in R_+}|\bal\cdot\bx|^{a_{\bal}}.
\]
A simple link can now be established through the equation
\begin{equation}\label{HCSJ^2.Ch1.rat}
\mu H\sca\mu^{-1}=-\sum_i(J_i\super{rat})^2+E_0\,,
\end{equation}
where $E_0$ is a constant and the constants $g_{\bal}$ are given by
\[
g_{\bal}^2=\|\bal\|^2a_{\bal}(a_{\bal}+2a_{2\bal}-1)\,.
\]
This equation is known to hold true~\cite{Go05} on the space
$L^2(\RR^N)^{\fW(R)}$ of $\fW(R)$-invariant square-integrable functions on
$\RR^N$, which is canonically isometric to our (reduced) Hilbert space
$L^2(C)$. Hence Dunkl operators appear in the rational ($\om=0$) case (the {\em
scattering Calogero model}\/) as first integrals of the system, and have the
physical interpretation of (conserved) asymptotic momenta. Furthermore,
Eq.~\eqref{HCSJ^2.Ch1.rat} partially accounts for the relevance of CS models in
harmonic analysis. External harmonic potentials can be introduced in this
picture using either deformed destruction operators $a_i=-\I(J_i\super{rat}+\om
x_i)$~\cite{Po92} or auxiliary differential operators~\cite{FGGRZ01}.

In the trigonometric case one can define an analogous set of commuting
differential-difference operators by
\begin{align}
J_i\super{trig}=\pd_{x_i}+\sum_{\bal\in
R_+}a_{\bal}\frac{\bal\cdot\bee_i}{1-\e^{-2\I\bal\cdot\bx}}\, (1-K_{\bal})\,.
\label{Dunkl.trig.Ch1}
\end{align}
The relation between the Sutherland system of type $R$ and the sum of squares
of these Dunkl operators goes along the lines of the rational
case~\eqref{HCSJ^2.Ch1.rat}, with gauge factor
\begin{equation}\label{mut.Ch1}
\mu(\bx)=\prod_{\bal\in R_+}\big|\sin(\bal\cdot\bx)\big|^{a_{\bal}}\,.
\end{equation}
It can be readily verified that the operators~\eqref{Dunkl.trig.Ch1} leave
invariant the subspace of polynomials in $\bz=(\e^{2\I x_1},\dots,\e^{2\I
x_N})$ of degree not higher than $k$, for all $k\in\NN_0$.

As a matter of fact, the trigonometric Dunkl operators~\eqref{Dunkl.trig.Ch1}
were introduced by Cherednik~\cite{Ch91,Ch94}, while a closely related set of
differential-difference operators has been previously considered by
Heckman~\cite{He91}. Several analogous families of operators are currently
known, both in the rational and trigonometric cases. For simplicity, we shall
understand by (generalized) \Emph{Dunkl operators} of type
$R$~\cite{He97,Ro03,FGGRZ01} any set of first order differential-difference
operators $J_i$ ($i=1,\dots,N$) in $N$ variables $\bz$ satisfying:
\begin{enumerate}
\item The operators $J_i$ map the space of polynomials $\CC[\bz]$
into itself and leave invariant finite-dimensional subspaces
$\cM^n\subset\CC[\bz]$.

\item The operators $\{J_i,K_{\bal}:1\leq i\leq N,\bal\in R_+\}$
span a degenerate Hecke algebra.
\end{enumerate}
Degenerate (or graded) Hecke algebras were defined independently by
Drinfeld~\cite{Dr86} and Lusztig~\cite{Lu89}. For the sake of completeness, let
us recall that the \emph{degenerate Hecke algebra} $(\HH,*)$ associated with
the system of positive roots $R_+\subset V$ and the parameters
$(a_{\bal})_{\bal\in R_+}$ is the linear space
$\HH=\cS(V^\CC)\otimes\CC[\fW(R)]$ endowed with the unique graded algebra
structure satisfying
\begin{enumerate}
\item $\cS(V^\CC)\otimes 1$ and $1\otimes\CC[\fW(R)]$ are subalgebras, so that
$\cS(V^\CC)$ and $\CC[\fW(R)]$ can be identified with their images in $\HH$.

\item If $\bal$ is a simple root, then
$\si_{\bal}*\bv-(\si_{\bal}\bv)*\si_{\bal}=-a_{\bal}\bal\cdot\bv$ for all
$\bv\in V^\CC$.
\end{enumerate}
Here $R_+$ is assumed to span $V$, $\cS(V^\CC)$ denotes the algebra of
symmetric tensors in the complex linear space $V^\CC$, and the grading in $\HH$
is in fact inherited from that of $\cS(V^\CC)$. Clearly one can identify
$T*w=T\otimes w$, where $T\in\cS(V^\CC)$ and $w\in\CC[\fW(R)]$.

As in Eq.~\eqref{HCSJ^2.Ch1.rat}, Dunkl operators happen to give rise to
(generalized) CS models through the addition of auxiliary operators, a
similarity transformation and a global change of variables. Full details can be
consulted in Refs.~\cite{FGGRZ01b,FGGRZ01} and in Chapter~\ref{Ch:NN}.

\section{Spin CS models}
\label{S:SpinCS}

We shall comment now on another benefit granted by the Dunkl operator formalism
that is essential for all the forthcoming sections: the definition of
\emph{spin CS models}. Since the study of (generalizations of) such models
could be rightly considered to be the core of this Ph.\ D.\ dissertation, we
shall have many opportunities in the following chapters to elaborate on
this topic. Our goal in this section is to define the basic operators and
present the fundamentals of the theory without providing any details on the
computational difficulties involved. This and other aspects shall be covered in
detail in Chapter~\ref{Ch:BCN}.

For the sake of simplicity, we shall restrict ourselves to the
non-exceptional root systems $A_N$ and $BC_N$, which are the only
ones that have been considered in the literature up to now. (In
fact, we shall make use of the root system $A_{N-1}$ embedded in
$\RR^N$, not of $A_N$. Nevertheless, we shall hereafter drop the
subscript $-1$ and regard the subscript $N$ as an abstract index.)

Let $\Sigma\approx\CC^{(2M+1)^N}$ be the Hilbert space of the
internal degrees of freedom of $N$ particles of spin
$M\in\frac12\NN$, and let
\begin{equation}\label{cBSi}
\cB_\Si=\big\{\ket{\bs}\equiv\ket{s_1,\dots,s_N}:s_i\in\{-M,-M+1,\dots,M\}\big\}
\end{equation}
be the canonical basis of $\Sigma$. Let us define the spin permutation and reversal operators by
\begin{subequations}\label{Ss}
\begin{align}
  S_{ij}\ket{s_1,\dots,s_i,\dots,s_j\dots
    s_N}&=\ket{s_1,\dots,s_j,\dots,s_i\dots s_N}\,,\label{Sij}
  \\
  S_i\ket{s_1,\dots,s_i,\dots s_N}&=\ket{s_1,\dots,-s_i,\dots s_N}\,,\label{Si}\\
  \tS_{ij}\ket\bs&=S_iS_jS_{ij}\ket\bs\,.\label{tSij}
\end{align}
\end{subequations}
One can easily verify that they satisfy the algebraic relations
\begin{gather*}
S_{ij}^2=1,\qquad S_{ij}S_{jk}=S_{ik}S_{ij}=S_{jk}S_{ik},\qquad
S_{ij}S_{kl}=S_{kl}S_{ij},\\
S_i^2=1,\qquad S_iS_j=S_jS_i,\qquad S_{ij}S_k=S_kS_{ij},\qquad
S_{ij}S_j=S_iS_{ij},
\end{gather*}
where the indices $i,j,k,l$ take distinct values in the range
$1,\dots,N$.

The one-dimensional representations $\chi:\fW(R)\to\{±1\}$ of the
Weyl group $\fW(R)$ are given by~\cite{Hu90}
\begin{equation*}
\si_{\bee_i-\bee_j}\overset{\chi_\ep}{\mapsto}\ep\,,\qquad\ep=±1
\end{equation*}
in the $A_N$ case and by
\begin{equation*}
\si_{\bee_i-\bee_j}\overset{\chi_{\Eps}}\mapsto\ep\,,\qquad\si_{\bee_i}\overset{\chi_{\Eps}}\mapsto\ep'\,,\qquad\ep,\ep'=±1
\end{equation*}
in the $BC_N$ case. For each of these one-dimensional representations $\chi$,
one can define the \Emph{spin CS model} with scalar counterpart~\eqref{Hsc} as
\begin{equation}\label{Hspin}
H_\chi=-\Delta+V_\chi\,,
\end{equation}
where the spin potentials $V_\chi$ are given by
\begin{subequations}\label{Vspin}
\begin{align}
V_{\ep}&=\sum_{i\neq j}\frac{a(a-\ep S_{ij})}{(x_i-x_j)^2}+\om^2r^2\,,\label{HAr}\\
V_{\ep}&=\sum_{i\neq j}\frac{a(a-\ep S_{ij})}{\sin^2(x_i-x_j)}\,,\label{HAt}\\
V_{\Eps}&=\sum_{i\neq j}\bigg[\frac{a(a-\ep
S_{ij})}{(x_i-x_j)^2}+\frac{a(a-\ep\tS_{ij})}{(x_i+x_j)^2}\bigg]
+\sum_i\frac{b(b-\ep'S_i)}{x_i^2}+\om^2r^2\,,\label{HBr}\\
V_{\Eps}&=\sum_{i\neq j}\bigg[\frac{a(a-\ep
S_{ij})}{\sin^2(x_i-x_j)}+\frac{a(a-\ep\tS_{ij})}{\sin^2(x_i+x_j)}\bigg]\notag\\
&\hphantom{{}={}\sum_{i\neq j}\bigg[\frac{a(a-\ep S_{ij})}{\sin^2(x_i-x_j)}}{}+\sum_i\frac{b(b-\ep'S_i)}{\sin^2x_i}+\sum_i
\frac{b'(b'-\ep'S_i)}{\cos^2x_i}\,.\label{HBt}
\end{align}
\end{subequations}
We shall frequently omit the subscript if there is no risk of
confusion. The extension to the hyperbolic case is
straightforward.

Let us introduce the coordinate permutation and sign reversal operators
\begin{subequations}\label{Ks}
\begin{align}
(K_{ij}f)(x_1,\dots,x_i,\dots,x_j,\dots,x_N)&=f(x_1,\dots,x_j,\dots,x_i,\dots,x_N)\,,\label{Kij}\\
(K_if)(x_1,\dots,x_i,\dots,x_N)&=f(x_1,\dots,-x_i,\dots,x_N)\,,\label{Ki}\\
\tK_{ij}f&=K_iK_jK_{ij}f\,,\label{tKij}
\end{align}
\end{subequations}
which are the reflection operators associated with the root systems $A_N$
and $BC_N$. One obviously has the group isomorphisms $\lan
K_{ij}\ran\approx\lan S_{ij}\ran\approx\fW(A_{N-1})$ and $\lan
K_i,K_{ij}\ran\approx\lan S_i,S_{ij}\ran\approx\fW(BC_N)$. We
shall denote by $\fK$, $\fS$ and $\fW$ respectively the
realizations of $\fW(R)$ generated by the operators~\eqref{Ks},
\eqref{Ss}, and
\begin{equation}\label{Pi}
\Pi_{ij}=K_{ij}S_{ij}\,,\qquad\Pi_i=K_iS_i\,.
\end{equation}
When notationally convenient, we shall regard these realizations
as unitary representations
\begin{subequations}
\begin{align}
K:w&\mapsto K_w\in\End(L^2)\,,\label{K.rep}\\
S:w&\mapsto S_w\in\End(\Si)\,,\label{S.rep}\\
\Pi:w&\mapsto \Pi_w\in\End(L^2\otimes\Si)\,,
\end{align}
\end{subequations}
with $w\in\fW(R)$. Moreover, we shall not distinguish between a character of
$\fW(R)$ and its isomorphic extension to $\fW,\fS$ or $\fK$.

The key property of the spin models~\eqref{Vspin} is their connection with
Dunkl operators. This connection is implemented via the \Emph{star mapping}
$\fU(\fK)\to\fU(\fS)$ associated with the one-dimensional representation $\chi$,
which is the antihomomorphism defined by
\begin{subequations}\label{star}
\begin{align}
K_{ij}&\mapsto (K_{ij})^*_{\Eps}=\ep S_{ij}\,,\\
K_i&\mapsto (K_i)^*_{\Eps}=\ep'S_i\,.\label{starb}
\end{align}
\end{subequations}
Here $\fU(\fg)$ stands for the (real) universal enveloping algebra of $\fg$,
and the second equation~\eqref{starb} is to be omitted in the $A_N$ case. We
shall also consider the extension of~\eqref{star} to $\fD\otimes\fK\to\fD\otimes\fS$
given by
\begin{equation*}
(D\otimes K)^*_{\Eps}=D\otimes(K)^*_{\Eps}\,.
\end{equation*}
Here $\fD$ is the space of (scalar) smooth differential operators, and $D$ and
$K$ belong to $\fD$ and $\fK$ respectively. It is easy to show that the spin
CS models~\eqref{Hspin} are obtained by applying the star mapping associated
with $\chi$ to the \Emph{exchange Hamiltonian}
\begin{equation}\label{HK}
\BH=-\Delta+\BV\,,
\end{equation}
with
\begin{subequations}\label{VK}
\begin{align}
\BV&=\sum_{i\neq j}\frac{a(a-K_{ij})}{(x_i-x_j)^2}+\om^2r^2\,,\label{HKAr}\\
\BV&=\sum_{i\neq j}\frac{a(a-K_{ij})}{\sin^2(x_i-x_j)}\,,\label{HKAt}\\
\BV&=\sum_{i\neq
j}\bigg[\frac{a(a-K_{ij})}{(x_i-x_j)^2}+\frac{a(a-\tK_{ij})}{(x_i+x_j)^2}\bigg]
+\sum_i\frac{b(b-K_i)}{x_i^2}+\om^2r^2\,,\label{HKBr}\\
\BV&=\sum_{i\neq
j}\bigg[\frac{a(a-K_{ij})}{\sin^2(x_i-x_j)}+\frac{a(a-\tK_{ij})}{\sin^2(x_i+x_j)}\bigg]\notag\\
&\hphantom{{}=\sum_{i\neq
j}\bigg[\frac{a(a-K_{ij})}{\sin^2(x_i-x_j)}}{}+\sum_i\frac{b(b-K_i)}{\sin^2x_i}+\sum_i
\frac{b'(b'-K_i)}{\cos^2x_i}\,.\label{HKBt}
\end{align}
\end{subequations}
Since these are essentially the operators obtained from the sum of the squares
of the Dunkl operators~\eqref{Dunkl.rat.Ch1} and~\eqref{Dunkl.trig.Ch1}
through a gauge transformation and a change of variables, the eigenvalue
problem for the spin Hamiltonian~\eqref{Hspin} can be effectively reduced to
the analogous problem for a quadratic polynomial in the Dunkl operators plus,
perhaps, some simple auxiliary operators.

The connection between CS models and Dunkl operators can also be used to define
generalizations of the above solvable Hamiltonians (with or without spin). Such
an operator $H$ is called \Emph{of CS type}. Its spectrum cannot be generally
computed by algebraic methods, but by construction~\cite{FGGRZ01,FGGRZ01b} $H$
remains partially solvable in the sense that it leaves invariant some known
finite-dimensional subspace of $L^2(\RR^N)\otimes\Sigma$; in this case, $H$ is
said to be \Emph{quasi-exactly solvable} (QES), and the eigenvalues that one
can obtain from the latter subspaces are called \Emph{algebraic}.

\begin{remark}
A particularly interesting situation~\cite{Tu88,Us94} arises when one can find
explicitly an infinite flag (i.e., a sequence
$\cM_1\subsetneq\cM_2\subsetneq\cdots\subset L^2(\RR^N)\otimes\Sigma$, where
$\cM_n$ is finite-dimensional) invariant under $H$. In this case, the
corresponding Hamiltonian $H$ is sometimes said to be \emph{exactly solvable in
the sense of Turbiner}~\cite{Tu88}. This expression is certainly misleading,
the latter condition does not imply by any means that algebraic eigenvalues
cover the whole point spectrum of $H$. In this dissertation we shall be
interested in this stronger type of quasi-exact solvability, and in fact all
the QES Hamiltonians that we shall encounter in Chapter~\ref{Ch:NN} will indeed
preserve an infinite-dimensional invariant flag.
\end{remark}

\section{Solvable spin chains}\label{S:HS}

Solvable spin chains have enjoyed a growing popularity in the last few years,
due in part to their novel applications to SUSY Yang--Mills and string
theories~\cite{MZ03,BC04,RV04,BS05,FKM05,Go05b}. The role of CS spin models in
the study of solvable spin chains was unveiled by
Polychronakos~\cite{Po92,Po94} in a couple of insightful papers from the mid
90s which elucidated the connection between the original Sutherland model and
the recently discovered Haldane--Shastry (HS) chain~\cite{Ha88,Sh88}. We shall
conclude this chapter with an overview of the essentials of the CS/HS
connection and the definition of some celebrated spin chains that shall be
frequently referred to in forthcoming chapters.

The \Emph{Heisenberg chain}~\cite{He28}, which was born as an
attempt to model ferromagnetic materials, describes $N$ particles on
a lattice with isotropic near-neighbor interactions independent of
the site. For particles of spin $M\in\frac12\NN$, the Hamiltonian of
the system is customarily written as
\begin{align}\label{Heis}
\ssH_{\mathrm{He}}&=\sum_i \bS_i\cdot\bS_{i+1}\,,
\end{align}
where $\bS_i$ is the (vector) spin operator of the particle $i$,
$\bS_{N+1}\equiv\bS_1$ and
\[
\bS_i\cdot\bS_j=\sum_{a=1}^{4M(M+1)}S_i^aS_j^a\,.
\]
For each fixed particle label $i$, we denote by $S_i^a$ ($a=1,\dots,4M(M+1)$)
a basis of the fundamental representation of $\su(2M+1)$, which we assume to
be orthonormal with respect to the bilinear form $\lan S,S'\ran=2\tr(SS')$. It
is straightforward to write the Hamiltonian~\eqref{Heis} in terms of the spin
exchange operators~\eqref{Sij} using that
\begin{equation}\label{SU(2M+1)inv}
S_{ij}=2\,\bS_i\cdot\bS_j+\frac1{2M+1}\,.
\end{equation}
Note that for spin $1/2$ particles we have
$\bS_i=\frac12(\sigma_i^1,\sigma_i^2,\sigma_i^3)$, where $\sigma_i^j$ is the $j$-th Pauli
matrix acting on the spin space of the $i$-th particle.

It is well known~\cite{Be31,Hu38,CP62} that the spin $1/2$ Heisenberg chain can
be solved exactly using the Bethe ansatz. Several (partially) solvable
generalizations of the Heisenberg chain~\eqref{Heis} with short-range
interactions (at most between next-to-nearest neighbors) have been subsequently
proposed in the literature. These include, in particular, the family of chains
with arbitrary spin and nearest-neighbors interactions polynomial in
$\bS_i\cdot\bS_{i+1}$ of Refs.~\cite{Ba82,Ta82}, as well as several models
whose ground state can be written in terms of ``valence
bonds''~\cite{MG69,MG69b,AKLT87}.

The \Emph{Haldane--Shastry model}~\cite{Ha88,Sh88} is the prime example of
solvable spin chain with long-range interactions. It describes an arrangement
of $N$ equally spaced spins on a circle which interact pairwise with a
strength proportional to the inverse square of their chord distance. The
Hamiltonian of the HS chain can be written as
\begin{equation}\label{Ch2.HS}
\ssH_{\mathrm{HS}}=\sum_{i<j} \sin^{-2}(\xi_i-\xi_j)(1-S_{ij})\,,
\end{equation}
where the positions of the sites are given by
\[
\xi_i=\frac{i\pi}N\,.
\]

This system quickly attracted considerable
attention~\cite{In90,Ha91,KA92,Fo92,SS93} due to its remarkable algebraic
properties and its connection with conformal field theory~\cite{FMS99} and the
symplectic Dyson model~\cite{Dy70}. The original interest of this model,
however, lies in the fact that Gutzwiller's variational wave function for the
Hubbard model~\cite{Gu63,GV87,GJR87}, which also coincides with the
one-dimensional version of the resonating valence bond state introduced by
Anderson~\cite{ABZH87} and with the one-dimensional version of the
Kalmeyer--Laughlin state~\cite{KL87}, becomes an exact eigenfunction of the HS
chain as the strength of the on-site interaction tends to infinity.

In the early nineties the integrability of the Hamiltonian~\eqref{Ch2.HS} was
rigorously proved and its Yangian symmetry~\cite{HHTBP92,FM93} was established.
Not surprisingly, the Yangian algebra~\cite{Dr85} also appears as the symmetry
algebra~\cite{BGHP93,AF98,Hi95} of the Sutherland model~\eqref{HscAt}. Although
the obvious relationship between this model and the chain~\eqref{Ch2.HS} was
already remarked by Shastry in his original paper, their connection was first
made explicit by Polychronakos through the so-called \Emph{freezing
trick}~\cite{Po93}. This technique \emph{formally} allows one to recover the HS
Hamiltonian from the Sutherland spin model~\eqref{HAt} (with $\ep=1$) in the
large coupling constant limit, which is clearly tantamount to the classical
limit $\hbar\downarrow0$. A rigorous proof of the classical limit and some
extensions and improvements will be presented in Chapters~\ref{Ch:BCN}
and~\ref{Ch:NNchain}.

A further application of the freezing trick enabled Polychronakos to
obtain the integrable spin chain with long-range interactions given
by the Hamiltonian
\begin{equation}\label{Ch2.PF}
\ssH_{\mathrm{PF}}=\sum_i(\xi_i-\xi_j)^{-2}(1-S_{ij})\,,
\end{equation}
where the sites $\xi_1<\cdots<\xi_N$ satisfy the equation
\begin{equation}\label{sites.PF}
\xi_i=\sum_{j\neq i}\frac1{\xi_i-\xi_j}\,.
\end{equation}
This system is usually referred to as the \Emph{Polychronakos--Frahm (PF)
  chain}, and relates to the spin Calogero model~\eqref{HAr} as the HS chain
does to the Sutherland model~\eqref{HAt}.

As remarked by Frahm~\cite{Fr93}, the sites of the PF chain are no longer
equally spaced, but Eq.~\eqref{sites.PF} shows that they are given by the zeros
of the $N$-th Hermite polynomial~\cite{Sz75}. In particular~\cite{CP78b}, this
implies that the density of sites for large $N$ is given by
\begin{equation}\label{rho.PF}
\rho^{\mathrm{PF}}_N(x)=\frac1{\pi N}\sqrt{2N-x^2}\,.
\end{equation}

\begin{remark}
  The CS/HS correspondence casts the original model of Haldane and Shastry
  into a much wider framework. Consequently, we shall say that a spin chain is
  \Emph{of HS type} if it can be derived from a spin Schrödinger operator of
  CS type via the freezing limit. In particular, both the HS and the PF chains
  are of HS type.

Interestingly, we shall see in Chapters~\ref{Ch:BCN}
and~\ref{Ch:NNchain} (cf.\ also~\cite{Po94,FG05}) that the chains of
HS type, featuring long-range position-dependent interactions, seem
to enjoy stronger solvability properties than those of Heisenberg
type, characterized by the short range and position independence of
the interactions.
\end{remark}

\chapter[HS spin chains of $BC_N$ type]{HS spin chains of $\boldsymbol{BC_N}$ type}
\label{Ch:BCN}

\section{Introduction}
\label{S:introHS}

In the previous chapter we saw that the connection between $A_N$ CS models and
spin chains of HS type has been thoroughly studied. In contrast, $BC_N$ spin
models~\cite{Ya95,Du98,FGGRZ01b,IS01,CC04} had received comparatively little
attention. Two main reasons were responsible for this fact. On one hand,
$BC_N$ HS chains depend nontrivially on free parameters (one in the rational
case and two in the trigonometric one). On the other hand, the sites of the
$BC_N$ chain are not equally spaced and no explicit description of their
positions is available.

Let us give a very brief review of the existing results on these models. The
integrability of the rational $BC_N$ HS chain was established by Yamamoto and
Tsuchiya~\cite{YT96} using Dunkl operators, but its spectrum has not been
computed so far. The HS trigonometric spin chain of $BC_N$ type was discussed
by Bernard, Pasquier and Serban~\cite{BPS95}, but only for spin $1/2$ and with
the assumption that the sites are equally spaced, which restricts the space of
free parameters in the model to just three particular values. Finkel et
al.~\cite{FGGRZ03} discussed the integrability of the hyperbolic HS chain of
$BC_N$ type, but did not examine its spectrum either.

The main result of this chapter is the exact computation of the spectrum of the
$BC_N$ HS chain. In fact, four chains associated with the $BC_N$ Sutherland model
are considered, two of which are essentially different. This computation is
based on a rigorous study of Polychronakos's ``freezing trick'', which is in
fact promoted to a ``freezing lemma'' in Section~\ref{S:chainBCN}.

The organization of this chapter is as follows. In
Sections~\ref{S:dynamical} to~\ref{S:sp.Hchi} we present a detailed
study of the $BC_N$ Sutherland spin model and prove its exact
solvability using Dunkl operators. In Section~\ref{S:chainBCN} we
introduce the $BC_N$ HS chains and compute their spectrum, whose
statistical properties we analyze carefully in the following
section. Finally, in Section~\ref{S:mag} we consider the addition of
a constant magnetic field to this picture.

The material presented in this chapter is based on
Refs.~\cite{EFGR05,EFGR05c,EFGR07b}.

\section[The spin $BC_N$ Sutherland model]{The spin $\boldsymbol{BC_N}$ Sutherland model}
\label{S:dynamical}

In this section we shall study in some detail the spin $BC_N$ CS models
associated with the one-dimensional representations $\chi_\Eps$ of $\fW(BC_N)$.
Physically, the parameter $\ep$ determines the ferromagnetic ($\ep=1$) or
antiferromagnetic ($\ep=-1$) behavior of the associated spin chains.

Let us recall that the potential $V_\Eps$ of the spin ${BC_N}$ Sutherland model
was given in Eq.~\eqref{HBt}. For reasons that will be apparent by the end of
this chapter, it is convenient to define the \Emph{$\boldsymbol{BC_N}$
Sutherland Hamiltonian} by the action of the differential operator
\begin{equation}\label{diff.op}
H_\Eps=-\Delta+V_\Eps
\end{equation}
on the ``reduced'' Hilbert space
\begin{equation}\label{Hilbert}
\cH=L^2(C)\otimes\Sigma\,.
\end{equation}
Here the Weyl alcove~\eqref{Walcove} reads
\[
C=\big\{\bx:0<x_1<\cdots<x_N<\pi/2\big\}\,.
\]
% We shall shortly define two self-adjoint operators associated
% to $H_{\Eps}$, namely $H^{\Eps}$ and $T_{\Eps}$.

For the sake of completeness, we shall prove rigorously that this choice of
domain is perfectly admissible from a mathematical point of view. To this end,
let us define the $C^2$ function $\mu$ as in Eq.~\eqref{mut.Ch1}, so that
\begin{equation}\label{mut}
\mu(\bx)=\prod_{i<j}\big|\sin(x_i-x_j)\ms\sin(x_i+x_j)\big|^a\prod_i\big|\sin
x_i\big|^b\prod_i\big|\cos x_i\big|^{b'}\,.
\end{equation}
Here we are assuming that $a,b,b'>1/2$. It can be readily verified that
\[
H\sca\mu=E_0\mu\,,
\]
where the \Emph{scalar Hamiltonian} $H\sca$ is defined on $L^2(C)$
by Eqs.~\eqref{Hsc}-\eqref{HscBt} and
\begin{equation}\label{E0}
E_0=N\big[\tfrac23(N-1)(2N-1)\ms
a^2+2(N-1)a(b+b')+(b+b')^2\big]\,.
\end{equation}
Since $\mu$ does not vanish on $C$, $\mu$ is the ground state
function of $H\sca$~\cite{LL01}.

\begin{proposition}\label{s-a}
If $a,b,b'>2$, $H_{\Eps}$ is essentially self-adjoint on $\cD\otimes\Sigma$,
where
\[
\cD=\mu\ms\CC[\e^{±2\I x_1},\dots,\e^{±2\I x_N}]\,.
\]
\end{proposition}
\begin{proof}
A theorem of Komori and Takemura~\cite{KT02} shows that $\cD$ is a core for the
operator $H\sca$~\eqref{HscBt}. Since $H_{\Eps}-H\sca$ is $H\sca$-bounded with
relative bound
\[
\max\{(a-1)^{-1},(b-1)^{-1},(b'-1)^{-1}\}\,,
\]
the Kato--Rellich theorem~\cite{RS75} shows that $H_{\Eps}$ is essentially
self-adjoint on $\cD\otimes\Sigma$.
\end{proof}

\begin{remark}
Although its proof is simple (in fact, a direct proof could have been provided
without much additional effort, and the conditions on $a,b,b'$ could have be
easily relaxed), the inverse-square singularity prevents us from obtaining
Proposition~\ref{s-a} from the usual theorems on self-adjointness~\cite{Si00}.
Interestingly, the proof of this result is essentially algebraic, and leans on
our precise knowledge of the eigenfunctions of the $BC_N$ scalar Sutherland
model.

Although it has some applications to CS models~\cite{BG01,BGG03,FTF05},
hereafter we shall not touch upon this topic, and by ``self-adjoint'' we shall
always mean ``formally self-adjoint''.
\end{remark}

Some words on the choice of Hilbert space are in order. Consider the orthogonal
projectors
\begin{subequations}\label{symmetrizers}
\begin{align}
\La_{\Eps}&=\frac1{|\fW|}\sum_{\Pi\in\fW}\chi_\Eps(\Pi)\ms\Pi\,,\label{La.chi}\\
\La_\ep&=\frac1{N!}\sum_{\Pi\in\lan\Pi_{ij}\ran}\chi_\ep(\Pi)\ms\Pi\,,\label{La.pm}
\end{align}
\end{subequations}
which we shall call \Emph{symmetrizers}, and let
\[
\TT^N=\{\bx:-\pi/2<x_i<\pi/2\}
\]
be the flat $N$-torus. Observe that the angle coordinates of the torus are
given by $2\bx$, so that every function in $C^0(\TT^N)$ will be $\pi$-periodic
in $x_i$. Define the ``physical'' Hilbert spaces of $N$ \emph{distinguishable}
particles as
\begin{align*}
\cH_0=L^2(\TT^N)\otimes\Sigma\,.
\end{align*}
It is a standard result in representation theory~\cite{Si95} that $\cH_0$ can
be decomposed into the direct sum
\begin{equation}\label{cH0.split}
\cH_0=\bigoplus_{\ep,\ep'=±1}\cH_{\Eps}
\end{equation}
of the auxiliary spaces
\[
\cH_{\Eps}=\La_{\Eps}\cH_0\,,
\]
which are given by the image of the
symmetrizers~\eqref{symmetrizers}. We shall also be interested in
the subspaces
\begin{align}
\cH_\ep&=\cH_{\ep,+}\oplus\cH_{\ep,-}\,,\label{cHep.split}
\end{align}
i.e., the \emph{fermionic} ($\ep=-1$) and \emph{bosonic} ($\ep=+1$)
sectors of $\cH_0$.

It is technically convenient to define other two self-adjoint operators defined
by the differential operator~\eqref{diff.op}. We shall denote by $H^{\Eps}$
(resp.\ $T_{\Eps}$) the Hamiltonian defined by the action of~\eqref{diff.op} on
$\cH_0$ (resp.\ $\cH_{\Eps}$). In the next proposition we summarize some
elementary results relating the ``physical'' Hamiltonians to the ``reduced''
ones.

\begin{proposition}\label{P:Hilbert}
The following statements hold:
\begin{enumerate}
\item $\fW$ maps $\cH_{\Eps}$ into itself.

\item $\fW$ is a symmetry group of the physical Hamiltonian
$H^{\Eps}$.

\item The decompositions~\eqref{cH0.split} and~\eqref{cHep.split}
are invariant under the physical Hamiltonians.

\item $H^{\Eps}|_{\cH_{\eta\eta'}}$ and $H_{\Eps}$ are isospectral, and the
corresponding eigenfunctions $\Psi\in\cH$ and
$\Phi\in\cH_{\eta\eta'}$ are related by
\begin{equation}\label{PhiPsi}
\Phi(\bx)=\chi_{\eta\eta'}(w)\ms S_w\Psi(w\bx)\,,
\end{equation}
where $w\in\fW(BC_N)$ is such that $w\bx\in C$.
\end{enumerate}
\end{proposition}

\begin{proof}
The first part of the lemma follows most easily from Lemma~\ref{easylemma} in
the next section. To prove the remaining points it is convenient to keep the
abstract root formulation, with $g_{\bal}^2=\|\bal\|^2a_{\bal}(a_{\bal}-1)$,
and use the representation~\eqref{S.rep}. Then one can write
\begin{align}
\Pi_w H^{\Eps}&=\Pi_w\bigg[-\Delta+\sum_{\bal\in
(BC_N)_+}\|\bal\|^2a_{\bal}(a_{\bal}-\ep_{\bal}S_{\bal})\,v(\bal\cdot\bx)\bigg]\nonumber\\
&=\bigg[-\Delta+\sum_{\bal\in
(BC_N)_+}\|\bal\|^2a_{\bal}(a_{\bal}-\ep_{\bal}S_{w\bal})\,v(w\bal\cdot\bx)\bigg]\Pi_w\nonumber\\
&=\bigg[-\Delta+\sum_{\bal\in
(BC_N)_+}\|\bal\|^2a_{\bal}(a_{\bal}-\ep_{\bal}S_{\bal})\,v(\bal\cdot\bx)\bigg]\Pi_w\\
&=H^{\Eps}\Pi_w\,,\label{comm}
\end{align}
so that $[H^{\Eps},\Pi_w]=0$. Since the decompositions~\eqref{cH0.split}
and~\eqref{cHep.split} are associated with a symmetry group of $H^{\Eps}$, they
are invariant under this operator.

Each eigenfunction $\Phi$ of $H^{\Eps}$ with energy $E$ belongs to some
subspace $\cH_{\eta\eta'}$ by the $\fW$ symmetry. The potential diverging
quadratically at $\pd C\subset\TT^N$, $\Phi$ must vanish at $\pd C$, and
therefore $\Psi=\Phi|_C$ must be an $L^2$ solution to the PDE
\begin{align*}
(H_{\Eps}-E)\Psi&=0\qquad\text{in }C\,,\\
\Psi|_{\pd C}&=0\,.
\end{align*}
Hence $\Psi$ is an eigenfunction of the self-adjoint operator $H_{\Eps}$ and
$\Phi$ can be recovered from $\Psi$ through Eq.~\eqref{PhiPsi}.
\end{proof}

%PENSAR BASE Y CONTINUAR Let us make explicitly the dependence of
%$H_{\Eps}$ on the constants $a_{\bal}$ by setting
%\[
%H_{\Eps}\equiv H_{\Eps}(a_{\bal})\,,
%\]
%and similarly with the gauge Hamiltonian~\eqref{HKBt}. Define an
%isomorphism $i_{\Eps}:\cH\to\cH_0^\chi$ as
%\[
%(i_\chi\Psi)(\bx)=\chi(\si)\ms S_\si\Psi(\si\bx)\,,
%\]
%where $\si\in\fW$ and $\si\bx\in C$. It is clear that the inverse of
%this map is simply the inclusion $C\subset\TT^N$. Now we can relate
%the spectra of the models that we have introduced previously as
%follows.
%
%\begin{proposition}\label{prop.spectra}
%The Hamiltonians $H_\chi$ and $H_\chi|_{\cH_0^{\chi'}}$ are
%isospectral, and
%\begin{equation}\label{HHbar}
%H_{\Eps}(a_{\bal})|_{\cH_0^{{\Eps}'}}=\BH_{\Eps}({\Eps}(\si_{\bal})\ms
%a_{\bal})|_{\cH_0^{{\Eps}'}}\,.
%\end{equation}
%\end{proposition}
%\begin{proof}
%The isospectrality of the models follows from the fact that
%$|\fW|^{-1}i_{\Eps}$ is unitary. Eq.~\eqref{HHbar} can be proved by
%noticing that COMPLETAR
%\end{proof}
%
%As a consequence of Proposition~\ref{prop.spectra}, the concrete
%choice of a domain $\cH_0^{{\Eps}'}$ is irrelevant. This is obvious:
%it simply states that the ``Weyl chambers'' $\cH_0^{{\Eps}'}$ are
%$|\fW|$ copies of $\cH$ that cannot be distingued by means of the
%$\fW$-symmetric operator $H_{\Eps}$. The formula~\eqref{HHbar} is
%nevertheless a key in the computation of the spectrum of $H_{\Eps}$.

\begin{corollary}\label{isospectrality}
$\spec(T_{\Eps})=\spec(H_{\Eps})$.
\end{corollary}

\section[$BC_N$ Dunkl operators]{$\boldsymbol{BC_N}$ Dunkl operators}
\label{S:DunklBCN}

In this section we shall make use of the Dunkl operator formalism to compute
the spectrum of the spin model $H_{\Eps}$ in closed form. Following
Refs.~\cite{Ya95,FGGRZ03} and Eq.~\eqref{Dunkl.trig.Ch1}, let us consider the
following set of commutative \Emph{Dunkl operators}
\begin{align}\label{Dunkl}
  J_i&=\iu\,\pa_{x_i}+a\sum_{j\neq i}\Big[(1-\iu\cot
  (x_i-x_j))\,K_{ij}+(1-\iu\cot(x_i+x_j))\,\tK_{ij}\Big]\nonumber\\
  &\qquad+\big[b\,(1-\iu\cot x_i)+b'\,(1+\iu\tan
  x_i)\big]K_i-2a\sum_{j<i}K_{ij}\,.
\end{align}
The following commutation relations with the exchange operators
$K_i,K_{ij}$ can be readily checked:
\begin{subequations}
\begin{align}
\label{KijJk} &[K_{ij}, J_k] = \begin{cases}
2a(K_{ik}-K_{jk})K_{ij}\,,\quad& \text{if }i<k<j\\[1mm]
0\,,& \text{otherwise}
\end{cases}\\[2mm]
&K_{ij}J_i-J_jK_{ij} = 2a\Big(1+\sum_{i<l<j}K_{ij}K_{il}\Big),\\[2mm]
&K_{ij}J_j-J_iK_{ij} = -2a\Big(1+\sum_{i<l<j}K_{ij}K_{jl}\Big),%\\[2mm]
\end{align}
\begin{align}
&[K_i,J_j]=2aK_{ij}(K_i-K_j)\,,\qquad
[K_j,J_i]=0\,,\\[2mm]
\label{KiJi}&\{K_i,J_i\}=2(b+b')+2a\sum_{l>i}K_{il}(K_i+K_l)\,,
\end{align}
\end{subequations}
Here we assume that $i<j$ and $k$ are distinct indices in the range
$1,\dots,N$.

As an application of the Dunkl operator formalism, let us show the complete
integrability of the spin Hamiltonian $H_{\Eps}$. The proof is based on the
following lemma, which has been essentially taken from Ref.~\cite{FGGRZ03}. The
definitions of the spaces $\fD,\fK$ and $\fS$ and of the star mapping were
given in Section~\ref{S:SpinCS}.

\begin{lemma}\label{easylemma}
Let $A,B\in\fD\otimes\fK$ and $T\in\fD\otimes\fS$. Then the
following statements hold:
\begin{enumerate}
\item $A\La_\Eps=A_\Eps^*\La_\Eps$.
\item If $T\La_\Eps=0$, then $T=0$.
\item If $B$ commutes with $\La_\Eps$, then
$(AB)^*_\Eps=A^*_\Eps B^*_\Eps$.
\item If $A,B$ commute with $\La_\Eps$, then
$[A,B]^*_\Eps=[A^*_\Eps,B^*_\Eps]$.
\item If $A$ commutes with $K_w$ for some $w\in\fW(BC_N)$, then
$A^*_\Eps$ commutes with $\Pi_w$.
\end{enumerate}
\end{lemma}
\begin{proof}
\begin{enumerate}
\item Let us write
\begin{equation}\label{A.lemma}
A=\sum_{\si}D_\si K_\si\,,
\end{equation}
where $\si$ takes values in $\fW(BC_N)$. By definition,
$K_\si\La_\Eps=\chi_\Eps(\si)\ms S_\si\La_\Eps$, so
\[
A\La_\Eps=\sum_\si\chi_\Eps(\si)\ms D_\si S_\si\La_\Eps=A^*_\Eps\La_\Eps\,.
\]
\item Write $T=\sum_{|\bn|<k,\si}B_{\si,\bn}(\bx)\ms\pd^{\bn} S_\si$,
where $\bn\in\NN_0^N$ and
$\pd^{\bn}=\pd_{x_1}^{n_1}\cdots\pd_{x_N}^{n_N}$. The statement
follows by choosing a finite set of functions
$\{\vp_{\bn,\si}\}\subset\cH_\Eps$ such that the matrix
$(S_\si\pd^{\bn}\vp_{\bn',\si'}(\bx))$ is nonsingular.

\item It follows from (i) and (ii) using that
\[
AB\La_\Eps=A\La_\Eps B=A^*_\Eps\La_\Eps B=A^*_\Eps
B^*_\Eps\La_\Eps\,.
\]

\item It is a direct consequence of (iii).

\item Let us denote by $\Ad_w$ the automorphism of
$\fW(BC_N)$ given by $\si\mapsto w\si w$. By hypothesis, and using
the notation~\eqref{A.lemma},
\[
A=K_wAK_w=\sum_\si K_w(D_\si)\ms K_{\Ad_w\si}\,,
\]
so $K_w(D_\si)=D_{\Ad_w\si}$ for each $\si$. Hence
\begin{align*}
\Pi_wA^*_\Eps\Pi_w&=\sum_\si\chi_\Eps(\si)\ms K_w(D_\si)\ms
S_{\Ad_w\si}\\
&=\sum_\si\chi_\Eps(\Ad_w\si)\ms D_{\Ad_w\si}\ms
S_{\Ad_w\si}
=A^*_\Eps\,,
\end{align*}
as we wanted to prove.
\end{enumerate}
\end{proof}

Let us define the operators
\begin{equation}\label{I_p}
I_p=\sum_iJ_i^{2p}
\end{equation}
acting on $L^2(C)$. It can be readily verified that $I_1$ coincides with the
exchange Hamiltonian~\eqref{HKBt}, so that the spin Hamiltonian~\eqref{HBt} can
be recovered as a differential operator from the relation
$H_\Eps=(I_1)^*_\Eps$. Since the Dunkl operators~\eqref{Dunkl} are self-adjoint
and commute among them, so do the positive operators $I_p$.

\begin{proposition}\label{Ip.inv}
The operators~\eqref{I_p} are $\fK$-invariant.
\end{proposition}
\begin{proof}
Let us prove that the elementary permutation $K_{i,i+1}$ commutes
with $I_p$. Eq.~\eqref{KijJk} shows that $[K_{i,i+1},J_j]=0$ for
$j\neq i,i+1$. Since
\begin{gather*}
K_{i,i+1}J_i^{2p}=J^{2p}_{i+1} K_{i,i+1}
+2a\sum_{r=0}^{2p-1} J_i^{2p-r-1} J^{r}_{i+1},\\
K_{i,i+1} J^{2p}_{i+1} =J_{i}^{2p} K_{i,i+1} -2a\sum_{r=0}^{2p-1} J_i^{2p-r-1}
J^{r}_{i+1},
\end{gather*}
it stems that $[K_{i,i+1},\sum J_i^{2p}]=0$. $K_N$ also commutes
with $I_p$ by virtue of the relations
\begin{gather*}
K_NJ_i=J_iK_N,\quad\text{if }i<N\,,\\
K_NJ_N^2=-J_NK_NJ_N+2(b+b')J_N=J_N^2K_N\,.
\end{gather*}
Here we have used that $\{K_N,J_N\}=2(b+b')$. Now it immediately follows that
\[
[K_i,I_p]=K_{iN}[K_{iN}K_i,I_p] = K_{iN}[K_NK_{iN},I_p]=0
\]
for all $i=1,\dots,N$.
\end{proof}

\begin{lemma}
The operator $(I_p)^*_\Eps$ is self-adjoint.
\end{lemma}
\begin{proof}
  Proposition~\ref{Ip.inv} shows that $I_p$ commutes with $\La_\Eps$.  Using
  Lemma~\ref{easylemma} and the self-adjointness of $I_p$ and $\Lambda_{\Eps}$, one
  can see that
\[
(I_p)^*_\Eps\La_\Eps=I_p\La_\Eps=(I_p
\La_\Eps)^\dagger=[(I_p)^*_\Eps\La_\Eps]^\dagger=[\La_\Eps
(I_p)_\Eps^*]^\dagger =[(I_p)_\Eps^*]^\dagger\La_\Eps\,,
\]
so $[(I_p)_\Eps^*]^\dagger=(I_p)_\Eps^*$ as a consequence of
Lemma~\ref{easylemma}.
\end{proof}

Thus we have proved the following
\begin{theorem}
  The self-adjoint operators $(I_p)_\Eps^*$ ($p\in\NN$) form a commuting family
  which includes the Hamiltonian $H_\Eps=(I_1)^*_\Eps$.
\end{theorem}

\section{Spectrum of the dynamical model}
\label{S:sp.Hchi}

There are two steps in the computation of the spectrum of $H_\Eps$.
First, a key observation is that the gauge Hamiltonian is given by
\begin{equation}\label{BH.J}
\BH=I_1=\sum_iJ_i^2
\end{equation}
and that $J_i$ leaves invariant an infinite flag
\[
\BcH^0\subset\BcH^1\subset\cdots
\]
of finite-dimensional subspaces. The action of $\BH$ is triangular in a
suitably chosen basis and the eigenvalues can be read off the diagonal
elements. Second, a careful analysis allows us to extend these results to the
spin Hamiltonian $H_\Eps$.

Let us begin with some definitions. Let us consider the flag
\[
\BcH^0\subset\BcH^1\subset\cdots\,,
\]
where
\begin{equation}\label{BcHk}
\BcH^k=\big\lan f_{\bn}:|\bn|\leq k\big\ran\,.
\end{equation}
Here $\bn\in\ZZ^N$, $|\bn|$ is defined as in Eq.~\eqref{bn} and
\begin{equation}\label{fn}
f_\bn(\bx)=\mu(\bx)\ms\e^{2\I\ms\bn\cdot\bx}\,,
\end{equation}
with the gauge factor $\mu$  given by Eq.~\eqref{mut} and
\[
\bn\cdot\bx=\sum_in_ix_i\,.
\]
Using Fourier analysis it is easy to see that the countable union
$\bigcup_{k=1}^\infty\BcH^k$ is dense in $L^2(\TT^N)$, and that one
can obtain a Hilbert basis $\BcB$ of this space by applying the
Gram--Schmidt procedure to the functions~\eqref{fn}.

Given a multiindex $\bn\in\ZZ^N$, we shall define the non-negative,
nonincreasing multiindex
\[
[\bn]=(|n_{\pi(1)}|,\dots,|n_{\pi(N)}|)\,,
\]
where the permutation $\pi\in S_N$ is chosen so that
\[
|n_{\pi(1)}|\geq\cdots\geq|n_{\pi(N)}|\,.
\]
The set of non-negative, nonincreasing multiindices will be denoted by
$[\ZZ^N]$. We shall order $\ZZ^N$ by defining $\bn\prec\bm$ if there exists a
number $i\geq1$ such that $[\bn]_j=[\bm]_j$ for all $j<i$ and $[\bn]_i<[\bm]_i$,
where $[\bn]_j$ denotes the $j$th component of $[\bn]$. This partial order can
be naturally extended to $\{f_\bn\}$ and hence to $\BcB$. We shall assume that
the ordering of $\BcB$ is compatible with this partial order.

In the following lemma we show that the order relations do in fact
hold on orbits of $\fK$.

\begin{lemma}\label{order}
For each $w\in\fW(BC_N)$, $K_w f_\bn=f_{w\bn}$ and $[w\bn]=[\bn]$. In
particular, there exists as element $K\in\fK$ such that $Kf_\bn=f_{[\bn]}$.
\end{lemma}
\begin{proof}
The first part is trivial from the fact that $w$ is a reflection, and thus
$\bn\cdot w\bx=w\bn\cdot\bx$. The second statement holds because $\fW(BC_N)$ is
generated by $K_{ij}$ and $K_i$, i.e., by permutations and sign reversals.
\end{proof}

Now let us introduce some more notation. Given $\bn\in[\ZZ^N]$ and
$k\in\NN_0$, we define
\begin{gather*}
\#(k)\equiv\#(k,\bn)=\card\{i:n_i=k\},\\
\ell(k)\equiv\ell(k,\bn)=\min\big(\{i:n_i=k\}\cup\{+\infty\}\big).
\end{gather*}
Now one can characterize the action of the Dunkl
operators~\eqref{Dunkl} on the functions~\eqref{fn} as follows.

\begin{proposition}\label{P:Jifn}
For each $\bn\in[\ZZ^N]$, the action of $J_i$ on the function
$f_\bn$ is given by
\[
J_if_\bn =\la_{i,\bn}f_\bn+\sum_{\ZZ^N\ni\ms\bm\prec\bn}
c^{\bm}_{i,\bn}f_{\bm}\,,
\]
where
\[
\la_{i,\bn} =
\begin{cases}
2 n_i+b+b'+2a\big(N+i+1-\#(n_i)-2\ell(n_i)\big),\quad& \text{if }n_i>0\,,\\[3pt]
-b-b'+2a(i-N)\,, &\text{if } n_i=0\,,
\end{cases}
\]
and $c^\bm_{i,\bn}\in\RR$.
\end{proposition}
\begin{proof}
Let $z_i=\e^{2\I x_i},\;y_{ij}=z_iz_j^{-1}$ y $w_{ij}=z_iz_j$. A
tedious but straightforward computation shows that
\begin{equation*}
\begin{aligned}
J_if_\bn& = 2f_\bn\Bigg[ n_i+a(N-1)+\frac12\,(b+b')+a \sum_{j< i}\left(
\frac{1-y_{ij}^{n_j-n_i}}{y_{ij}-1}+\frac{1-w_{ij}^{1-n_i-n_j}}{w_{ij}-1}
\right)\\
&+a\sum_{j> i}\left(
\frac{1-y_{ij}^{1+n_j-n_i}}{y_{ij}-1}+\frac{1-w_{ij}^{1-n_i-n_j}}{w_{ij}-1}
\right) +b\,
\frac{1-z_i^{1-2n_i}}{z_i-1}-b' \frac{1+z_i^{1-2n_i}}{z_i+1}
\Bigg]\,.
\end{aligned}
\end{equation*}
It can be readily verified that all the multiindices appearing in this formula
are smaller than or equal to $\bn$. In particular, the value of $\la_{i,\bn}$
can be computed from the constant terms in the above formula.
\end{proof}

\begin{corollary}\label{C:Jifn}
For all $\bn\in\ZZ^N$,
\[
J_if_\bn=\sum_{\ZZ^N\ni\ms\bm\preceq\bn}\gamma^\bm_{i,\bn}f_\bm
\]
for some real constants $\gamma^\bm_{i,\bn}$.
\end{corollary}
\begin{proof}
Let $K\in\fK$ be defined as in Lemma~\ref{order}. Then
$J_iKf_\bn=J_if_{[\bn]}$ is given by Proposition~\ref{P:Jifn}, and
the result follows from the commutation relations
\eqref{KijJk}--\eqref{KiJi} and the invariance of the partial order
under the Weyl group (cf.~Lemma~\ref{order}).
\end{proof}

\begin{proposition}\label{P:BHfn}
For all $\bn\in\ZZ^N$ there exist some real constants $c_\bn^\bm$
such that
\[
\BH f_\bn=E_\bn f_\bn+\sum_{\ZZ^N\ni\ms\bm\prec\bn} c_\bn^\bm
f_\bm\,,
\]
where
\begin{equation}\label{En}
E_\bn=\sum_i\big(2[\bn]_i+b+b'+2a(N-i)\big)^2
\end{equation}
only depends on $[\bn]$.
\end{proposition}
\begin{proof}
By Proposition~\ref{Ip.inv}, $\BH=I_1$ is $\fK$-invariant, so $\BH
f_\bn=K\BH Kf_\bn$. By Lemma~\ref{order}, one can assume that
$Kf_\bn=f_{[\bn]}$. By Proposition~\ref{P:Jifn},
\begin{align*}
\BH
f_\bn&=K\BH f_{[\bn]}=K\sum_iJ_i^2f_{[\bn]}\\
&=\sum_i\la_{i,[\bn]}^2f_\bn+\sum_{i,\,\bm\prec
\bn}\la_{i,[\bn]}c_{i,[\bn]}^{\bm}Kf_{\bm}+\sum_{i,\,\bm\prec
\bn}c_{i,[\bn]}^{\bm}KJ_if_{\bm}\,.
\end{align*}
By Lemma~\ref{order} and Corollary~\ref{C:Jifn}, it suffices to show
that $\sum_i\la_{i,[\bn]}^2$ is given by Eq.~\eqref{En}.

There is no loss of generality in assuming that $\bn\in[\ZZ^N]$. To compute the
sum, let us gather those components of the multiindex that have the same
value. Thus, take those components of the non-negative, nonincreasing
multiindex $\bn$ such that $n_{k-1}\neq n_k=\cdots =n_{k+p}\neq n_{k+p+1}$. Then
$\ell(n_{k+j})=k$ and $\#(n_{k+j})=p+1$ for all $0\leq j\leq p$. If $n_k\neq0$,
\[
\la_{k+j,\bn}=2 n_{k+j}+b+b'+2a(N-k-p+j)= 2n_{k+p-j}+b+b'
+2a\big(N-(k+p-j)\big)\,,
\]
so that
\begin{equation}\label{eqprop}
\sum_{i=k}^{k+p}\la_{i,\bn}^2=\sum_{i=k}^{k+p}\big(2n_i+b+b'+2a(N-i)
\big)^2\,.
\end{equation}
If $n_k=0$, Proposition~\ref{P:Jifn} shows directly that
Eq.~\eqref{eqprop} also holds in this case.
\end{proof}

\begin{corollary}\label{sp.BH}
The spectrum of $\BH$ as a self-adjoint operator on $L^2(\TT^N)$ is given by
$\{E_\bn:\bn\in\ZZ^N\}$. Thus the degeneracy of each eigenvalue is at least
$|\fW|=2^NN!$.
\end{corollary}
\begin{proof}
It follows from the density of the flag~\eqref{BcHk} and the fact
that $\BH|_{\BcH^k}$ is a triangular matrix in the basis $\BcB$ with
diagonal elements given by $E_\bn$.
\end{proof}

\begin{remark}
Here we loosely (but safely) regard the spectra of the models
$H_\Eps$ as ``sets with multiplicities''. We shall use this
convention whenever appropriate.
\end{remark}

Our goal now is to extend Corollary~\ref{sp.BH} to the spin and
spinless Hamiltonians $H_\Eps$ and $H\sca$. For simplicity, let us
start with the latter, considered as an operator on $L^2(C)$.

\begin{theorem}\label{sp.Hsc}
$\spec(H\sca)=\big\{E_\bn:\bn\in[\ZZ^N]\big\}$.
\end{theorem}
\begin{proof}
By definition of the Weyl alcove, $|\fW|^{-1}$ times the inclusion
$C\subset\TT^N$ yields a unitary transformation $U$ from $L^2(C)$ to the space
of $\fK$-invariant functions $L^2(\TT^N)^\fK$. Its inverse is given by $|\fW|$
times the symmetric extension $\psi(w\bx)=\psi(\bx)$ (for all $w\in\fW(BC_N)$).
Since the action of $H\sca$ and $\BH$ agree on $\fK$-invariant functions,
$\BH|_{L^2(\TT^N)^\fK}$ and $H\sca$ are isospectral.

For each $\bn\in\ZZ^N$, the orbit $\fW(BC_N)\cdot\bn$ intersects
$[\ZZ^N]$ exactly once, at $[\bn]$. By Lemma~\ref{order}, this
implies that
\[
\big\{f_\bn:|\bn|\leq k,\;\bn\in[\ZZ^N]\big\}
\]
is a basis of $(\BcH^k)^\fK$, yielding a Hilbert basis of
$L^2(\TT^N)^\fK$ via the Gram--Schmidt procedure. By
Proposition~\ref{P:BHfn}, the matrix of $\BH|_{(\BcH^k)^\fK}$ is
triangular in an ordered basis and its eigenvalues are given by
$E_\bn$, where now $\bn\in[\ZZ^N]$.
\end{proof}

The computation of the spectrum of $H_\Eps$ is similar, although the
characterization of the bases is considerably more involved.

\begin{lemma}\label{basis.cHchi}
A basis of $\La_\Eps(\BcH^k\otimes\Si)$ is given by
\begin{equation}\label{cBEps}
\cB^k_\Eps=\big\{\La_\Eps(f_\bn\ket\bs):\bn\in[\ZZ^N],\:\ket\bs\in\cB^{\bn,\ep,\ep'}_\Si\big\}\,,
\end{equation}
where
\begin{align*}
  \cB^{\bn,\ep,\ep'}_\Si=\left\{\ket\bs\in\cB_\Si\,
    \left|
      \begin{matrix}
        &s_i-s_j\geq\delta_{-1,\ep}\hfill&\text{
          \rm when }\,n_i=n_j
        \text{ \rm and } i<j\hfill\\[1mm]
        &s_i\geq\tfrac12\delta_{-1,\ep'}\hfill&\text{ \rm when }\,n_i=0\hfill
      \end{matrix}
    \right.
  \right\}\,.
\end{align*}
\end{lemma}
\begin{proof}
Since
\begin{equation}\label{aux1}
\Pi_w\La_\Eps\Psi=\chi_\Eps(w)\La_\Eps\Psi\,,
\end{equation}
it is clear that
\[
\big\{\La_\Eps(f_\bn\ket\bs):\bn\in[\ZZ^N],\:\ket\bs\in\cB_\Si\big\}
\]
must span the whole $\La_\Eps(\BcH^k\otimes\Si)$. Moreover, a state
of the form $\La_\Eps(f_\bn\ket\bs)$ ($\ket\bs\in\cB_\Si$) vanishes
if and only if some of the following conditions hold:
\begin{enumerate}\label{conditions}
\item $n_i=n_j$, $s_i=s_j$ and $\ep=-1$.
\item $n_i=n_j=0$, $s_i=-s_j$ and $\ep=\ep'=-1$.
\item $n_i=0$, $s_i=0$ and $\ep'=-1$.
\end{enumerate}
It is easy to check that the above states do not vanish and are in fact
linearly independent modulo $\ker\La_\Eps$. Furthermore, it is immediate to see
that for any element $\La_\Eps(f_\bn\ket\bs)$ not satisfying any of the above
conditions one can use permutations and sign reversals to construct an element
$w\in\fW(BC_N)$ in the stabilizer of $\bn$ such that
$\Pi_w[\La_\Eps(f_\bn\ket\bs)]$ belongs to the basis~\eqref{cBEps}. By
~Eq.~\eqref{aux1}, the claim follows.
\end{proof}
\begin{corollary}
For each element $\La_\Eps(f_\bn\ket\bs)$ of the basis $\cB^k_\Eps$,
\begin{equation*}
  \#(n_i)\leq\begin{cases}
    2M+1\quad& \text{if }\;n_i>0\\[1mm]
    M_{\ep'}\hfill&
    \text{if }\;n_i=0\,,
\end{cases}\label{cond1}
\end{equation*}
where
\begin{equation}\label{Mep}
M_+=\lfloor M\rfloor+1\,,\qquad M_-=\lceil M\rceil\,,
\end{equation}
and $\lfloor x\rfloor$ and $\lceil x\rceil$ respectively denote the integer
part of $x$ and the smallest integer greater than or equal to $x$.
\end{corollary}

\begin{theorem}\label{sp.Hchi}
$\spec(H_\Eps)=\big\{E_\bn:\bn\in[\ZZ^N],\;\ket\bs\in\cB^{\bn,\ep,\ep'}_\Si\big\}$.
\end{theorem}
\begin{proof}
By Proposition~\ref{P:Hilbert}, $H_\Eps$ and $H^\Eps$ are
isospectral. We can assume that the basis~\eqref{cBEps} of
$\La_\Eps(\BcH^k\otimes\Si)$ is ordered according to the partial
order $\prec$. By Lemma~\ref{easylemma},
\[
H^\Eps\La_\Eps(\psi\ket\bs)=\La_\Eps\big((\BH\psi)\ket\bs\big)
\]
for each $\psi\in L^2(\TT^N)$. Now the same arguments used in the
proofs of Corollary~\ref{sp.BH} and Theorem~\ref{sp.Hsc} show that
the action of $H_\Eps$ in the basis given by Lemma~\ref{basis.cHchi}
is triangular with diagonal elements $E_\bn$.
\end{proof}

\section{The spin chain}
\label{S:chainBCN}

We define the \Emph{$\boldsymbol{BC_N}$ spin chain Hamiltonian} as
\begin{multline}\label{ssH}
8\,\ssH_\Eps=\sum_{i\neq j}\big[\sin^{-2}(\xi_i-\xi_j)\ms(1-\ep
S_{ij})+\sin^{-2}(\xi_i+\xi_j)\ms(1-\ep \tS_{ij})\big]\\
+\sum_i\big(\be\ms\sin^{-2}\xi_i+\be'\ms\cos^{-2}\xi_i\big)\ms(1-\ep'
S_i)\,,
\end{multline}
where the spin sites $0<\xi_1<\cdots<\xi_N<\pi/2$ satisfy
\begin{equation}\label{sites.BC}
\be'\tan \xi_i-\be\cot \xi_i=\sum_{j\neq
i}\big[\cot(\xi_i-\xi_j)+\cot(\xi_i+\xi_j)\big]\,.
\end{equation}
\begin{remark}
A cursory glance at the spin chain~\eqref{Ch2.HS} reveals that the
Hamiltonian~\eqref{ssH} is in fact the appropriate $BC_N$ version of
the celebrated HS chain.
\end{remark}

In this section we shall study the Hamiltonian~\eqref{ssH} and compute its
spectrum by taking the large coupling constant limit of the model~\eqref{HBt}.
To this end, let us rescale the parameters of the $BC_N$ Sutherland Hamiltonian
as
\begin{equation}\label{bb'}
b=a\be\,,\qquad b'=a\be'\,.
\end{equation}
The ground state function~\eqref{mut} can thus be written as
\[
\mu=\e^{a\la}\,,
\]
where the smooth function
\begin{equation}\label{lat}
\la(\bx)=\sum_{i<j}\big[\log\sin(x_i-x_j)+\log\sin(x_i+x_j)\big]
+\sum_i\big[\be\log\sin x_i+\be'\log\cos x_i\big]
\end{equation}
is independent of the coupling parameter $a$. It is immediate to show that
Eq.~\eqref{sites.BC} merely states that $\bxi=(\xi_1,\dots,\xi_N)\in C$ is a
critical point of $\la$. In the following proposition we shall prove an
existence and uniqueness result for the solutions of this equation.

\begin{proposition}\label{crit.BC}
The function $\la\in L^2(C)\cap C^\infty(C)$ has a unique critical
point $\bxi$ in $C$, which is a hyperbolic maximum.
\end{proposition}
\begin{proof}
Since $C$ is bounded and $\la$ tends to $-\infty$ at $\pd C$, $\la$ must attain
its maximum in $C$. Uniqueness shall follow easily from the fact that the
Hessian $D^2\la=(\pd^2\la/\pd x_i\pd x_j)$ is negative definite in the convex
domain $C$. To show the definiteness of $D^2\la$, let us compute
\begin{gather*}
\frac{\pd^2\la}{\pd x_i^2}=-\sum_{j\neq
i}\big[\sin^{-2}(x_i-x_j)+\sin^{-2}(x_i+x_j)\big]-\be\sin^{-2}x_i-\be'\cos^{-2}x_i<0\,,\\
\frac{\pd^2\la}{\pd x_i \pd x_j}=-\sin^{-2}(x_i-x_j)+\sin^{-2}(x_i+x_j)\,.
\end{gather*}
By Gerschgorin's theorem~\cite[Theorem 15.814]{GR00}, the eigenvalues of
$D^2\la(\bx)$ lie in
\[
\bigcup_i\bigg[\frac{\pd^2\la}{\pd x_i^2}-\ga_i,\frac{\pd^2\la}{\pd
x_i^2}+\ga_i\bigg]\,,
\]
with
\[
\ga_i=\sum_{j\neq i}\bigg|\frac{\pd^2\la}{\pd x_i x_j}\bigg|\,.
\]
As $\pd_{x_i}^2\la+\ga_i\leq-\be\sin^{-2}x_i-\be'\cos^{-2}x_i<0$,
$D^2\la(\bx)$ is negative definite. Hence, any critical point of
$\la$ must be a hyperbolic maximum.

To complete the proof of uniqueness, let us suppose that $\bxi_1$ and $\bxi_2$
are two critical points of $\la$. Consider the segment
$\boldsymbol\ga:t\in[0,1]\mapsto t\bxi_2+(1-t)\bxi_1\in C$ and set
$f=\la\circ\boldsymbol\ga$. Clearly $f'(0)=f'(1)=0$, so there must exist
$t\in(0,1)$ such that
\[
f''(t)=\lan\boldsymbol{\ga}'(t),D^2\la(\boldsymbol\ga(t))\ms\boldsymbol{\ga}'(t)\ran=0\,,
\]
contradicting the fact that $D^2\la$ is negative definite in $C$.
\end{proof}
\begin{remark}
It can be proved~\cite{HK06,Si83} that $\bxi$ is also a (global)
minimum of the part of the potential $V\sca$ quadratic in $a$,
namely
\begin{multline*}
U(\bx)=\sum_{i\neq j}\sin^{-2}(x_i-x_j)+\sin^{-2}(x_i+x_j)
+\be^2\sum_i\sin^{-2}x_i+\be'^2\sum_i\cos^{-2}x_i\,.
\end{multline*}
\end{remark}

Let us now consider the positive function
\begin{equation}\label{vp.BC}
\vp_0=\frac{\mu^2}{\|\mu\|^2}=\frac{\e^{2a\ms\la}}{\int_C\e^{2a\ms\la}}\in
L^1(C)\cap C^\infty(C)\,.
\end{equation}
It is intuitively obvious that $\vp_0$ concentrates near its maximum
$\bxi$ as $a\to\infty$. More precise, one can prove the following

\begin{lemma}\label{delta.BC}
Let $\bxi$ be the unique maximum of $\la$. Then $\vp_0$ converges to
$\delta_\bxi$ in the sense of distributions and falls off
exponentially fast away from $\bxi$.
\end{lemma}
\begin{proof}
We shall show that
\[
\lim_{a\to\infty}\int_C\vp_0\ms\phi=\phi(\bxi)
\]
for any $\phi\in C^\infty_0(C)$. By continuity, for any
$\varepsilon>0$ there exists $\de>0$ such that
\[
\la(\bxi)-\la(\bx)<\varepsilon\,,\qquad\big|\phi(\bxi)-\phi(\bx)\big|<\varepsilon\,,
\]
for all $\bx$ in the ball $B$ centered at $\bxi$ and of radius
$\de$. Furthermore, one can assume that $\la|_{C\backslash
B}\leq-\frac c2\de^2$, with $c\geq c_0>0$. Since the ground state
function $\mu$ is only defined up to a multiplicative constant and
the claim obviously applies to constant functions, we can assume
that $\la(\bxi)=\phi(\bxi)=0$ and $\max_C|\phi\ms|\leq1$. The method
of steepest descent ensures that
\[
\|\mu\|^2=\int_C\e^{2a\la}=\bigg(\frac{\pi}{a}\bigg)^{N/2}\frac{1+o(1)}{\det
[D^2\la(\bxi)]}
\]
as $a\to\infty$, so that
\begin{equation}\label{int.C-B}
\int_{C\backslash B}\vp_0=\|\mu\|^{-2}\int_{C\backslash
B}\e^{2a\la}\leq C_1a^{N/2}\e^{-ac\de^2}
=\cO(a^{N/2}\e^{-ac\de^2})\,.
\end{equation}
Hence
\begin{align*}
\lim_{a\to\infty}\bigg|\int_C\vp_0\ms\phi\,\bigg|&\leq\vep\lim_{a\to\infty}\int_B\vp_0+\lim_{a\to\infty}
\int_{C\backslash B}\vp_0\ms|\phi\ms|\\
&\leq \vep+\lim_{a\to\infty}\cO(a^{N/2}\e^{-ac\de^2})\\&=\vep\,,
\end{align*}
for any $\vep>0$.
\end{proof}

\begin{proposition}\label{deltan.BC}
Let $\psi$ be a normalized eigenfunction of $H\sca$. Then
$\vp=|\psi|^2\to\delta_\bxi$ as $a\to\infty$ and $\vp$ falls off
exponentially fast away from $\bxi$.
\end{proposition}
\begin{proof}
It is well known~\cite{HO87,He87,Op88,Op88b} that $g=\mu^{-1}\psi$
is a hypergeometric polynomial in $\bz=(\e^{2\I x_1},\dots,\e^{2\I
x_N}\!)$ associated with the $BC_N$ root system. Therefore,
$\|g\|^{-2}|g|^2$ is polynomially bounded in $a$ and
\[
\int_{C\backslash B}\vp\to0
\]
exponentially fast as $a\to\infty$ on account of
Lemma~\ref{delta.BC}. Since $\vp$ is obviously positive and
$\int_C\vp=1$, the previous argument shows that
\[
\lim_{a\to\infty}\int_C\vp\ms\phi=\phi(\bxi)
\]
for all $\phi\in C^\infty_0(C)$, as claimed.
\end{proof}
\begin{remark}
Finer control of the decay could have been obtained if
needed~\cite{Ag82,HS84}.
\end{remark}

We shall use the above lemma to relate the spectrum of the $BC_N$
chain with those of the $BC_N$ scalar and spin Sutherland models.
The precise relationship can be introduced by means of the partition
function of these systems, i.e., the trace of the exponential of
minus their Hamiltonian operators. In the large coupling constant
limit, we obtain an exact formula for the eigenvalues of the $BC_N$
chain Hamiltonian.

To this end, let us write the spin differential operator~\eqref{HBt}
as
\begin{equation}\label{Hh}
H_\Eps=H\sca+8a\ms\ssh_\Eps\,,
\end{equation}
where
\begin{multline}\label{ssh.BC}
\ssh_\Eps(\bx)=\sum_{i\neq j}\big[\sin^{-2}(x_i-x_j)\ms(1-\ep
S_{ij})+\sin^{-2}(x_i+x_j)\ms(1-\ep \tS_{ij})\big]\\
+\sum_i\big(\be\ms\sin^{-2}x_i+\be'\ms\cos^{-2}x_i\big)\ms(1-\ep'
S_i)
\end{multline}
is an $\End(\Si)$-valued multiplication operator and
\begin{equation}\label{ssHh.BC}
\ssH_\Eps=\ssh_\Eps(\bxi)\,.
\end{equation}

\begin{gloss}
The essence of Polychronakos's freezing trick is that the
eigenfunctions of $H_\Eps$ become sharply peaked at $\bxi$ as
$a\to\infty$, so that
\[
H_\Eps\Psi(\bx)\simeq
H\sca\Psi(\bx)+8a\ms\ssh_\Eps(\bxi)\Psi(\bx)=(H\sca+8a\ms\ssH_\Eps)\Psi(\bx)\,,
\]
in view of Eqs.~\eqref{Hh} and~\eqref{ssHh.BC}. This idea suggests
that the eigenvalue equation
\[
(H_\Eps-E_\Eps)\Psi=0
\]
should yield approximate eigenfunctions of $H\sca$ and $\ssH_\Eps$,
i.e.,
\[
(H\sca-E\sca)\Psi\simeq0\,,\qquad (\ssH_\Eps-\ssE_\Eps)\Psi\simeq0
\]
for some eigenvalues $E\sca,\ssE_\Eps$. This leads to
Polychronakos's heuristic formula~\cite{Po94}
\[
E_\Eps\simeq E\sca+8a\ms\ssE_\Eps
\]
relating the spectra of $H_\Eps$, $H\sca$ and $\ssH_\Eps$. A major
drawback of this relation is that it does not determine a priori
which eigenvalues of the spin and scalar models should be combined
to obtain an eigenvalue of the spin chain in the large coupling
constant limit. This difficulty can be
overcome~\cite{Po94,EFGR05,FG05} by summing over the spectra of
these operators and promoting the above relation to the more precise
formula~\eqref{Z} relating the partition functions of $H_\Eps$,
$H\sca$ and $\ssH_\Eps$ respectively. In what follows, we provide a
full proof of this technique and apply it to compute the spectrum of
the $BC_N$ chain.
\end{gloss}

The \Emph{partition function} of a self-adjoint operator $H$ acting on a
Hilbert space $\cH$ is the map $Z_H:\RR^+\to\RR^+$ defined by
\begin{equation}\label{partition}
Z_H(T)=\tr_{\cH_1}\e^{-H/T}=\sum_{\psi\in\cB}(\psi,\e^{-H/T}\psi)
\end{equation}
provided that the above sum converges (i.e., that $\e^{-H/T}$ is
trace class for all $T>0$.) Here $\cB$ is an orthonormal basis of
$\cH$. Clearly the above definition amounts to setting
\[
Z_H(T)=\sum_{E\in\spec(H)}\e^{-E/T}
\]
when $H$ has a pure point spectrum. We shall use the shorthand notation
$Z_\Eps\equiv Z_{H_\Eps}$, $Z\sca\equiv Z_{H\sca}$ and $\ssZ_\Eps\equiv
Z_{\ssH_\Eps}$ and define the equivalence relation
\begin{equation}\label{sim}
A\sim B\iff\lim_{a\to\infty}\frac AB=1\,.
\end{equation}
Moreover, \emph{in the rest of this chapter we shall set $E_0=0$} by
adding a constant term to the Hamiltonians $H_\Eps$ and $H\sca$. The
relation~\eqref{Hh} is not modified by this rescaling.

\begin{lemma}[Freezing trick]\label{freezing}
For all $T>0$,
\begin{equation}\label{Z}
\ssZ_\Eps(T)=\frac{\lim\limits_{a\to\infty}Z_\Eps(8aT)}{\lim\limits_{a\to\infty}Z\sca(8aT)}\,.
\end{equation}
\end{lemma}
\begin{proof}
Let $\cB\sca=\{\psi\equiv\psi(\bx;a)\}\subset L^2(C)$ and
$\cB_{\mathrm{spin}}=\{\ket\bs\}\subset\Si$ be orthonormal
eigenbases of $H\sca$ and $\ssH_\Eps$ respectively, and consider the
orthonormal basis of $\cH$ given by the tensor product of these
bases, i.e., $\cB=\{\Psi\equiv\psi\ket\bs\}$. As we have set
$E_0=0$,
\[
Z_\Eps(8aT)=\sum_{E_\Eps\in\spec(H_\Eps)}\e^{-E_\Eps/8aT}=
Z_{H_\Eps/8a}(T)\,.
\]
By Theorem~\ref{sp.Hchi}, the eigenvalues of $H_\Eps$ with $E_0=0$ depend
linearly on $a$, and hence it is not difficult to check that the sum defining
$Z_{H_\Eps/8a}$ converges uniformly on compact sets. In particular, the limit
\[
\lim_{a\to\infty}Z_\Eps(8aT)
\]
exists and depends continuously on $T\in\RR^+$. The same result
holds for the scalar Hamiltonian $H\sca$.

Let us estimate the value of $(\Psi,\e^{-H_\Eps/8aT}\Psi)$. By continuity and
the fact that $\bxi$ is the unique maximum of $\la$, for any $\vep>0$ there
exist $\vep_1>0$  and an open neighborhood $B$ of $\bxi$ such that
\begin{equation}\label{hHep}
-\vep<\ssh_\Eps-\ssH_\Eps<\vep
\end{equation}
in $B$ and $\la(\bxi)-\la>\vep_1$ in $C\backslash B$. As $H_\Eps\geq0$, it
follows from Proposition~\ref{deltan.BC} that
\[
\int_{C\backslash B}\lan\Psi,\e^{-H_\Eps/8aT}\Psi\ran\leq
\int_{C\backslash B}|\Psi|^2\to0
\]
as $a\to\infty$. Therefore,
\begin{equation}\label{Psiinfty}
\lim_{a\to\infty}(\Psi,\e^{-H_\Eps/8aT}\Psi)=\lim_{a\to\infty}\int_B\lan\Psi,\e^{-H_\Eps/8aT}\Psi\ran\,,
\end{equation}
and similarly for $H\sca+ca$ ($c\in\RR$) since it only differs from
$H_\Eps$ by $a$ times a relatively bounded operator. Moreover,
Eq.~\eqref{hHep} implies that
\[
\e^{-\frac\vep
T}\int_B\lan\Psi,\e^{-\frac{H\sca}{8aT}}\e^{-\frac{\ssH_\Eps}T}\Psi\ran<
\int_B\lan\Psi,\e^{-\frac{H_\Eps}{8aT}}\Psi\ran <\e^{\frac\vep
T}\int_B\lan\Psi,\e^{-\frac{H\sca}{8aT}}\e^{-\frac{\ssH_\Eps}T}\Psi\ran\,,
\]
and hence
\begin{equation}\label{lim}
\lim_{a\to\infty}(\Psi,\e^{-H_\Eps/8aT}\Psi)=
\lan\bs|\e^{-\ssH_\Eps/T}\ket\bs\,\lim_{a\to\infty}(\psi,\e^{-H\sca/8aT}\psi)
\end{equation}
for each $\Psi\in\cB$, on account of Eq.~\eqref{Psiinfty}. As we have uniform
convergence in compact sets we can exchange limits, so that
\begin{align*}
\lim_{a\to\infty}Z_\Eps(8aT)&=\lim_{a\to\infty}\sum_{\Psi\in\cB}(\Psi,\e^{-H_\Eps/8aT}\Psi)\\
&=\sum_{\Psi\in\cB}\lim_{a\to\infty}(\Psi,\e^{-H_\Eps/8aT}\Psi)\\
&=\sum_{\ket\bs\in\cB_{\mathrm{spin}}}\lan\bs|\e^{-\ssH_\Eps/T}\ket\bs
\sum_{\psi\in\cB\sca}\lim_{a\to\infty}(\psi,\e^{-H\sca/8aT}\psi)\\
&=\ssZ_\Eps(T)\,\lim_{a\to\infty}Z\sca(8aT)\,,
\end{align*}
as claimed.
\end{proof}

%
%The following proposition shows a slightly different approach to the
%freezing trick.
%
%\begin{corollary}\label{wlim}
%$\wlim\limits_{a\to\infty}\e^{-H_\Eps/8aT}=\e^{-\ssH_\Eps/T}\wlim\limits_{a\to\infty}
%\e^{-H\sca/8aT}$.
%\end{corollary}
%\begin{proof}
%First let us show that the above weak limits exist. Let
%$\Psi,\Phi\in\cH$ have unit norm and consider the ``matrix element''
%\[
%(\Phi,\e^{-H_\Eps/8aT}\Psi)=\int_C\lan\Phi,\e^{-H_\Eps/8aT}\Psi\ran\,.
%\]
%As the eigenvalues of $H_\Eps$ are linear in $a$ (recall that we
%have set $E_0=0$), the limit
%\[
%\lim_{a\to\infty}(\Psi',\e^{-H_\Eps/8aT}\Psi)
%\]
%clearly exists for all $\Psi,\Psi'\in\cB$. It is a standard
%result~\cite{RS80} that this implies the existence of the weak limit
%$\wlim_{a\to\infty}\e^{-H_\Eps/8aT}$. Since all the arguments in the
%proof remain valid if we replace the expectation values
%$(\Psi,\e^{-H_\Eps/8aT}\Psi)$ by the matrix elements
%$(\Psi',\e^{-H_\Eps/8aT}\Psi)$, where $\Psi,\Psi'\in\cB$, the
%corollary follows.
% Pero la base varía!
%\end{proof}

We shall now compute the partition functions of $H_\Eps$ and $H\sca$
and obtain the spectrum of the spin chain~\eqref{ssH} using the
freezing trick. A first observation is that the
eigenvalues~\eqref{En} can be written in a more compact form as
\begin{equation}\label{En.simple}
E_\bn=a^2E_0+8a\sum_in_i(\Bbe+N-i)+\cO(1)\,,
\end{equation}
where $\bn\in[\ZZ^N]$, $\Bbe=\frac12(\be+\be')$ and $E_0$ was
defined in Eq.~\eqref{E0}.

\begin{proposition}\label{Zsca}
Set $q=\e^{-1/T}$. Then
\[
\lim_{a\to\infty}Z\sca(8aT)=\prod_i\Big[1-q^{i[\Bbe+N-\tfrac12(i+1)]}\Big]^{-1}\,.
\]
\end{proposition}
\begin{proof}
By Theorem~\ref{sp.Hsc} and Eq.~\eqref{En.simple},
\[
Z\sca(8aT)=\sum_{\bn\in[\ZZ^N]}\e^{-E_\bn/8aT}\sim\sum_{\bn\in[\ZZ^N]}
\prod_iq^{n_i(\Bbe+N-i)}\,.
\]
Defining $p_i=n_i-n_{i+1}$ for $1\leq i\leq N-1$ and $p_N=n_N$, we
have
\[
\prod_i q^{n_i(\Bbe+N-i)}=\prod_{i\leq j}q^{p_j(\Bbe+N-i)}=\prod_j
q^{p_j\sum\limits_{i=1}^j(\Bbe+N-i)}=\prod_j q^{j
p_j[\Bbe+N-\tfrac12(j+1)]}\,,
\]
and hence
\begin{align}
Z\sca(aT)&\sim \sum_{\mathbf p\in\NN_0^N}\,\prod_i q^{i
p_i[\Bbe+N-\tfrac12(i+1)]}
=\prod_i \sum_{p_i\geq0} q^{i p_i[\Bbe+N-\tfrac12(i+1)]}\notag\\
&=\prod_i\bigg[1-q^{i[\Bbe+N-\tfrac12(i+1)]}\bigg]^{-1}\,.
\end{align}
\end{proof}

The computation of the partition function of the spin model is
considerably more involved. First, let us define the set of
\Emph{partitions} of a natural number $n$ as
\[
\fP_n=\bigcup_{r=1}^n\big\{\bk\in\NN^r:|\bk|=n\big\}\,,
\]
where $|\bk|=k_1+\dots +k_r\,$. We find it convenient to represent each positive
nonincreasing multiindex $\bn\in[\ZZ^N]$ as
\begin{equation}\label{nm}
\bn = \big(\overbrace{\vphantom{1}m_1,\dots,m_1}^{k_1},
\overbrace{\vphantom{1}m_2,\dots,m_2}^{k_2},\dots,
\overbrace{\vphantom{1}m_r,\dots,m_r}^{k_r}\big),
\end{equation}
where $m_1>m_2>\cdots>m_r\geq0$ and $k_i=\#(m_i)$. In particular,
$\bk\in\fP_N\cap\NN^r$ and condition~\eqref{cond1} is satisfied if
$\ep=-1$.
\begin{example}
$\fP_3=\big\{(3),(2,1),(1,2),(1,1,1)\big\}$. For
$\bn=(6,3,3,3,2,1,1)\in[\ZZ^7]$ one has $\bm=(6,3,2,1)\in[\ZZ^4]$
and $\bk=(1,3,1,2)\in\fP_7\cap\NN^4$.
\end{example}

\begin{proposition}\label{P:Z.BC}
Given $\bk\in\fP_N$, set
\begin{equation}\label{Nj}
N_j\equiv N_j(\bk)=\Big(\sum_{i=1}^jk_i\Big)\Big(\Bbe
+N-\frac12-\frac12\sum_{i=1}^jk_i\Big)\,,
\end{equation}
where $1\leq j\leq r$ and $r\equiv r(\bk)$ is the length of $\bk$.  Then one can write
the large $a$ limit of the partition function of the Hamiltonian~\eqref{HBt}
as
\begin{multline}\label{Z.BC}
  \lim_{a\to\infty}Z_{\ep\ep'}(8aT)=\sum_{\bk\in\fP_N}\bigg\{
  \prod_{j=1}^{r-1}\bigg[{\binom{2M+1+\de_{1,\ep}(k_j-1)}{k_j}}
\,\frac{q^{N_j}}{1-q^{N_j}}\bigg]\\
×\bigg[{\binom{M_{\ep'}+\de_{1,\ep}(k_r-1)}{k_r}
+\binom{2M+1+\de_{1,\ep}(k_r-1)}{k_r}}
\,\frac{q^{N_r}}{1-q^{N_r}}\bigg]\bigg\}\,.
\end{multline}
\end{proposition}
\begin{proof}
It follows from Eq.~\eqref{nm} that the
eigenvalues~\eqref{En.simple} can be written as
\begin{align*}
E_\bn&\sim8a\sum_{i=1}^rm_i\sum_{j=
    k_1+\cdots+k_{i-1}+1}^{k_1+\cdots+k_{i-1}+k_i}(\Bbe+N-j)\\
  &=8a\sum_{i=1}^r m_i
  k_i\bigg(\Bbe+N-\frac12-\frac{k_i}2-\sum_{j=1}^{i-1}k_j\bigg)\\
  &\equiv8a\sum_{i=1}^r m_i\nu_i\,,
\end{align*}
since we have set $E_0=0$. The degeneracy $d_\Eps(\bn)\equiv
d_\Eps(\bk,\bm)$ of each eigenvalue $E_\bn$ is given by the cardinal
of the set $\cB^{\bn,\ep,\ep'}_\Si$ introduced in
Lemma~\ref{basis.cHchi}. It is not difficult to check that
$d_\Eps(\bk,\bm)$ solely depends on whether $m_r=0$ or not, and is
in fact given by
\begin{align*}
d_{\ep\ep'}(\bk,\bm)&=\begin{cases}
\prod\limits_{j=1}^r\binom{2M+1+\de_{1,\ep}(k_j-1)}{k_j}\equiv d_\Eps^+(\bk)\,,\qquad & \text{if }m_r>0\,,\\[2mm]
\binom{M_{\ep'}+\de_{1,\ep}(k_r-1)}{k_r}\prod\limits_{j=1}^{r-1}
\binom{2M+1+\de_{1,\ep}(k_j-1)}{k_j}\equiv d_\Eps^0(\bk) &\text{if
}m_r=0\,.
\end{cases}
\end{align*}
The partition function can thus be written as
\begin{align*}
Z_{\ep\ep'}(8aT)&=\sum_{\bn\in[\ZZ^N]}d_\Eps(\bn)\ms\e^{-E_\bn/8aT}\\
&\sim\sum_{\bk\in\fP_N}
  \sum_{m_1>\cdots>m_r\geq0} d_{\ep\ep'}(\bk,\bm)
  \prod_{i=1}^r q^{m_i\nu_i}\\
  &{}=\sum_{\bk\in\fP_N}\Big[
  \sum_{m_1>\cdots>m_r>0}d_{\ep\ep'}^+(\bk)\prod_{i=1}^r q^{m_i\nu_i}
  +%\sum_{\bk\in\fP_N}
  \sum_{m_1>\cdots>m_{r-1}>0} d_{\ep\ep'}^0(\bk)
  \prod_{i=1}^{r-1} q^{m_i\nu_i}\Big]\,.
\end{align*}
Since
\begin{align}
\sum_{m_1>\cdots>m_s>0}\prod_{i=1}^s q^{m_i\nu_i}
&=\sum_{\bp\in\NN^s}\prod_{i=1}^s q^{\nu_i \sum\limits_{j=i}^s p_j}
=\sum_{\bp\in\NN^s}\prod_{i=1}^s\prod_{j=i}^s q^{p_j\nu_i}\notag\\
&=\sum_{\bp\in\NN^s}\prod_{j=1}^s q^{p_j\sum\limits_{i=1}^j\nu_i}
=\prod_{j=1}^s\sum\limits_{p_j>0}q^{p_j\sum\limits_{i=1}^j\nu_i}\notag\\
&={\ds\prod\limits_{j=1}^s\frac{q^{N_j}}{1-q^{N_j}}}\,,\label{sum.mi}
\end{align}
we easily obtain Eq.~\eqref{Z.BC}.
\end{proof}

From Propositions~\ref{Zsca} and~\ref{P:Z.BC} we immediately derive the
following theorem, which is the main result of this chapter.

\begin{theorem}\label{T:ssZ.BC}
The partition function of the spin chain~\eqref{ssH} is given by
\begin{multline}
  \ssZ_{\ep\ep'}(T)=\prod_{i=1}^N
  \bigg[1-q^{i[\Bbe+N-\tfrac12(i+1)]}\bigg]\\
  \sum_{\bk\in\fP_N}\bigg\{
  \bigg[\binom{M_{\ep'}+\de_{1,\ep}(k_r-1)}{k_r}+{\binom{2M+1+\de_{1,\ep}(k_r-1)}{k_r}}
  \frac{q^{N_r}}{1-q^{N_r}}\bigg]\\
  \prod_{j=1}^{r-1}\bigg[{\binom{2M+1+\de_{1,\ep}(k_j-1)}{k_j}}\,\frac{q^{N_j}}{1-q^{N_j}}\bigg]\bigg\}\,.
\label{Zchainfinal}
\end{multline}
\end{theorem}
\begin{remark}
The eigenvalues $\ssE$ of the spin chain Hamiltonian and their degeneracies
$d(\ssE)\equiv d_\Eps(\ssE)$ can be recovered by identifying
\begin{equation}\label{sse.ident}
\ssZ_\Eps(T)=\sum_{\ssE\in\ms\spec(\ssH_\Eps)} d(\ssE)\ms q^\ssE\,,
\end{equation}
where the sum runs over distinct eigenvalues.
\end{remark}

\section{Statistical properties of the spectrum}
\label{S:stat.BCN}

Several remarkable properties of the spectrum of the spin
chain~\eqref{ssH} can be inferred from Theorem~\ref{T:ssZ.BC}. First
of all, for half-integer spin the partition
function~\eqref{Zchainfinal} does not depend on $\ep'$, since in
this case $M_±=M+\frac12$. Hence the spectrum of the spin chain is
independent of $\ep'$ when $M$ is a half-integer, a property that is
not apparent from the expression of the Hamiltonian~\eqref{ssH}.
Secondly, all the denominators $1-q^{N_j}$, $1\leq j\leq r$,
appearing in Eq.~\eqref{Zchainfinal} are included as factors in the
product in the first line. Hence the partition
function~\eqref{Zchainfinal} can be rewritten as
\begin{equation}\label{Zsimple}
\ssZ_{\ep\ep'}(T)=\sum_{\bde\in\{0,1\}^N}d_{\ep\ep'}(M,\bde)\,q^{\ssE_\bde}\,,
\end{equation}
where $\ssE_\bde$ is given by
\begin{equation}\label{ssEbde}
\ssE_\bde=\sum_{i=1}^Ni\ms\de_i\big(\Bbe+N-\frac12\,(i+1)\big)\,,
\end{equation}
and the degeneracy factor $d_{\ep\ep'}(M,\bde)$ is a polynomial of
degree $N$ in $M$. Therefore,

\begin{proposition}\label{several.BC}
For all values of $\Bbe$, $\ep$, $\ep'$ and $M$, the following
statements hold:
\begin{enumerate}
\item For half-integer spin, $\ssH_\Eps$ and $\ssH_{\ep,-\ep'}$ are
isospectral.

\item
$\spec(\ssH_\Eps)\subset\big\{\ssE_\bde:\bde\in\{0,1\}^N\big\}$, and
it exactly coincides with this set (of cardinal at most $2^N$) for
%generic values of $\Bbe$ and
sufficiently large $M$.

\item Set $\ssE_{\max}=\tfrac16\,N(N+1)(2N+3\Bbe-2)$. Then
\begin{equation}\label{-ep-ep'}
\ssH_\Eps=\ssE_{\max}-\ssH_{-\ep,-\ep'}
\end{equation}
\end{enumerate}
\end{proposition}
\begin{proof}
The first two statements follow from the above discussion. Let us prove the
last claim. First, observe that the above equation holds with the constant
\begin{multline}\label{Emax.xi}
4\ms\ssE_{\max}=\sum_{i\neq
j}\big[\sin^{-2}(\xi_i-\xi_j)+\sin^{-2}(\xi_i+\xi_j)\big]
+\sum_i\big(\be\ms\sin^{-2}\xi_i+\be'\ms\cos^{-2}\xi_i\big)
\end{multline}
by the definition of the spin chains. Since $\ssH_\Eps\geq0$ and
$\ker(\ssH_\Eps)=\La_{\ep\ep'}(\Si)$ is nontrivial for sufficiently high $M$,
$\ssE_{\max}$ must coincide with the highest possible eigenvalue of
$\ssH_{-\ep,-\ep'}$, which can be easily obtained using Eq.~\eqref{ssEbde}:
\[
\ssE_{\max}=\sum_{i=1}^N i\big(\Bbe+N-\frac12\,(i+1)\big)
=\frac16\,N(N+1)(2N+3\Bbe-2)\,.
\]
Note, however, that in the antisymmetric case some of the possible energies may
not be attained if $M$ is kept fixed and $N$ increases. For instance, it is
obvious that the kernel $\ker(\ssH_{-,\ep})=\La_{-,\ep}(\Si)$ becomes trivial
in this case.
\end{proof}
\begin{remark}
Eq.~\eqref{-ep-ep'} allows us to restrict our attention to the
ferromagnetic chains $\ssH_{-,±}$ without loss of generality.
Moreover, $\ssE_{\max}=\max\spec(\ssH_\Eps)$.
\end{remark}

We shall now present several concrete examples where we analyze the
spectrum of the chains by means of Theorem~\ref{T:ssZ.BC}.

\begin{example}\label{ex1}
The structure of Eq.~\eqref{Zchainfinal} makes it straightforward to
compute the spectrum of the spin chains for any fixed number of
particles as a function of the spin. For instance, for $N=3$ sites
and integer $M$ the eigenvalues of the spin chain $\ssH_{--}$ are
$0,\,\Bbe+2,\,2\Bbe+3,\,3\Bbe+3,\,3\Bbe+5,\,4\Bbe+5,\,
5\Bbe+6,\,6\Bbe+8$, with respective degeneracies
\begin{align*}
&\tfrac16\,M(M-1)(M-2),\;\tfrac56\,M(M^2-1),\;\tfrac16\,M(M+1)(11M-2),\\[1mm]
&\tfrac16\,M(M+1)(7M-4),\;\tfrac16\,M(M+1)(7M+11),\;\tfrac16\,M(M+1)(11M+13),\\[1mm]
&\tfrac56\,M(M+1)(M+2),\;\tfrac16\,(M+1)(M+2)(M+3).
\end{align*}
Note that in this case $\spec(H_{--})=\{\ssE_\bde:\bde\in\{±1\}\}$ for $M\geq
3$, in agreement Proposition~\ref{several.BC}. The above results have been
numerically checked for $M=1$ and a few values of $\be$ and $\be'$ by
representing the operators $S_{ij}$ and $S_i$ as $27× 27$ matrices. The
obtained results are in complete agreement with those listed above.
\end{example}

For a fixed value of the spin $M$, it is not apparent how to derive an explicit
formula expressing the eigenvalues and their multiplicities in terms of the
number of particles $N$. We shall next present two concrete examples for the
cases $M=1/2$ and $M=1$.

\begin{example}[Spin 1/2]\label{ex2}
By Proposition~\ref{several.BC}, $\ssH_{-+}$ and $\ssH_{--}$ are isospectral.
We have computed the partition function $\ssZ_{-,±}$ for up to 20 particles.
For instance, for $N=6$ the antiferromagnetic spin chain eigenvalues and their
corresponding multiplicities (denoted by subindices) are
\begin{align*}
&(9\Bbe+32)_2,\: (10\Bbe+36)_2,\: (11\Bbe+38)_2,\: (11\Bbe+41)_4,\:
(12\Bbe+38)_1,\:
(12\Bbe+43)_6,\\
&(13\Bbe+43)_3,\: (13\Bbe+46)_6,\: (14\Bbe+46)_4,\: (14\Bbe+50)_4,\:
(15\Bbe+47)_2,\:
(15\Bbe+50)_3,\\
& (15\Bbe+55)_6,\: (16\Bbe+51)_2,\: (16\Bbe+55)_5,\:
(17\Bbe+53)_1,\: (17\Bbe+56)_4,\:
(18\Bbe+58)_3,\\
& (19\Bbe+61)_2,\: (20\Bbe+65)_1,\: (21\Bbe+70)_1.
\end{align*}
The number of energy levels increases rapidly with the number of particles $N$.
For example, if $N=10$ the number of distinct eigenvalues (for generic values
of $\Bbe$) is $136$, while for $N=20$ this number becomes $7756$. It is
therefore convenient to plot the eigenvalues $\ssE$ and their corresponding
degeneracies $d(\ssE)$, as is done in Fig.~\ref{spin12N10ed} for $N=10$
particles. Note that Eq.~\eqref{ssEbde} implies that when $\Bbe\gg N$ the
levels cluster around integer multiples of $\Bbe$. In fact, for all $N$ up to
$20$ we have observed that these integers take \emph{all} values in a certain
range $j_0,j_0+1,\ldots,N(N+1)/2$. For example, in the case $N=6$ presented
above, $j_0=9$.
\begin{figure}[t]
\centering \psfrag{e}{$\ssE$} \psfrag{d}{$d(\ssE)$}
\includegraphics[width=11cm]{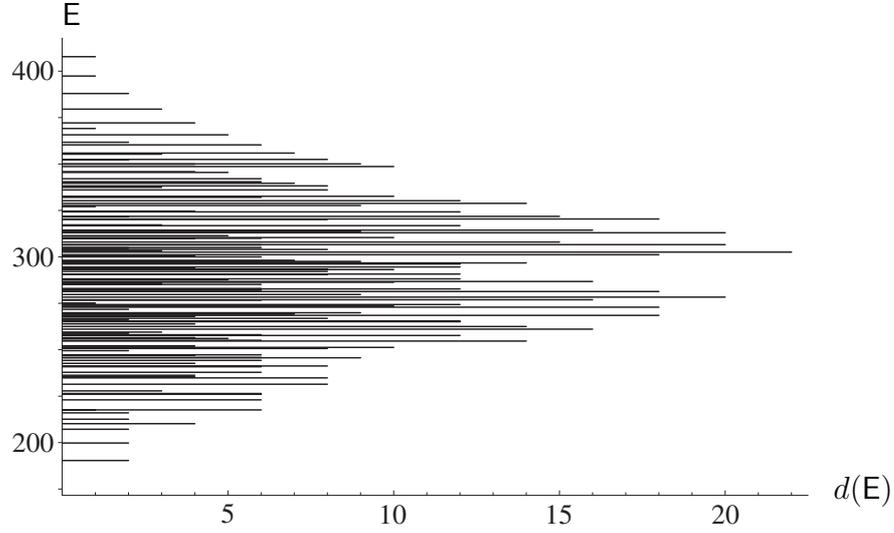}
\begin{quote}
\caption{Eigenvalues $\ssE$ and degeneracies $d(\ssE)$ of the spin
$1/2$ chain $\ssH_{-,±}$ for $N=10$ particles and $\Bbe=\sqrt
2$.\label{spin12N10ed}}
\end{quote}
\end{figure}
\end{example}

\begin{example}[Spin 1]
We have computed the partition functions $\ssZ_{-,±}$ of the spin chains
$\ssH_{-,±}$ with spin $M=1$ for up to $15$ particles. As remarked in the
previous section, for integer $M$ the partition functions $\ssZ_{-,±}$ are
expected to be essentially different. This is immediately apparent from
Fig.~\ref{spin1N10ed}, where the energy spectra of the even and odd spin chains
$\ssZ_{-,±}$ with $\Bbe=\sqrt 2$ for $N=10$ particles are graphically
compared. However, the standard deviation of the energy is exactly the same for
both chains. This rather unexpected result will be relevant in the ensuing
discussion of the level density (see Conjecture~\ref{conj2} below). We also
note that, just as for spin $1/2$, for $N$ up to (at least) $15$ and
\mbox{$\Bbe\gg N$} the energy levels cluster around an equally spaced set of
nonnegative integer multiples of $\Bbe$.
\begin{figure}[t]
\centering\psfrag{e}{$\ssE$} \psfrag{d}{$d(\ssE)$}
\includegraphics[width=11cm]{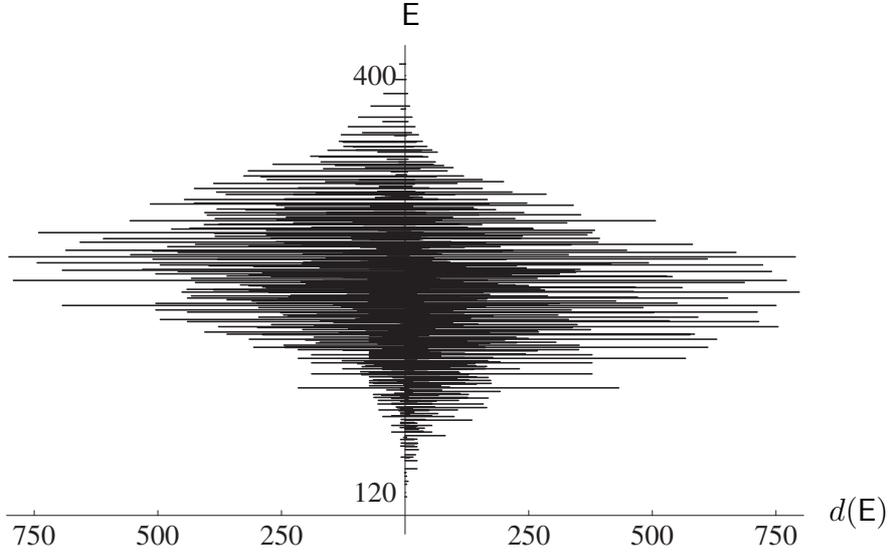}
\begin{quote}
\caption{Comparison of the energy levels $\ssE$ and degeneracies $d(\ssE)$ of
the spin $1$ chains $\ssH_{--}$ (left) and $\ssH_{-+}$ (right) for $N=10$
particles and $\Bbe=\sqrt 2$.\label{spin1N10ed}}
\end{quote}
\end{figure}
\end{example}

The numerical explorations that we have carried off naturally give
rise to several conjectures that we shall now discuss in detail.

\begin{conjecture}\label{conj1}
For $\Bbe\gg N$, the eigenvalues cluster around an equally spaced
set of levels of the form $j\ms\Bbe$, with
$j=j_0,j_0+1,\ldots,N(N+1)/2$
\end{conjecture}
\begin{remark}
For sufficiently large values of the spin $M$ this conjecture (with
$j_0=0$) follows directly from Eq.~\eqref{ssEbde}. Numerical
calculations for a wide range of values of $N$ and $M$ fully
corroborate the above conjecture.
\end{remark}

\begin{conjecture}\label{conj2}
For $N\gg 1$, the level density follows a Gaussian distribution
\end{conjecture}

\begin{figure}[t]
\centering\psfrag{F}{$F_{\cN}(\ssE),\,F(\ssE)$} \psfrag{e}{$\ssE$}
\includegraphics[width=11cm]{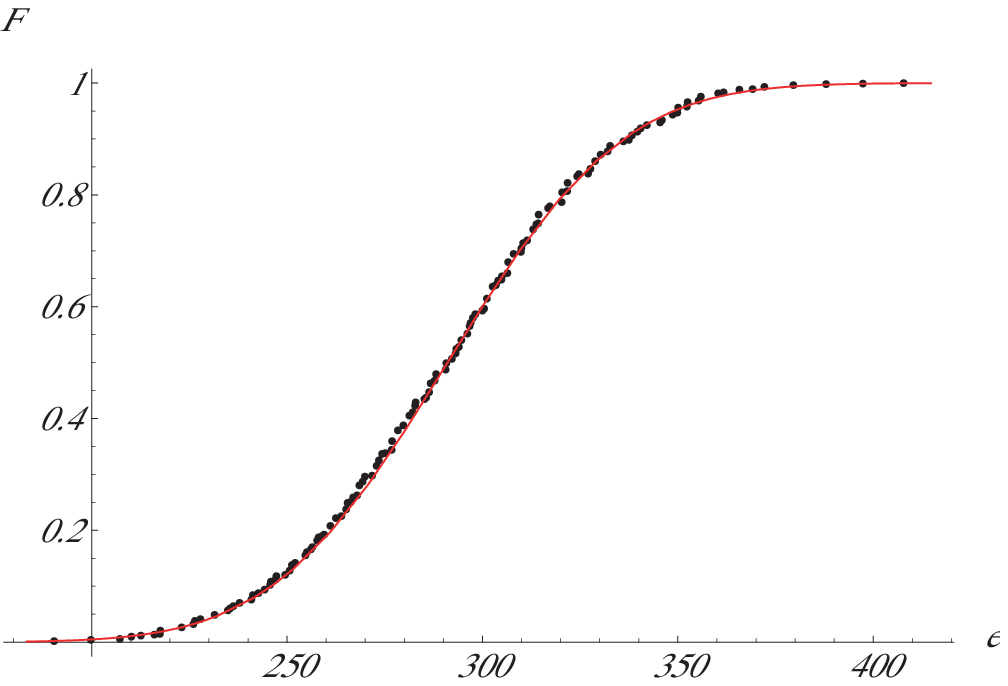}
\begin{quote}
\caption{Distribution functions $F_{\cN}(\ssE)$ (continuous line)
and $F(\ssE)$ (at its discontinuity points) for $\Bbe=\sqrt 2$,
$M=1/2$, and $N=10$.\label{normalspin12N10}}
\end{quote}
\end{figure}

More precisely, we claim that the number of eigenvalues (counting
their degeneracies) in an interval $I\subset\RR$ is approximately
given by
\begin{equation}\label{Numlevels}
(2M+1)^N \int_I \cN(\ssE;\mu,\si)\ms\dd\ssE\,,
\end{equation}
where
\begin{equation}\label{gaussian}
\cN(\ssE;\mu,\si)=\frac1{\si\sqrt{2\pi}}\,\e^{-\frac{(\ssE-\mu)^2}{2\si^2}}
\end{equation}
is the normal (Gaussian) distribution with parameters $\mu$ and
$\sigma$ respectively equal to the mean and standard deviation of
the spectrum of the spin chain. Although the shape of the plots in
Figs.~\ref{spin12N10ed} and~\ref{spin1N10ed} make this conjecture
quite plausible, for its precise numerical verification it is
preferable to compare the distribution function
\begin{equation}\label{FcN}
F_{\cN}(\ssE)=\int_{-\infty}^\ssE \cN(\sse;\mu,\si)\dd\sse
\end{equation}
of the Gaussian probability density with its discrete analog
\begin{equation}\label{F}
F(\ssE)=(2M+1)^{-N}\sum_{\ssE\geq\sse\in\spec(H_\Eps)} d(\sse)\,,
\end{equation}
where $d(\sse)\equiv d_\Eps(\sse)$ denotes the degeneracy of the eigenvalue
$\sse$. Indeed, our computations for a wide range of values of $M$ and
$N\gtrsim 10$ are in total agreement with the latter conjecture for all four
chains~\eqref{ssH}. This is apparent, for instance, in the case $\Bbe=\sqrt 2$,
$M=1/2$, and $N=10$ presented in Fig.~\ref{normalspin12N10}. The agreement
between the distribution functions~\eqref{FcN} and~\eqref{F} improves
dramatically as $N$ increases. In fact, their plots are virtually
indistinguishable for $N\gtrsim 15$.

It is well known in this respect that a Gaussian level density is a
characteristic feature of the embedded Gaussian ensemble (EGOE) in Random
Matrix Theory~\cite{MF75}. It should be noted, however, that the EGOE applies
to a system of $N$ particles with up to $n$-body interactions ($n<N$) in the
high dilution regime $N\to\infty$, $\ka\to\infty$ and $N/\ka\to 0$, where $\ka$
is the number of one-particle states. Since in our case $\ka=2M+1$ is fixed,
the fact that the level density is Gaussian does not follow from the above
general result.

If Conjecture~\ref{conj2} is true, the spectrum for large $N$ is
completely characterized by the parameters $\mu$ and $\si$ through
the Gaussian law~\eqref{gaussian}. It is therefore of great interest
to compute these parameters in closed form as functions of $N$ and
$M$. To this end, let us write
\begin{equation}\label{hmm}
\ssH_{-,±}=\sum_{i\neq
j}\Big[h_{ij}(1+S_{ij})+\tih_{ij}(1+\tS_{ij})\Big] +\sum_i h_i(1\mp
S_i)\,,
\end{equation}
where the constants $h_{ij}$, $\tih_{ij}$ and $h_i$ can be easily
read off from Eq.~\eqref{ssH}. We shall denote by $\mu_±$ the
average energy of $\ssH_{-,±}$.

\begin{proposition}\label{P:mu.BC}
If the spin $M$ is an integer,
\begin{gather}
\mu_-=\frac{M+1}{6(2M+1)}\,N(N+1)(2N+3\Bbe-2)\label{mum}\,,\\
\mu_+=\frac1{2M+1}\,\big[(M+1)\ssE_{\max}-2\ms\Si_1\big]\label{mup}\,,
\end{gather}
with $\Si_1=\sum_ih_i$. If it is a half-integer,
\begin{equation}\label{mupmhalf}
\mu_{±}=\frac1{2M+1}\,\big[(M+1)\ssE_{\max}-\Si_1\big]\,.
\end{equation}
\end{proposition}
\begin{proof}
We shall begin with the case of integer spin. Using the formulas for the traces
of the spin operators given in Table~\ref{traces} we immediately obtain
\begin{align*}
\mu_-&=(2M+1)^{-N}\tr\ssH_{--}\notag\\
&=\frac{2(M+1)}{2M+1}\,\Big[\sum_{i\neq j}(h_{ij}+\tih_{ij})+\sum_i
h_i\Big]
=\frac{M+1}{2M+1}\ssE_{\max}\notag\\
&=\frac{M+1}{6(2M+1)}\,N(N+1)(2N+3\Bbe-2)\,.
\end{align*}
Here we have used Proposition~\ref{several.BC} and
Eq.~\eqref{Emax.xi} to compute the sum.
\begin{table}[t]
\centering\caption{Traces of products of the spin
operators.}\label{traces}
\begin{tabular}{lll}\hline
  \vrule height 13pt depth 6pt width0pt Operator & Trace (integer $M$)
  & Trace (half-integer $M$)\\ \hline
  \vrule height 13pt depth 6pt width0pt
  $S_i\ms$ & $(2M+1)^{N-1}$ & 0\\ \hline
  \vrule height 13pt depth 6pt width0pt
  $S_{ij}\ms$,\, $\tS_{ij}$ & $(2M+1)^{N-1}$ & $(2M+1)^{N-1}$\\
\hline
\vrule height 13pt depth 6pt width0pt
$S_iS_j$ & $(2M+1)^{N-2+2\de_{ij}}$ & $(2M+1)^N\de_{ij}$\\ \hline
\vrule height 13pt depth 6pt width0pt
$S_{ij}S_k\ms$,\, $\tS_{ij}S_k\ms$ & $(2M+1)^{N-2}$ & 0\\ \hline
\vrule height 13pt depth 6pt width0pt
$S_{ij}\tS_{kl}$ & $(2M+1)^{N-2}$ &
$(2M+1)^{N-2}(1-\de_{ik}\de_{jl})(1-\de_{il}\de_{jk})$\\ \hline
\vrule height 13pt depth 6pt width0pt
$S_{ij}S_{kl}\ms$,\, $\tS_{ij}\tS_{kl}$ &
$(2M+1)^{N-2+2\de_{ik}\de_{jl}+2\de_{il}\de_{jk}}$ &
$(2M+1)^{N-2+2\de_{ik}\de_{jl}+2\de_{il}\de_{jk}}$\\ \hline
\end{tabular}
\end{table}
On the other hand, the average energy $\mu_{+}$ of the chain
$\ssH_{-+}$ is given by
\begin{align*}
\mu_+&=(2M+1)^{-N}\tr\ssH_{-+}\notag\\
&=\frac{2(M+1)}{2M+1}\,\sum_{i\neq j}(h_{ij}+\tih_{ij})
+\frac{2M}{2M+1}\,\sum_i h_i\notag\\
&=\frac1{2M+1}\,\big[(M+1)\ssE_{\max}-2\ms\Si_1\big]\,,\qquad\text{$M$
integer,}\label{mup}
\end{align*}
where $\Si_1=\sum_i h_i$.

When $M$ is a half-integer, $\mu_+=\mu_-$ due to
Proposition~\ref{several.BC}. A similar calculation shows that the
formulas for the traces of the spin operators in Table~\ref{traces}
yield the expression~\eqref{mupmhalf} for the mean energy.
\end{proof}

Let us turn now to the (squared) standard deviation of the spectrum
of $\ssH_{-,±}$, which is given by
\[
\si_±^2=\frac{\tr(\ssH_{-,±}^2)}{(2M+1)^N}
-\frac{(\tr\ssH_{-,±})^2}{(2M+1)^{2N}}\,.
\]
A long but straightforward calculation using the formulas in
Table~\ref{traces} yields the following

\begin{proposition}\label{P:std}
For integer spin, the standard deviation is given by
\begin{equation}
\si_±^2=\frac{4M(M+1)}{(2M+1)^2}\,\Si_2\,,\label{sigmapm}
\end{equation}
with $\Si_2=2\sum\limits_{i\neq
j}(h_{ij}^2+\tih_{ij}^2)+\sum\limits_ih_i^2$. If $M$ is a
half-integer, then
\begin{equation}\label{sipmhalf}
\si_±^2=\frac{4M(M+1)}{(2M+1)^2}\,\Big[\Si_2+\frac{\Si_3}{M(M+1)}\Big]\,,
\end{equation}
where $\Si_3=\frac14\,\sum\limits_i h_i^2-\sum\limits_{i\neq
j}h_{ij}\tih_{ij}$.
\end{proposition}

Since $\Si_1$, $\Si_2$ and $\Si_3$ do not depend on $M$, the above propositions
completely determine the dependence of $\mu_±$ and $\si_±$ on the spin. A
nontrivial consequence of Proposition~\ref{P:std} is the equality of the
standard deviation of the energy for the even and odd antiferromagnetic chains
(for half-integer spin, this trivially follows from the fact that the even and
odd chains have the same spectrum). This result is quite surprising, since for
integer spin the energy spectra of the chains $\ssH_{-,±}$ are essentially
different, as shown in Fig.~\ref{spin1N10ed}.

To analyze the dependence on the number of particles of the spectrum
we still need to evaluate $\Si_1$, $\Si_2$ and $\Si_3$ as functions
of $N$. Although it is not at all apparent how to compute these
quantities in closed form, in view of
Eqs.~\eqref{mum}--\eqref{mupmhalf} one is lead to formulate a
natural

\begin{conjecture}\label{conj3}
The average energy $\mu_±$ and its squared standard deviation
$\si_±^2$ depend polynomially on $N$.
\end{conjecture}

In fact, since $\ssE_{\max}$ is a polynomial of degree $3$ in $N$ by
Proposition~\ref{several.BC}, it follows that the degrees in $N$ of $\mu_±$
and $\si_±^2$ cannot exceed $3$ and $6$, respectively. The latter conjecture
and this fact allow us to determine the quantities $\Si_i(N)$ by evaluating
$\mu_±$ and $\si_±^2$ for $N=2,\dots,8$ and $M=1/2,1$ using
Theorem~\ref{T:ssZ.BC} (cf.~Eqs.~\eqref{mup}, \eqref{sigmapm}
and~\eqref{sipmhalf}). Hence one immediately proves the following
\begin{proposition}\label{P:Si}
If Conjecture~\ref{conj3} is true, then
\begin{align}
\Si_1&=8\Si_3=\frac N4\,(2\Bbe+N-1)\,,\notag\\
\Si_2&=\frac{N}{144}\ms \big[\,2(2N^2+3N+13)\ms\Bbe^{\,2}+
(N-1)(5N^2+7N+20)\ms\Bbe\notag\\
&\hphantom{{}=\frac{4N}9\,\big[\,2(2N^2+3N}%
{}+\frac15\,(N-1)(8N^3+3N^2+13N-12)\ms\big]\,.\label{Sigmas}
\end{align}
\end{proposition}

These expressions, together with Eqs.~\eqref{mum}--\eqref{sipmhalf},
completely determine $\mu_±$ and $\si_±$ for all values of $M$
and $N$. It has been numerically verified that the resulting
formulas yield the exact values of $\mu_±$ and $\si_±$ computed
from the partition function~\eqref{Zchainfinal} for a wide range of
values of $M$ and $N$. This provides a very solid confirmation of
Conjecture~\ref{conj3}. Let us mention, in closing, that formulas
analogous to~\eqref{mum}--\eqref{sipmhalf} expressing the mean and
standard deviations of the energy for the ferromagnetic chains
$\ssH_{+,±}$ can be immediately deduced from the previous
expressions and Eq.~\eqref{-ep-ep'}.

\section{Models with constant magnetic fields}
\label{S:mag}

In this section we shall briefly show how a constant magnetic field
can be added into the picture that we have presented in the previous
sections. For the sake of simplicity, \emph{we restrict ourselves to
the case $M=1/2$, $\ep=-1$} (``one-dimensional electrons''), and
define the \Emph{magnetic $\boldsymbol{BC_N}$ Hamiltonian} as
\begin{equation}\label{BC.mag}
H_\ep\mg=H_{-,\ep}-8aB\sum_iS_i\,.
\end{equation}

\begin{gloss}
The motion of a (three-dimensional) electron in the presence of static magnetic
and electric fields $\bB=\grad\land\bA$ and $\mathbf E=-\grad V$ in $\RR^3$ is
governed by the Pauli operator
\[
H=(\I\grad+e\bA)^2-2e\boldsymbol{\bB}\cdot\bS+eV\,,
\]
which is the nonrelativistic limit (in the resolvent norm topology) of the
Dirac operator~\cite{Th92}. Here $\bx=(x,y,z)$, $e$ is the electric charge of
the electron, $\bS=\frac12\boldsymbol\si$ is the spin operator of the particle
and $\boldsymbol\si=(\si_1,\si_2,\si_3)$ are the Pauli matrices. If $\bB$ is
constant, one can take $\bA=\frac12\bx\land\bB$, so that
\[
H=-\Delta+\I\ms e\ms\bB\cdot\bx\land\grad+\frac{e^2}4\|\bx\land\bB\|^2-2e\bB\cdot\bS+eV\,.
\]
Since we are interested in a one-dimensional scenario, we drop the second and
third summands in order to preserve the symmetries $\pd_y$ and $\pd_z$. This is
a priori a good approximation to the problem when the angular momentum and the
charge is small and the electric potential is strongly binding.

Hence one is led to define the magnetic $BC_N$ Hamiltonian for spin
$1/2$ particles as
\[
H_{-,\ep}\mg=H_\Eps-2e\sum_i\bB\cdot\bS_i.
\]
As the exchange operator $S_{ij}$ is written in terms of the $\SU(2)$-invariant
product as in Eq.~\eqref{SU(2M+1)inv}, in this model of one-dimensional
particles there is only one privileged direction $\bee$, namely, the one which
implements the Weyl reflections in spin space via
\[
S_i=2\bee\cdot\bS_i\,.
\]
Therefore it is natural to choose $\bB$ parallel to this direction, so that
\[
2e\sum_i\bB\cdot\bS_i=e\|\bB\|\sum_iS_i\,.
\]
If we now rescale $e\|\bB\|$ to $8aB$ as in Eqs.~\eqref{bb'} we
arrive at Eq.~\eqref{BC.mag}.
\end{gloss}

Since $H_\Eps$ and $\sum_iS_i$ commute, they possess a common basis of
eigenfunctions. Let $\bn\in[\ZZ^N]$ and $\ket\bs\in\cB^{\bn,-,\ep}_\Si$ (cf.\
Eq.~Lemma~\ref{basis.cHchi}), and define
\begin{align*}
\rho(\bk)&=\card\big\{i:k_i=1\big\}\,,\\
\rho_+(\bn,\bs)&=\card\big\{i:s_i=\tfrac12,\;n_i\neq0,\;\#(n_i)=1\big\}\,,
\end{align*}
where $\bk$ is defined as in Eq.~\eqref{nm}. Then one can easily
prove the following

\begin{theorem}\label{sp.Zmag}
Let $\cB\equiv\cB^{-,\ep,\bn}_\Si$. Then
\[
\spec(H_\ep\mg)=\big\{E_\bn+8aB\ms\big[\rho(\bk)-2\rho_+(\bn,\bs)-\de_{n_N,0}(1+\ep)\big]:
\bn\in[\ZZ^N],\;\ket\bs\in\cB\big\}\,.
\]
\end{theorem}
\begin{proof}
Let $\Si_1=\CC\ket{\frac12}\oplus\CC\ket{{-\frac12}}$ be the one-particle spin
space, so that $\Si=\Si_1^{\otimes N}$, and define the basis of one-particle
states
\[
\ket±=\ket{\tfrac12}±\ket{{-\tfrac12}}\in\Si_1\,.
\]
Clearly $S_1\ket±=±\ket±$. Now define the one- and
two-particle operators
\begin{align*}
\La^{(1)}_\ep&:\ket{{±\tfrac12}}\mapsto±\ket{\ep}\,,\\
\La^{(2)}_\ep&:\ket{{±\tfrac12,\mp\tfrac12}}\mapsto\ket{\ep,-\ep}\,,
\end{align*}
with $\ep=±$. By Propositions~\ref{P:Hilbert} one can consider the
action of the differential operator~\eqref{BC.mag} on
$\La_{-,\ep}(L^2(\TT^N)\otimes\Si)$. Theorem~\ref{sp.Hchi} states
that $H_{-,\ep}$ is diagonal in the basis~\eqref{cBEps}. Expressing
the nonnegative, nonincreasing multiindex $\bn$ in terms of $\bm$
and $\bk$ as in Eq.~\eqref{nm}, one can use Conditions~(i)--(iii) in
Lemma~\ref{basis.cHchi} to readily verify that
\begin{align*}
\big\lan\La_{-,\ep}(f_\bn\ket\bs):\ket\bs\in\cB\big\ran=
\big\lan\Psi_{\bn,\bs}:\ket\bs\in\cB\big\ran\,.
\end{align*}
Here we have set
\[
\Psi_{\bn,\bs}=\La_{-,\ep}\bigg(f_\bn\bigotimes_{j=1}^r\bv_\ep(s_{\ka_j+1},k_j,m_j)\bigg)\,,
\]
with
\[
\ka_j\equiv\ka_j(\bk)=\sum_{i=1}^{j-1}k_i
\]
and
\begin{align*}
\bv_\ep(±\tfrac12,1,m)&=
\begin{cases}
\ket{±}\qquad&\text{if }\;m\neq0\,,\\
\ket{\ep}&\text{if }\;m=0\,,
\end{cases}\\
\bv_\ep(±\tfrac12,2,m)&=\ket{+}\otimes\ket-\,,
\end{align*}
so that for each $j=1,2$ one has
\[
\sum_{i=1}^jS_i\bv_\ep(±\tfrac12,j,m)=
\begin{cases}
±\delta_{j,1}\qquad&\text{if }\;m\neq0\,,\\
\ep &\text{if }\;m=0\,.
\end{cases}
\]
Since $\sum_iS_i$ commutes with the symmetrizer and
$m_j=n_{\ka_j+1}$, one immediately derives from these equations that
\begin{align*}
\sum_iS_i\Psi_{\bn,\bs}&=\Psi_{\bn,\bs}\bigg[\sum_{\eta=±1}\eta\card\big\{i:s_i=\tfrac\eta2,\;
n_i\neq0,\;\#(n_{\ka_i+1})=1\big\}+\ep\de_{n_N,0}\bigg]\\
&=\big[2\rho_+(\bn,\bs)-\rho(\bk)+\de_{n_N,0}(1+\ep)\big]\Psi_{\bn,\bs}\,,
\end{align*}
i.e., $\sum_iS_i$ is diagonal in this basis. As the action of
$H_{-,\ep}$ and $\BH$ agree on $\Psi_{\bn,\bs}$ and, on account of
Proposition~\ref{P:BHfn}, the action of the latter operator on
$\{f_\bn\}$ is triangular with diagonal elements $E_\bn$, the result
follows.
\end{proof}

The spin chain associated with the magnetic Hamiltonian~\eqref{BC.mag}
is
\begin{equation}\label{ssHmag}
\ssH_\ep\mg=\ssH_{-,\ep}-B\sum_iS_i\,.
\end{equation}
It is straightforward to show that Lemma~\ref{freezing} also
applies, mutatis mutandis, to the magnetic
Hamiltonian~\eqref{BC.mag}. Hence we shall compute its spectrum by
means of the freezing trick. To this end, let us denote by
$Z_\ep\mg$ and $\ssZ_\ep\mg$ the partition functions
of~\eqref{BC.mag} and~\eqref{ssHmag}, and again we shall set the
ground state energy $E_0$ equal to 0. We shall freely use the
notation of Section~\ref{S:chainBCN}.

\begin{proposition}\label{Zmag}
Let
\[
\fB_N=\bigcup_{j=1}^N\big\{\bk\in\{1,2\}^j:|\bk|=N\big\}=\fP_N\cap\bigcup_{j=1}^N\{1,2\}^j
\]
the set of partitions of $N$ by $\{1,2\}$. Then
\begin{multline}
\lim\limits_{a\to\infty}Z_\ep\mg(8aT)=\sum_{\bk\in\fB_N}\big(q^B+q^{-B}\big)^{\rho(\bk)}
\prod_{j=1}^{r-1}\frac{q^{N_j}}{1-q^{{N_j}}}
\bigg(\frac{q^{N_r}}{1-q^{{N_r}}}+\de_{k_r,1}\frac{q^{\ep
B}}{q^B-q^{-B}}\bigg)\,.
\end{multline}
\end{proposition}
\begin{proof}
From Proposition~\ref{P:Z.BC} one immediately obtains
\[
Z_\ep\mg(8aT)\sim\sum_{\bk\in\fB_N}\sum_{m_1>\cdots>m_r\geq0}
q^{\sum_{i=1}^rm_i\nu_i}\sum_{\ket\bs\in\cB}
q^{B[2\rho_+(\bn,\bs)-\rho(\bk)]}\,,
\]
so that the sum over $\ket\bs$ depends on whether $m_r>0$ or not. If this is
the case,
\begin{equation}\label{card.eq}
\card\big\{\ket\bs\in\cB:\rho_+(\bn,\bs)=\varrho\big\}=\binom{\rho(\bk)}\varrho
\end{equation}
for any $0\leq\varrho\leq\rho(\bk)$, so
\begin{align*}
\sum_{\ket\bs\in\cB}q^{B[2\rho_+(\bn,\bs)-\rho(\bk)]}&=q^{-B\rho(\bk)}\sum_{\varrho=0}^{\rho(\bk)}\binom{\rho(\bk)}
\varrho q^{2B\varrho}=\big(q^B+q^{-B}\big)^{\rho(\bk)}\,.
\end{align*}
If, on the contrary, $m_r=0$, then $k_r=1$ and it is not difficult to see that
the formula~\eqref{card.eq} becomes
\begin{equation*}
\card\big\{\ket\bs\in\cB:\rho_+(\bn,\bs)=\varrho\big\}=\binom{\rho(\bk)-1}{\varrho}\,,
\end{equation*}
where $0\leq\varrho\leq\rho(\bk)-1$. Hence in this case
\begin{align*}
\sum_{\ket\bs\in\cB}q^{B[2\rho_+(\bn,\bs)-\rho(\bk)+1]}&=q^{-B[\rho(\bk)-1]}\sum_{\varrho=0}^{\rho(\bk)-1}
\binom{\rho(\bk)-1}{\varrho}
q^{2B\varrho}=\big(q^B+q^{-B}\big)^{\rho(\bk)-1}\,,
\end{align*}
and therefore
\[
Z_\ep\mg(8aT)\sim\sum_{\bk\in\fB_N}\big(q^B+q^{-B}\big)^{\rho(\bk)}
\sum_{m_1>\cdots>m_{r-1}>0}q^{\sum\limits_{i=1}^{r-1}m_i\nu_i}
\bigg(\frac{q^{\sum_{i=1}^rm_i\nu_i}}{1-q^{\sum_{i=1}^rm_i\nu_i}}+\frac{\de_{k_r,1}q^{\ep
B}}{q^B-q^{-B}}\bigg)\,.
\]
Taking Eqs.~\eqref{Nj} and~\eqref{sum.mi} into account we obtain the
desired formula.
\end{proof}

The freezing trick formula~\eqref{Z} and the scalar partition function (cf.
Proposition~\ref{Zsca}) then yield the following

\begin{theorem}
  When $M=1/2$, the partition function of the spin chain~\eqref{ssHmag} is
  given by
\begin{multline*}
\ssZ_\ep\mg(T)=\prod_i\big[1-q^{i[\Bbe+N-\frac12(i+1)]}\big]\\
×\sum_{\bk\in\fB_N}\big(q^B+q^{-B}\big)^{\rho(\bk)}
\prod_{j=1}^{r-1}\frac{q^{N_j}}{1-q^{{N_j}}}
\bigg(\frac{q^{N_r}}{1-q^{{N_r}}}+\de_{k_r,1}\frac{q^{\ep
B}}{q^B-q^{-B}}\bigg)\,.
\end{multline*}
\end{theorem}

\chapter{Spin models with NN interactions}
\label{Ch:NN}

\section{Introduction}

In this chapter we analyze in detail three QES $N$-body spin models which can
be regarded as cyclic versions of (generalized) CS models, as constructed by
Finkel et al.~\cite{FGGRZ01}. As this kind of systems feature only
near-neighbors interactions, we shall generically call them \emph{NN models}.
There are two reasons that make this kind of systems very promising from a
physical point of view. First, some of them are related to the short-range
Dyson model in random matrix theory~\cite{BGS99}. Second, the HS chains
associated with these models occupy an interesting intermediate position
between the Heisenberg chain (short-range, position-independent interactions)
and the usual HS chains (long-range, position-dependent interactions). We
shall consider this latter aspect in detail in the next chapter.

The study of these systems was initiated a few years ago, when Jain and Khare
presented a novel class of partially solvable models in which each particle
interacted only with its nearest and next-to-nearest neighbors~\cite{JK99}. In
a subsequent paper~\cite{AJK01}, Auberson, Jain and Khare discussed a
generalization of these models to the $BC_N$ root system and to higher
dimensions. The latter papers, however, left open some important issues, such
as the exact or quasi-exact solvability of these models, the derivation of
general explicit formulas for their eigenfunctions, or even the existence of
similar models for particles with spin. The last question was first addressed
by Deguchi and Ghosh~\cite{DG01}, who introduced spin $1/2$ extensions of the
scalar models of Jain and Khare and used the supersymmetric approach to obtain
some eigenfunctions. All these authors solely managed to construct a few exact
solutions, with trivial spin dependence. Moreover, their approach was by no
means systematic.

Our construction of NN models is based on a nontrivial modification of the
Dunkl operator technique and yields fully explicit formulas for a wide variety
of spin eigenfunctions. The presentation of these results is organized as
follows.  In Section~\ref{sec.mods} we define the Hamiltonians of the spin
$N$-body models which are the subject of this chapter
(cf.~Eqs.~\eqref{Hep}-\eqref{Vs}), and show that they can be expressed in
terms of suitable differential operators with near-neighbors exchange terms.
Section~\ref{sec.spaces} is entirely devoted to the characterization of
certain finite-dimensional spaces of polynomial spin functions invariant under
the latter operators. In Section~\ref{sec.eig} we show that the eigenvalue
problems for the Hamiltonians of the models~\eqref{Vs} restricted to their
invariant spaces reduce to finding the polynomial solutions of a corresponding
system of partial differential equations. By completely solving the latter
problem, we obtain several (infinite) families of eigenfunctions of the
models~\eqref{Vs} in closed form.

The material presented in this chapter is taken from
Refs.~\cite{EFGR05b,EFGR06,EFGR07}.

\section{The NN models}
\label{sec.mods}

In this chapter we shall study in detail the $N$-body \Emph{NN spin
models} defined by the Schrödinger operators
\begin{equation}\label{Hep}
H_\ep=-\Delta+V_\ep\,,
\end{equation}
where $\ep=0,1,2$ and
\begin{subequations}\label{Vs}
\begin{align}
&\hspace*{-0.1em}V_0=\omega^2
r^2+\sum_i\frac{2a^2}{(x_i-x_{i-1})(x_i-x_{i+1})}+
\sum_i\frac{2a(a-S_{i,i+1})}{(x_i-x_{i+1})^2}\,,\label{V0}\\
&\hspace*{-0.1em}V_1=\omega^2 r^2+\sum_i\frac{b(b-1)}{x_i^2}
+\sum_i\frac{8a^2x_i^2}{(x_i^2-x_{i-1}^2)(x_i^2-x_{i+1}^2)}\notag\\
&\hspace{13em}{}+4a\sum_i\frac{x_i^2+x_{i+1}^2}{{(x_i^2-x_{i+1}^2)}^2}
(a-S_{i,i+1})\,,\label{V1}\\[1mm]
&\hspace*{-0.1em}V_2=2a^2\sum_i\cot(x_i-x_{i-1})\cot(x_i-x_{i+1})
+\sum_i\frac{2a(a-S_{i,i+1})}{\sin^{2}(x_i-x_{i+1})}\,,\label{V2}
\end{align}
\end{subequations}
where we identify $x_0\equiv x_N$ and $x_{N+1}\equiv x_1$. As we did in the
study of CS models, we shall assume $a,b>\frac12$ and take into account the
inverse-square singularities to consider the above Hamiltonians as self-adjoint
operators on
\[
\cH_\ep=L^2(C_\ep)\otimes\Si\,,
\]
where the ``reduced'' configuration spaces are given by
\begin{subequations}\label{Cep}
\begin{align}
C_0&=\big\{\bx:x_1<\cdots< x_N\big\}\,,\label{C0}\\
C_1&=\big\{\bx:0<x_1<\cdots <x_N\big\}\,,\\
C_2&=\big\{\bx:0<x_{i+1}-x_i<\pi,\;\forall\,i<N\big\}\,.
\end{align}
\end{subequations}
\begin{remark}
Unlike CS models, the Hamiltonians defined by~\eqref{Vs} on $\cH_\ep$ and on
$L^2(\RR^N)\otimes\Si$ (or $L^2(\TT^N)\otimes\Si$ for the third potential) are
a priori different. This is due to the fact that the Weyl symmetry present in
CS models has been replaced in the NN case by the cyclic group
\begin{align}\label{fZ}
\fZ&=\big\lan\Pi_{12}\cdots\Pi_{N1}\big\ran\\
&\approx\big\lan\si_{\bee_1-\bee_2}\cdots\si_{\bee_N-\bee_1}\big\ran\notag\,,
\end{align}
and this latter group does not tessellate $\RR^N$ (or $\TT^N$). Nonetheless,
all the eigenfunctions presented in Theorems~\ref{thm.H0}, \ref{thm.H1}
and~\ref{thm.H2} can be clearly extended to eigenfunctions of $H_\ep$ on the
``unreduced'' Hilbert space since all our developments are essentially
algebraic.
\end{remark}

We shall also make use of the \Emph{scalar reductions} of these models, which
are the self-adjoint operators on $L^2(C_\ep)$ given by
\begin{equation}
\Hsc_\ep=H_\ep|_{S_{i,i+1}\to1}\,.
\end{equation}
The identity
\[
H_\ep(\psi\ket\bs)=(\Hsc_\ep\psi)\ket\bs\,,
\]
with $\ket\bs$ a symmetric spin vector, is obvious.

The models~\eqref{Hep} share a common property that is ultimately
responsible for their partial solvability, namely that each
Hamiltonian $H_\ep$ is related to a scalar differential-difference
operator involving near-neighbors exchange operators. In fact,
consider the operators
\begin{equation}\label{Tep}
T_\ep=\sum_iz_i^\ep\pa_i^2+2a\sum_i\frac1{z_i-z_{i+1}}\,(z_i^\ep\pa_i-z_{i+1}^\ep\pa_{i+1})
-2a\sum_i\frac{\vartheta_\ep(z_i,z_{i+1})}{(z_i-z_{i+1})^2}\,(1-K_{i,i+1}),
\end{equation}
where $\pa_i\equiv\pa_{z_i}$, $z_{N+1}\equiv z_1$, and
\[
\vartheta_0(x,y)=1\,,\qquad \vartheta_1(x,y)=\frac12\,(x+y)\,,\qquad
\vartheta_2(x,y)=xy\,.
\]
Each Hamiltonian $H_\ep$ is related to a linear combination
\begin{equation}\label{BHep}
\BH_{\!\ep}=c\,T_\ep+c_- J^-+c_0 J^0+E_0
\end{equation}
of its corresponding operator $T_\ep$ and the first-order
differential operators
\begin{equation}\label{J-J0}
J^-=\sum_i \pa_i\,,\qquad J^0=\sum_i z_i\pa_i
\end{equation}
through the star mapping~\eqref{star}, a change of variables and a
gauge transformation. More precisely,
\begin{equation}\label{Hepstar}
H_\ep=\mu\ms\BH_{\!\ep}^*\big|_{z_i=\ze(x_i)}\ms\mu^{-1}\,,\qquad
\ep=0,1,2,
\end{equation}
where the constants $c$, $c_-$, $c_0$, $E_0$, the gauge factor $\mu$, and the
change of variables $\ze$ for each model are listed in
Table~\ref{table:params}. Here and in what follows \emph{we write the star
mapping and the symmetrizer as $^*\equiv\,^*_+$ and $\La\equiv\La_+$,} since
for the sake of concreteness we shall restrict our attention to the symmetric
case.

\begin{table}[t]
\caption{Parameters, gauge factor and change of
variable.}\label{table:params}
\begin{center}
\begin{tabular}{lccc}\hline
\BStrut & $\ep=0$ & $\ep=1$ & $\ep=2$\\ \hline
\BStrut $c$ & $-1$ & $-4$ & $4$\\ \hline
\BStrut $c_-$ & $0$ & $-2(2b+1)$ & $0$\\ \hline
\BStrut $c_0$ & $2\om$ & $4\om$ & $4(1-2a)$\\ \hline
\BStrut $E_0$ & $N\omega(2a+1)$ & $N\omega(4a+2b+1)$ & $2Na^2$\\
\hline
\BStrut $\mu(\bx)\quad$ &
$\;\e^{-\frac\omega2\,r^2}\prod\limits_i|x_i-x_{i+1}|^a\;$ &
$\;\e^{-\frac\omega2\,r^2}\prod\limits_i{|x_i^2-x_{i+1}^2|}^a\,x_i^b\;$
& $\;\prod\limits_i\sin^a|x_i-x_{i+1}|\;$\\ \hline
\BStrut $\ze(x)$ & $x$ & $x^2$ & $\e^{±2\iu x}$\\ \hline
\end{tabular}
\end{center}
\end{table}

{}From Lemma~\ref{easylemma} and Eq.~\eqref{Hepstar} it follows that
if $\Phi(\bz)\in\La\big(C^{\infty}\otimes\Si\big)$ is a symmetric
(formal) eigenfunction of $\BH_{\!\ep}$, then
\begin{equation}\label{PsimuPhi}
\Psi(\bx)=\mu(\bx)\Phi(\bz)|_{z_i=\ze(x_i)}
\end{equation}
is a (formal) eigenfunction of $H_{\ep}$ with the same eigenvalue.
In the next section we shall construct a flag $\cH^{\ms
0}_\ep\subset\cH^{\ms 1}_\ep\subset\cdots$ of finite-dimensional
subspaces of $\La\big(\CC[\bz]\otimes\Si\big)$ invariant under each
$\BH_{\!\ep}$. We will show that the problem of diagonalizing
$\BH_{\!\ep}$ in each subspace $\cH^{\ms n}_\ep$ is equivalent to
the computation of the polynomial solutions of a system of linear
differential equations. We shall completely solve this problem,
thereby obtaining several infinite families of eigenfunctions
of~$H_{\!\ep}$ for each $\ep$. {}From the expressions for the change
of variable and the gauge factor in Table~\ref{table:params}, and
the fact that the functions $\Phi$ in Eq.~\eqref{PsimuPhi} are in
all cases polynomials, it immediately follows that the
eigenfunctions thus obtained are in fact normalizable.

\begin{gloss}
The above procedure parallels the construction of (generalized) CS
models of $A_N$ type~\cite{FGGRZ01}, which also applies to the
$BC_N$ case~\cite{FGGRZ01b}. Each operator $T_\ep$ is the NN analog
of the sum of squares of one of the three known families of $A_N$
Dunkl operators. E.g., the operator $T_0$ is to be compared with the
sum of the squares of the Dunkl operators~\eqref{Dunkl.rat.Ch1}
associated with the $A_N$ system, i.e.,
\[
T\super{CS}_0=\sum_i(J_i\super{rat})^2=\sum_i\pd_i^2+2a\sum_{i<j}\frac1{z_i-z_j}\,(\pa_i-\pa_j)
-2a\sum_{i<j}\frac1{(z_i-z_j)^2}\,(1-K_{ij})\,.
\]
Hence, one can regard $H_0$ as the proper NN analog of the Calogero model from
the point of view of quasi-exact solvability. In contrast with the usual CS
models, however, the physical Hamiltonians $H_\ep$ associated with the cyclic
differential-difference operators $T_\ep$ include a three-body term which
turns out to be crucial in the analysis of the solvability properties of these
models.

Two natural questions arise from the above discussion. First, it should be
ascertained whether the full Dunkl operator approach can be extended to the NN
setting or one is actually obliged to do without it. Secondly, one could wonder
whether one would be able to use the above construction to prove the
quasi-exact solvability of the models~\eqref{Hep} as in the case of CS models.
The answer to both questions is negative. Since $\fZ$ is not a Coxeter group,
there are no known procedures to construct a set of Dunkl operators associated
with this problem. Moreover, a couple of technical details prevents the usual
approach to CS systems to work successfully in the NN context. Let us elaborate
on this.

The proof of the quasi-exact solvability of the Hamiltonian obtained
from $T_0\super{CS}$ (or a linear combination similar
to~\eqref{Hepstar}) through a gauge transformation is based on two
facts:
\begin{enumerate}
\item $T_0\super{CS}$ leaves invariant the space of polynomials
\begin{equation*}
\cP^n=\big\{f\in\CC[\bz]:\deg f\leq n\big\}
\end{equation*}
for any $n\in\NN_0$, where here and in what follows we denote by $\deg f$ the
degree in $\bz$ of the polynomial $f$.

\item $T_0\super{CS}$ commutes with the symmetrizer $\La$, which satisfies
$K_{ij}\La=S_{ij}\La$.
\end{enumerate}
These conditions immediately imply that
\[
\big(T_0\super{CS}\big)^*\La\big(\cP^n\ket\bs\big)
=\La\big[(T_0\super{CS}\cP^n)\ket\bs\big]\subset
\La\big(\cP^n\ket\bs\big)
\]
for any $n\in\NN_0$, $\ket\bs\in\Si$. Whereas $T_0$ certainly satisfies the
first condition, it does not commute with $\La$, as it is not invariant under
arbitrary permutations of the particles. It is also clear that any attempt of
replacing $\La$ by the symmetrizer under the cyclic group~\eqref{fZ} is doomed
to fail, as this symmetry is not enough to exchange $K_{i,i+1}$ by
$S_{i,i+1}$, and thus $\BH_\ep$ by $H_\ep$.

Hence, the extension of the CS methods to the context of NN models is not at
all trivial, and in fact $H_\ep$ is not a priori guaranteed to admit
finite-dimensional invariant subspaces of
$\mu\ms\La\big(\CC[\bz]\otimes\Si\big)$. We shall next show that this is the
case nonetheless.
\end{gloss}

\section{The invariant spaces}\label{sec.spaces}

In this section we shall prove that each operator $T_\ep$ leaves
invariant a flag $\cT^{\ms 0}_\ep\subset\cT^{\ms
1}_\ep\subset\cdots$, where $\cT^{\ms n}_\ep$ is a
finite-dimensional subspace of $\La(\cP^n\otimes\Si)$. This result
will then be used to construct a corresponding invariant flag
\mbox{$\cH^{\ms 0}_\ep\subset\cH^{\ms 1}_\ep\subset\cdots$} for the
operator $\BH_{\!\ep}$, where $\cH^{\ms n}_\ep\subset\cT^{\ms
n}_\ep$ for all $n$.

Let us first introduce the following two sets of elementary
symmetric polynomials:
\[
\si_k=\sum_i z_i^k\,,\qquad
\tau_k=\sum\limits_{i_1<\cdots<i_k}z_{i_1}\cdots z_{i_k}\,; \qquad
k=1,\dots,N\,.
\]
It is well known~\cite{Ma95} that any symmetric polynomial in $\bz$ can be
expressed as a polynomial in either $\bsi\equiv(\si_1,\dots,\si_N)$ or
$\btau\equiv(\tau_1,\dots,\tau_N)$.

We shall denote by $2aX_\ep$ the terms of $T_\ep$ linear in
derivatives, that is
\[
X_\ep=\sum_i\frac1{z_i-z_{i+1}}\,(z_i^\ep\pa_i-z_{i+1}^\ep\pa_{i+1})\,.
\]
In the next lemma we show that each vector field $X_\ep$ leaves
invariant a corresponding flag
$\cX_\ep^0\subset\cX_\ep^1\subset\cdots$ of finite-dimensional
subspaces of the space $\La(\CC[\bz])=\CC[\bsi]=\CC[\btau]$ of
symmetric polynomials in $\bz$. If $f$ is a function of the
symmetric variables $\si_1,\si_2,\si_3,\tau_{N-1},\tau_N$, we shall
use from now on the convenient notation
\[
f_k=\begin{cases}
\pa_{\si_k}f\,,\quad & k=1,2,3,\\[1mm]
\pa_{\tau_k}f\,,\quad & k=N-1,N.
\end{cases}
\]

\begin{lemma}
For each $n=0,1,\dots$, the operator $X_\ep$ leaves invariant the
linear space $\cX_\ep^n$, where
\[
\cX_0^n=\CC[\si_1,\si_2,\si_3]\cap\cP^n,\quad
\cX_1^n=\CC[\si_1,\si_2,\tau_N]\cap\cP^n,\quad
\cX_2^n=\CC[\si_1,\tau_{N-1},\tau_N]\cap\cP^n.
\]
\end{lemma}
\begin{proof}
Let us first consider the vector field $X_0$. Since
\[
X_0\si_k=k\sum_iz_i^{k-1}X_0z_i=k\bigg(\sum_i\frac{z_i^{k-1}}{z_i-z_{i+1}}
-\sum_i\frac{z_i^{k-1}}{z_{i-1}-z_i}\bigg)=
\begin{cases}
0\,, & k=1\,,\\
2N\,, & k=2\,,\\
6\si_1\,, & k=3\,,
\end{cases}
\]
if $f\in\cX^n_0$ we have
\begin{subequations}\label{Xf}
\begin{equation}\label{X0f}
X_0f=2(Nf_2+3\si_1 f_3)\in\cX^n_0\,.
\end{equation}
The proof for the remaining two cases follows from the analogous
formulas
\begin{align}
X_1f&=Nf_1+4\si_1f_2\,,& f\in\cX^n_1\,;\label{X1f}\\
X_2f&=2\si_1f_1+N(\tau_{N-1}f_{N-1}+\tau_Nf_N)\,,&
f\in\cX^n_2\,.\label{X2f}
\end{align}
\end{subequations}
\end{proof}
\begin{remark}
It should be noted that these flags cannot be trivially enlarged,
since, e.g.,
\begin{align*}
\frac14\,X_0\si_4&=2\si_2+\sum_iz_iz_{i+1}\,,\\
\frac13\,X_1\si_3&=2\si_2+\sum_iz_iz_{i+1}\,,&
X_1\tau_{N-1}&=\tau_N\sum_i({z_iz_{i+1}})^{-1}\,,\\
\frac12\,X_2\si_2&=2\si_2+\sum_iz_iz_{i+1}\,,&
X_2\tau_{N-2}&=N\tau_{N-2}-\tau_N\sum_i(z_iz_{i+1})^{-1}
\end{align*}
are not symmetric polynomials.
\end{remark}
We note that the restriction of $T_\ep$ to
$\cX^n_\ep\subset\La(\CC[\bz])$ obviously satisfies
\begin{equation}\label{Tep-res}
T_\ep|_{\cX^n_\ep}=\sum_iz_i^\ep\pa_i^2+2aX_\ep\,.
\end{equation}
The second-order terms of the operator~\eqref{Tep-res}, however, do
not preserve the corresponding space $\cX^n_\ep$, unless one imposes
the additional restrictions specified in the following proposition:
\begin{proposition}\label{prop.cS}
For each $n\in\NN_0$, the operator $T_\ep$ leaves invariant the
linear space $\cS_\ep^n$, where
\begin{align*}
\cS_0^n&=\{f\in\cX_0^n: f_{33}=0\}\,,\\
\cS_1^n&=\{f\in\cX_1^n: f_{22}=f_{NN}=0\}\,,\\
\cS_2^n&=\{f\in\cX_2^n: f_{11}=f_{N-1,N-1}=0\}\,.
\end{align*}
\end{proposition}
\begin{proof}
Let us begin with the operator $T_0$. If $f\in\cX_0^n$, an
elementary computation shows that
\begin{equation}\label{paif0}
\pa_if=f_1+2z_if_2+3z_i^2f_3\\
\end{equation}
and therefore
\begin{multline}\label{T20f}
\sum_i \pa_i^2f=N(f_{11}+2f_2)+
2(2f_{12}+3f_3)\si_1\\
{}+2(3f_{13}+2f_{22})\si_2 +12f_{23}\si_3+9f_{33}\si_4\,.
\end{multline}
{}From the previous formula and Eq.~\eqref{X0f} it follows that
$T_0f\in\cS_0^n$ whenever $f\in\cS_0^n$. Similarly, if $f\in\cX^n_1$
we have
\begin{equation}\label{paif1}
  \pa_if=f_1+2z_if_2+z_i^{-1}{\tau_N}f_N\,,
\end{equation}
so that
\begin{multline}\label{T21f}
\sum_i z_i\pa_i^2f=(f_{11}+2f_2)\si_1+4f_{12}\si_2+4f_{22}\si_3\\
{}+2Nf_{1N}\tau_N+4f_{2N}\si_1\tau_N+f_{NN}\tau_{N-1}\tau_N\,,
\end{multline}
which together with Eq.~\eqref{X1f} implies that $T_1f\in\cS^n_1$
for all $f\in\cS^n_1$. Finally, if $f\in\cX^n_2$ then
\begin{equation}\label{paif2}
\pa_if=f_1+\big(z_i^{-1}{\tau_{N-1}}-{z_i^{-2}}{\tau_N}\big)f_{N-1}+z_i^{-1}{\tau_N}f_N
\end{equation}
and hence
\begin{multline}\label{T22f}
  \sum_i
  z_i^2\pa_i^2f=f_{11}\si_2+2f_{1,N-1}(\si_1\tau_{N-1}-N\tau_N)\\[-1mm]
  +2f_{1N}\si_1\tau_N+f_{N-1,N-1}\big[(N-1)\tau_{N-1}^2-2\tau_{N-2}\tau_N\big]\\[2mm]
  +2(N-1)f_{N-1,N}\tau_{N-1}\tau_N+Nf_{NN}\tau_N^2\,.
\end{multline}
The statement follows again from the previous equation and
Eq.~\eqref{X2f}.
\end{proof}
The last proposition implies that each operator $T_\ep$ preserves
``trivial'' symmetric spaces $\cS^n_\ep\otimes\La(\Si)$ spanned by
factorized states. The main theorem of this section shows that in
fact the latter operator leaves invariant a flag of nontrivial
finite-dimensional subspaces of $\La(\cP^n\otimes\Si)$. Before
stating this theorem we need to make a few preliminary definitions.
Given a spin state $\ket\bs\in\Si$, we set
\begin{equation}\label{sisij}
\ket{\bs_i}=\frac1{N!}\sum_{\substack{\pi\in
S_N\\\pi(1)=i}}\pi\ms\ket\bs\,,\qquad\quad
\ket{\bs_{ij}^±}=\frac1{N!}\!\! \sum_{\substack{\pi\in
S_N\\\pi(1)=i,\pi(2)=j}}\!\!\pi\ms(1± S_{12})\ms\ket\bs\,,
\end{equation}
where $S_N$ is the symmetric group on $N$ elements and we identify
an abstract permutation $\pi$ with its realization as a permutation
of the particles' spins. {}From Eq.~\eqref{sisij} we have
\begin{equation}\label{Lafs}
\La\big(f(z_1)\ket\bs\big)=\sum_if(z_i)\ket{\bs_i}\,,\qquad
\La\big(g_±(z_1,z_2)\ket\bs\big)=\sum_{i<j}g_±(z_i,z_j)\ket{\bs^±_{ij}}\,,
\end{equation}
where the last identity holds if $g_±(x,y)=± g_±(y,x)$. We
also define the subspace
\begin{equation}\label{Sip}
\Si'=\Big\{\,\ket\bs\in\Si:{\textstyle\sum\limits_i}\ket{\bs^+_{i,i+1}}\in\La(\Si)\,\Big\}\subset\Si\,.
\end{equation}
\begin{theorem}\label{thm.1}
Let
\begin{align*}
\cT^n_0 &=\big\langle
f(\si_1,\si_2,\si_3)\La\ket\bs,g(\si_1,\si_2,\si_3)\La(z_1\ket\bs),
h(\si_1,\si_2)\La(z_1^2\ket\bs),\\
&\hspace{4em}\Th(\si_1,\si_2)\La(z_1z_2\ket{\bs'}),w(\si_1,\si_2)\La(z_1z_2(z_1-z_2)\ket{\bs})
: f_{33}=g_{33}=0\big\rangle\,,\\[1mm]
\cT^n_1 &=\big\langle f(\si_1,\si_2,\tau_N)\La\ket\bs,
g(\si_1,\tau_N)\La(z_1\ket\bs)
:{}f_{22}=f_{NN}=g_{NN}=0\big\rangle\,,\\[1mm]
\cT^n_2&=\big\langle
f(\si_1,\tau_{N-1},\tau_N)\La\ket\bs,g(\tau_{N-1},\tau_N)\La(z_1\ket\bs),\\
%&\hspace{14.2em}{}
&\hspace{5.5em}\tau_Nq(\si_1,\tau_N)\La(z_1^{-1}\ket\bs):f_{11}=f_{N-1,N-1}=g_{N-1,N-1}=q_{11}=0\big\rangle\,,
\vrule depth6pt width0pt
\end{align*}
where $\ket\bs\in\Si$, $\ket{\bs'}\in\Si'$, $\deg f\leq n$, $\deg
g\leq n-1$, $\deg h\leq n-2$, $\deg\Th\leq n-2$, $\deg w\leq n-3$,
$\deg q\leq n-N+1$. Then $\cT_\ep^n$ is invariant under $T_\ep$ for
all $n\in\NN_0$.
\end{theorem}
\begin{proof}
By Proposition~\ref{prop.cS} it suffices to show that $T_\ep$ maps
$\cT_\ep^n/\big(\cS^n_\ep\otimes\La(\Si)\big)$ into $\cT_\ep^n$. We
shall first deal with the operator $T_0$. Consider the states of the
form $g\La(z_1\ket\bs)$, with $g\in\cS_0^{n-1}$. Since
\[
(\pa_l-\pa_{l+1})z_i=\frac1{z_l-z_{l+1}}\,(1-K_{l,l+1})z_i\,,\qquad\forall\:i,l,
\]
we have
\[
T_0(gz_i)=(T_0g)z_i+2\pa_ig\,.
\]
Calling
\begin{equation}\label{Phik}
\Phi^{(k)}\equiv\Phi^{(k)}(\ket\bs)=\La(z_1^k\ket\bs),\qquad k\in\ZZ\,,
\end{equation}
from Eqs.~\eqref{paif0} and~\eqref{Lafs} we obtain
\begin{equation}\label{Phi01}
T_0\big(g\Phi^{(1)}\big)=\sum_iT_0(gz_i)\ket{\bs_i}=
(T_0g)\Phi^{(1)}+2\sum_{k=1}^3kg_k\Phi^{(k-1)}\in\cT_0^{n-2}\,.
\end{equation}
Similarly, if $h(\si_1,\si_2)\in\cS_0^{n-2}$, the identity
\[
(\pa_l-\pa_{l+1})z_i^2=\frac1{z_l-z_{l+1}}\,(1-K_{l,l+1})z_i^2+(z_l-z_{l+1})
(\de_{li}+\de_{l,i-1})\,,\qquad\forall\:i,l
\]
implies that
\[
T_0(hz_i^2)=(T_0h)z_i^2+4z_i\pa_ih+2(2a+1)h\,,
\]
and therefore
\begin{align}
T_0\big(h\Phi^{(2)}\big)&=\sum_iT_0(hz_i^2)\ket{\bs_i}\notag\\
&=(T_0h+8h_2)\Phi^{(2)}+4h_1\Phi^{(1)}+2(2a+1)h\Phi^{(0)}\label{Phi02}
\end{align}
belongs to $\cT_0^{n-2}$ on account of Eqs.~\eqref{X0f}
and~\eqref{T20f}. On the other hand, from the equality
\[
(\pa_l-\pa_{l+1})z_iz_j=\frac1{z_l-z_{l+1}}\ms(1-K_{l,l+1})z_iz_j-(z_l-z_{l+1})
\de_{j,i+1}\de_{l,i},\quad\forall\:i<j,\forall\:l
\]
it follows that
\[
T_0(\tilde
hz_iz_j)=(T_0\Th)z_iz_j+2(z_i\pa_j\Th+z_j\pa_i\Th)-2a\Th\de_{j,i+1}\,.
\]
Setting
\begin{equation}\label{tPhi2}
\tPhi^{(2)}\equiv\tPhi^{(2)}(\ket\bs)=\La(z_1z_2\ket\bs)
\end{equation}
and using again Eqs.~\eqref{paif0} and~\eqref{Lafs} we then have
\begin{align}
T_0\big(\Th\tPhi^{(2)}\big)&=\sum_{i<j}T_0(\tilde hz_iz_j)\ket{\bs^+_{ij}}\notag\\
&=(T_0\Th+8\Th_2)\tPhi^{(2)}+2\Th_1\La\big[(z_1+z_2)\ket\bs\big]
-2a\Th\sum_i\ket{\bs^+_{i,i+1}}\,.\label{tPhi02}
\end{align}
Since
$\La\big[(z_1+z_2)\ket\bs\big]=\La\big[z_1(1+S_{12})\ket\bs\big]$,
the RHS of Eq.~\eqref{tPhi02} belongs to $\cT_0^{n-2}$ if and only
if $\ket\bs\in\Si'$. The last type of states generating the module
$\cT_0^n$ are of the form $w(\si_1,\si_2)\hPhi^{(3)}$, where
\begin{equation}\label{hPhi3}
\hPhi^{(3)}\equiv\hPhi^{(3)}(\ket\bs)=\La(z_1z_2(z_1-z_2)\ket\bs)\,.
\end{equation}
{}From the equality
\begin{multline*}
\frac1{z_l-z_{l+1}}\,(\pa_l-\pa_{l+1})\big[z_iz_j(z_i-z_j)\big]
=\frac1{(z_l-z_{l+1})^2}\,(1-K_{l,l+1})\big[z_iz_j(z_i-z_j)\big]
\\+(\de_{l,i-1}+\de_{li})z_j-(\de_{l,j-1}+\de_{lj})z_i,\qquad\forall\:i<j,\forall\:l
\end{multline*}
it follows that
\begin{multline}
T_0\big[w z_iz_j(z_i-z_j)\big]=(T_0w)z_iz_j(z_i-z_j)+2z_j(2z_i-z_j)\pa_iw\\
-2z_i(2z_j-z_i)\pa_jw-2(2a+1)(z_i-z_j)w\,.
\end{multline}
Using again Eqs.~\eqref{paif0} and~\eqref{Lafs} we obtain
\begin{multline}
T_0\big(w\hPhi^{(3)}\big)=\sum_{i<j}T_0\big(wz_iz_j(z_i-z_j)\big)\ket{\bs^-_{ij}}
=(T_0w+12w_2)\hPhi^{(3)}\\
+2w_1\La\big[(z_1^2-z_2^2)\ket\bs\big]
-2(2a+1)w\La\big[(z_1-z_2)\ket\bs\big]\,.\label{hPhi03}
\end{multline}
Since
$\La\big[(z_1^k-z_2^k)\ket\bs\big]=\La\big[z_1^k(1-S_{12})\ket\bs\big]$,
the RHS of the latter equation clearly belongs to $\cT_0^{n-2}$.
This shows that $T_0(\cT_0^n)\subset\cT_0^{n-2}\subset\cT_0^n$.

Consider next the action of the operator $T_1$ on a state of the
form $g(\si_1,\tau_N)\Phi^{(1)}$, with $g\in\cS_1^{n-1}$. From the
identity
\[
(z_l\pa_l-z_{l+1}\pa_{l+1})z_i=\frac12\,\frac{z_l+z_{l+1}}{z_l-z_{l+1}}\,(1-K_{l,l+1})z_i
+\frac12\,(z_l-z_{l+1})(\de_{l,i}+\de_{l,i-1})\,,\quad\forall\:i,l
\]
we easily obtain
\[
T_1(gz_i)=(T_1g)z_i+2z_i\pa_ig+2ag\,,
\]
and therefore, by Eqs.~\eqref{X1f}, \eqref{paif1} and~\eqref{T21f},
\begin{equation}\label{Phi11}
T_1\big(g\Phi^{(1)}\big)=\sum_i T_1(gz_i)\ket{\bs_i}=
(T_1g+2g_1)\Phi^{(1)}+2(ag+\tau_Ng_N)\Phi^{(0)}\in\cT_1^{n-1}\,.
\end{equation}
Thus $T_1(\cT_1^n)\subset\cT_1^{n-1}\subset\cT_1^n$, as claimed.

Consider, finally, the operator $T_2$. If
$g(\tau_{N-1},\tau_N)\in\cS^{n-1}_2$, the identity
\[
(z_l^2\pa_l-z_{l+1}^2\pa_{l+1})z_i=\frac{z_lz_{l+1}}{z_l-z_{l+1}}(1-K_{l,l+1})z_i
+z_i(z_l-z_{l+1})(\de_{l,i}+\de_{l,i-1}),\quad\forall\,i,l
\]
yields
\[
T_2(gz_i)=(T_2g)z_i+2z_i^2\pa_ig+4az_ig\,,
\]
and hence, by Eq.~\eqref{paif2},
\begin{align}
T_2(g\Phi^{(1)})&=\sum_i T_2(gz_i)\ket{\bs_i}\notag\\
&=\big[T_2g+2(\tau_{N-1}g_{N-1}+\tau_Ng_N+2ag)\big]\Phi^{(1)}
-2\tau_Ng_{N-1}\Phi^{(0)}\label{Phi21}
\end{align}
clearly belongs to $\cT_2^n$ on account of Eqs.~\eqref{X2f}
and~\eqref{T22f}. The last type of spin states we need to study are
of the form $\hq\Phi^{(-1)}$, where $\hq\equiv\tau_Nq(\si_1,\tau_N)$
with $q_{11}=0$. Since
\[
(z_l^2\pa_l-z_{l+1}^2\pa_{l+1})z_i^{-1}
=\frac{z_lz_{l+1}}{z_l-z_{l+1}}\,(1-K_{l,l+1})z_i^{-1}\,,\qquad\forall\:i,l,
\]
we obtain
\[
T_2\big(\hq z_i^{-1}\big)=(T_2\hq)z_i^{-1}-2\pa_i\hq+2\hq
z_i^{-1}\,,
\]
and thus, by Eqs.~\eqref{paif2} and~\eqref{Lafs},
\begin{equation}\label{Phi2-1}
T_2\big(\hq\Phi^{(-1)}\big)=\sum_i T_2(\hq z_i^{-1})\ket{\bs_i}
=(T_2\hq-2\tau_N\hq_N+2\hq)\Phi^{(-1)}-2\hq_1\Phi^{(0)}\,.
\end{equation}
{}From Eqs.~\eqref{X2f} and~\eqref{T22f}, it follows that the RHS of
the previous equation belongs to $\cT_2^n$. Hence
$T_2(\cT_2^n)\subset\cT_2^n$, which concludes the proof.
\end{proof}
\begin{remark}\label{r.NN.1/2}
We have chosen to allow a certain overlap between the different types of states
spanning the spaces $\cT^n_\ep$. For instance, if $\ket s$ is symmetric a state
$g(\si_1,\si_2,\si_3)\La(z_1\ket s)\in\cT_0^n$ is also of the form
$f(\si_1,\si_2,\si_3)\La\ket s$. Less trivially, in the case of spin $1/2$ the
identity
\[
\hPhi^{(3)}(\ket\bs)=\frac2N\big(\si_1\Phi^{(2)}(\ket\bs)-\si_2\Phi^{(1)}(\ket\bs)\big)\,,
\qquad S_{12}\ket\bs=-\ket\bs\,,
\]
where $\Phi^{(k)}$ and $\hPhi^{(3)}$ are respectively defined in
Eqs.~\eqref{Phik} and~\eqref{hPhi3}, implies that the states of the form
$w(\si_1,\si_2)\hPhi^{(3)}$ in the space $\cT_0^n$ can be expressed in terms of
the other generators of this space.
\end{remark}

\begin{corollary}\label{cor.1}
For each $\ep=0,1,2$, the gauge Hamiltonian $\BH_\ep$ leaves
invariant the space $\cH^n_\ep$ defined by
\begin{equation*}
\cH^n_0=\cT^n_0\,,\qquad \cH^n_1=\cT^n_1\big|_{f_N=g_N=0}\,,\qquad
\cH^n_2=\cT^n_2\,.
\end{equation*}
\end{corollary}
\begin{proof}
We shall begin by showing that each space $\cT_\ep^n$ is invariant
under the operator $J^0$. Note first that
\begin{equation}\label{J0Phis}
J^0\Phi^{(j)}=j\ms\Phi^{(j)}\,,\qquad
J^0\tPhi^{(2)}=2\ms\tPhi^{(2)}\,,\qquad
J^0\hPhi^{(3)}=3\ms\hPhi^{(3)}\,;\qquad j\in\ZZ\,,
\end{equation}
where the states $\Phi^{(j)}$, $\tPhi^{(2)}$ and $\hPhi^{(3)}$ are
defined in Eqs.~\eqref{Phik}, \eqref{tPhi2} and~\eqref{hPhi3},
respectively. Using Eqs.~\eqref{paif0}, \eqref{paif1}
and~\eqref{paif2} one can immediately establish the identities
\begin{subequations}\label{J0fs}
\begin{align}
J^0f&=\si_1f_1+2\si_2f_2+3\si_3f_3\,,\qquad & &\forall f(\si_1,\si_2,\si_3)\,,\label{J0f0}\\
J^0f&=\si_1f_1+2\si_2f_2+N\tau_Nf_N\,,\qquad & &\forall f(\si_1,\si_2,\tau_N)\,,\label{J0f1}\\
J^0f&=\si_1f_1+(N-1)\tau_{N-1}f_{N-1}+N\tau_Nf_N\,,\qquad & &\forall
f(\si_1,\tau_{N-1},\tau_N)\,. \label{J0f2}
\end{align}
\end{subequations}
{}From Eqs.~\eqref{J0Phis}-\eqref{J0fs} and the fact that $J^0$ is a
derivation it follows that $J^0$ leaves invariant the spaces
$\cT^n_\ep$ for all $\ep=0,1,2$. This implies that $\BH_\ep$
preserves $\cT^n_\ep$ for $\ep=0,2$, since the coefficient $c_-$
vanishes in these cases (cf.~Table~\ref{table:params}). On the other
hand, for $\ep=1$ the coefficient $c_-$ is nonzero, and thus we have
to consider the action of the operator $J^-$ on the space $\cT^n_1$.
We now have
\begin{equation}\label{JmPhis}
J^-\Phi^{(j)}=j\ms\Phi^{(j-1)}\,,\qquad j\in\ZZ\,,
\end{equation}
and, from Eq.~\eqref{paif1},
\begin{equation}\label{Jmf1}
J^-f=Nf_1+2\si_1f_2+\tau_{N-1}f_N\,,\qquad\forall
f(\si_1,\si_2,\tau_N)\,.
\end{equation}
Hence $J^-$ leaves invariant the subspace $\cH^n_1$ of $\cT^n_1$
defined by the restrictions $f_N=g_N=0$. {}From the obvious identity
$T_1\big(f\Phi^{(0)}\big)=\big(T_1f\big)\Phi^{(0)}$ and
Eq.~\eqref{Phi11}, together with~\eqref{X1f}, \eqref{Tep-res}
and~\eqref{T21f}, it follows that the operator $T_1$ also
preserves~$\cH^n_1$. Likewise, Eqs.~\eqref{J0Phis} and~\eqref{J0f1}
imply that $\cH^n_1$ is invariant under $J^0$, and hence under the
gauge Hamiltonian $\BH_1$.
\end{proof}

Theorem~\ref{thm.1} characterizes the invariant space $\cT_0^n$ in
terms of the subspace $\Si'\subset\Si$ in Eq.~\eqref{Sip} that we
shall now study in detail. In fact, from the definition of the
invariant space $\cT_0^n$ it follows that we can consider without
loss of generality the quotient space $\Si'/{\sim}$, where
$\ket\bs\sim\ket{\tbs}$ if
$\La(z_1z_2\ket{\bs})=\La(z_1z_2\ket{\tbs})$. For instance, from
Eq.~\eqref{sisij} it immediately follows that if $\ket{\bs}\in\Si'$
and $\pi\in S_N$ is a permutation such that $\pi(i)\in\{1,2\}$ for
$i=1,2$, then $\pi\ket{\bs}$ belongs to $\Si'$ and is equivalent to
$\ket\bs$.

In the rest of this section, we shall denote $\ket{\bs^+_{ij}}$
simply as $\ket{\bs_{ij}}$ for the sake of conciseness. From
Eq.~\eqref{sisij} it easily follows that any symmetric state belongs
to $\Si'$, since
\begin{equation}\label{AsLas}
\sum_i\ket{\bs_{i,i+1}}=\frac2{N-1}\,\ket\bs\,,\quad\text{for all
}\ket\bs\in\La(\Si)\,.
\end{equation}
On the other hand, if $\ket\bs\in\La(\Si)$ the corresponding state
$h(\si_1,\si_2)\La(z_1z_2\ket\bs)$ is a trivial (factorized) state.
We shall next show that the reciprocal of this statement is also
true, up to equivalence.
\begin{lemma}
For every $\ket\bs\in\Si$, $\La(z_1z_2\ket{\bs})$ is a factorized
state if and only if $\ket\bs\sim\La\ket\bs$.
\end{lemma}
\begin{proof}
Suppose that
\[
\La(z_1z_2\ket{\bs})=\ket{\hat\bs}\sum_{i<j}c_{ij}z_iz_j
\]
is a factorized state. Since the LHS of the previous formula is
symmetric,
 $c_{ij}=c$ for all $i,j$ and $\ket{\hat\bs}\in\La(\Si)$. By absorbing the constant
$c$ into $\ket{\hat\bs}$ we can take $c=1$ without loss of
generality, and therefore
\[
\La(z_1z_2\ket{\bs})=\sum_{i<j}z_iz_j\ket{\bs_{ij}}=\tau_2\ket{\hat\bs}
\quad\implies\quad \ket{\bs_{ij}}=\ket{\hat\bs}\,,\qquad
i,j=1,\dots\,,N.
\]
From Eq.~\eqref{Lafs} with $f(z_1,z_2)=1$ it then follows that
\[
\La\ket\bs =
\sum_{i<j}\ket{\bs_{ij}}=\frac12\,N(N-1)\ket{\hat\bs}\,.
\]
Setting $\ket{\bs_0}=\ket\bs-\La\ket\bs$ and using the previous
identity we obtain
\[
  \La(z_1z_2\ket{\bs_0})=\La(z_1z_2\ket{\bs})-\frac{2\ms\tau_2}{N(N-1)}\,\La\ket\bs
  =\ket{\hat\bs}\tau_2-\frac{2\ms\tau_2}{N(N-1)}\,\La\ket\bs=0\,.
\]
Hence $\ket\bs\sim\La\ket\bs$, as claimed.
\end{proof}
By the previous lemma, it suffices to characterize the nonsymmetric
states in $\Si'$. To this end, let us introduce the linear operator
$A:\Si\to\Si$ by
\begin{equation}\label{A}
A\ket\bs=\sum_i\ket{\bs_{i,i+1}}.
\end{equation}
Given an element $\ket\bs\equiv\ket{s_1\dots s_N}$ of the canonical
basis of $\Si$, we shall also denote by $\{s^1,\dots,s^n\}$ the set
of distinct components of $\bs\equiv(s_1,\dots,s_N)$, and by $\nu_i$
the number of times that $s^i$ appears among the components of
$\bs$. For instance, if $\ket\bs=\ket{{-2},0,1,-2,1}$, then we can
take $s^1=-2$, $s^2=0$, $s^3=1$, so that $\nu_1=\nu_3=2$, $\nu_2=1$.
Consider the spin states $\ket{\chi_i(\bs)}\equiv\ket{\chi_i}$,
$i=1,\dots,n$, given by
\begin{subequations}\label{chis}
\begin{align}
\ket{\chi_i}&= \nu_i(\nu_i-1)\ket{s^is^i\dots}-\sum_{\substack{1\leq
j,k\leq n\\j,k\neq i}}\nu_j(\nu_k-\de_{jk})
\ket{s^js^k\dots}\,,\qquad &\nu_i>1\,,\label{chia}\\[1mm]
\ket{\chi_i}&=\sum_{\substack{1\leq j\leq n\\j\neq i}}\nu_j\,
\big(\ket{s^is^j\dots}+\ket{s^js^i\dots}\big)\,,\qquad
&\nu_i=1\,.\label{chib}
\end{align}
\end{subequations}
Here we have adopted the following convention: an ellipsis inside a
ket stands for an arbitrary ordering of the components in $\bs$ not
indicated explicitly. Note that the states~\eqref{chis} are defined
only up to equivalence, and that
$\ket{\chi_i(\bs)}=\ket{\chi_i(\pi\bs)}$ for any permutation $\pi\in
S_N$.
\begin{proposition}\label{P:basic}
Given a basic spin state $\ket\bs$, the associated spin states
$\ket{\chi_i(\bs)}$ are all in $\Si'/{\sim}$.
\end{proposition}
\begin{proof}
Consider first a state $\ket{\chi_i}$ of the type~\eqref{chia}.
Using the definition of the operator $A$ in Eq.~\eqref{A} we obtain
\begin{multline}\label{Achia}
N!A\ket{\chi_i}=2\nu_i(\nu_i-1)\sum_l\sum_{\pi\in S_{N-2}}
\pi\ket{\dots \underset{\underset{l}{\downarrow}}{s}^is^i\dots}\\
-2\sum_l\sum_{\substack{1\leq j,k\leq n\\j,k\neq i}}\sum_{\pi\in
S_{N-2}} \nu_j(\nu_k-\de_{jk})\pi\ket{\dots
\underset{\underset{l}{\downarrow}}{s}^js^k\dots}\,,
\end{multline}
where the permutations $\pi$ act only on the $N-2$ spin components
specified by an ellipsis. On the other hand, we have
\begin{multline}\label{Las}
N\cdot N!\La\ket{\bs}=\sum_l\sum_{\pi\in S_{N-2}}\nu_i(\nu_i-1)
\pi\ket{\dots \underset{\underset{l}{\downarrow}}{s}^is^i\dots}\\
{}+2\sum_l\sum_{\substack{1\leq j\leq n\\j\neq i}}\sum_{\pi\in
S_{N-1}}
\nu_j\pi\ket{\dots \underset{\underset{l}{\downarrow}}{s}^j\dots}\\
-\sum_l\sum_{\substack{1\leq j,k\leq n\\j,k\neq i}}\sum_{\pi\in
S_{N-2}} \nu_j(\nu_k-\de_{jk})\pi\ket{\dots
\underset{\underset{l}{\downarrow}}{s}^js^k\dots}\,.
\end{multline}
Comparing Eqs.~\eqref{Achia} and \eqref{Las} we obtain
\begin{align}
A\ket{\chi_i}&=2\bigg(N\La\ket\bs-\frac2{N!}\,\sum_{\substack{1\leq
j\leq n\\j\neq i}}\nu_j\sum_l
\sum_{\pi\in S_{N-1}}\pi\ket{\dots \underset{\underset{l}{\downarrow}}{s}^j\dots}\bigg)\notag\\
&=2\bigg(N-2\sum_{\substack{1\leq j\leq n\\j\neq
i}}\nu_j\bigg)\La\ket\bs =2(2\nu_i-N)\La\ket\bs\,.\label{chiainSip}
\end{align}
This shows that any state of the form~\eqref{chia} belongs to
$\Si'/{\sim}$. Suppose next that $\nu_i=1$, so that $\ket{\chi_i}$
is given by Eq.~\eqref{chib}. Since
\begin{equation}\label{chibinSip}
A\ket{\chi_i}=\frac2{N!}\, \sum_l\sum_{\substack{1\leq j\leq
n\\j\neq i}}\sum_{\pi\in S_{N-2}} \nu_j\pi\big(\ket{\dots
\underset{\underset{l}{\downarrow}}{s}^is^j\dots} +\ket{\dots
\underset{\underset{l}{\downarrow}}{s}^js^i\dots}\big)=4\La\ket\bs\,,
\end{equation}
it follows that in this case $\ket{\chi_i}$ is also in
$\Si'/{\sim}$\,.
\end{proof}
\begin{remark}
Just as symmetric spin states, cf.~Eq.~\eqref{AsLas}, the states
$\ket{\chi_i}$ satisfy the relation
\begin{equation}\label{AchiLachi}
A\ket{\chi_i}=\frac2{N-1}\,\La\ket{\chi_i}\,.
\end{equation}
Indeed, if $\nu_i>1$, from Eqs.~\eqref{chia} and~\eqref{chiainSip}
we have
\begin{align*}
\La\ket{\chi_i}&=\bigg[\nu_i(\nu_i-1) +\sum_{\substack{1\leq j\leq
n\\j\neq i}}\nu_j
-\sum_{\substack{1\leq j,k\leq n\\j,k\neq i}}\nu_j\nu_k\bigg]\La\ket{\bs}\\
&=\big[\nu_i(\nu_i-1)+N-\nu_i-(N-\nu_i)^2\big]\La\ket{\bs}\\
&=(N-1)(2\nu_i-N)\La\ket{\bs}=\frac{N-1}2\,A\ket{\chi_i}\,.
\end{align*}
On the other hand, if $\nu_i=1$ Eqs.~\eqref{chib}
and~\eqref{chibinSip} imply that
\[
\La\ket{\chi_i}=2\bigg(\sum_{\substack{1\leq j\leq n\\j\neq
i}}\nu_j\bigg)\La\ket{\bs}
=2(N-1)\La\ket{\bs}=\frac{N-1}2\,A\ket{\chi_i}\,.
\]
\end{remark}
\begin{example}\label{ex1b}
We shall now present all the states of the form~\eqref{chis} for
spin $1/2$. In this case, up to a permutation the basic state
$\ket\bs$ is given by
\begin{equation}\label{nu}
\ket\bs=
\ket{\overbrace{\vphantom{|}{+}\dots+}^{\nu}\overbrace{\vphantom{|}-\dots-}^{N-\nu}\:}\,.
\end{equation}
If $\nu$ is either $0$ or $N$, then $n=1$ and thus $\ket{\chi_1}$ is
of the type~\eqref{chia} and proportional to $\ket\bs$. If $\nu=1$,
then $n=2$ and we can take (dropping inessential factors)
\[
\ket{\chi_1}=\frac12\,\big(\ket{{+-}\cdots}+\ket{{-+}\cdots}\big)\sim\ket{{+-}\cdots}\,,\qquad
\ket{\chi_2}=\ket{{--}\cdots}\,.
\]
Although the states $\ket{\chi_1}$ and $\ket{\chi_2}$ are linearly
independent, the combination $2\ket{\chi_1}+(N-2)\ket{\chi_2}$ is
equivalent to a symmetric state. In the case $\nu=N-1$ the states
$\ket{\chi_i}$ are obtained from the previous ones by flipping the
spins. Finally, if $2\leq \nu\leq N-2$ then $n=2$ and the states
$\ket{\chi_i}$ are now given by
\begin{equation}\label{chi1}
\ket{\chi_1}=-\ket{\chi_2}=\nu(\nu-1)\ket{{++}\cdots}
-(N-\nu)(N-\nu-1)\ket{{--}\cdots}\,.
\end{equation}
\end{example}
According to the previous example, for spin $1/2$ there are exactly
$n-1$ independent states of the form~\eqref{chis} associated with each
basic state $\ket\bs$, up to symmetric states. We shall see next
that this fact actually holds for arbitrary spin:
\begin{proposition}\label{P:basic2}
Let $\ket\bs$ be a basic spin state. If $n$ is the number of
distinct components of $\bs$, there are exactly $n-1$ independent
states of the form~\eqref{chis} modulo symmetric states.
\end{proposition}
\begin{proof}
Let $p$ be the number of distinct components $s^i$ of $\bs$ such
that $\nu_i>1$. A straightforward computation shows that the
combination
\begin{equation}\label{sumchis}
\sum_{i=1}^n\ket{\chi_i}\sim(2-p)\sum_{i,j=1}^n\nu_i(\nu_j-\de_{ij})\ket{s^is^j\cdots}
\sim(2-p)N(N-1)\La\ket\bs
\end{equation}
is equivalent to a symmetric state. Suppose first that $p\neq2$. It
is immediate to check that in this case the set $\{\ket{\chi_i}:
i=1,\dots,n\}$ is linearly independent. If a linear combination
$\sum_{i=1}^nc_i\ket{\chi_i}$ is equivalent to a symmetric state
$\ket{\hat\bs}$, then $\ket{\hat\bs}$ must be proportional to
$\La\ket\bs$, so that we can write
\[
\sum_{i=1}^nc_i\ket{\chi_i}\sim\la(2-p)N(N-1)\La\ket\bs\,.
\]
Hence $\sum_{i=1}^n(c_i-\la)\ket{\chi_i}\sim0$, and the linear
independence of the states $\ket{\chi_i}$ implies that $c_i=\la$ for
all $i$. On the other hand, if $p=2$ the set $\{\ket{\chi_i}:
i=1,\dots,n\}$ is linearly dependent on account of
Eq.~\eqref{sumchis}, but removing one of the two states with
$\nu_i>1$ clearly yields a linearly independent set. It is also
obvious from the coefficients of the states $\ket{s^is^i\dots}$ that
no linear combination $\sum_{i=1}^nc_i\ket{\chi_i}$ can be
equivalent to a nonzero symmetric state.
\end{proof}
The next natural question to be addressed is whether the states of
the form~\eqref{chis} span the space $\Si'/{\sim}$ up to symmetric
states:
\begin{proposition}\label{prop.span}
The space $\big(\Si'/\La(\Si)\big)/{\sim}$ is spanned by states of
the form \eqref{chis}.
\end{proposition}
\begin{proof}
For conciseness, we present the proof of this result only for the
case $M=1/2$. Let $\Si_\nu$ denote the subspace of $\Si$ spanned by
basic spin states with $\nu$ ``${+}$'' spins, and set
$\Si'_\nu=\Si'\cap\Si_\nu$. Since the operators $A$ and $\La$
involved in the definition~\eqref{Sip} of $\Si'$ clearly preserve
$\Si_\nu$, it suffices to show that the states $\ket{\chi_i(\bs)}$
with $\bs$ given by~\eqref{nu} span the space $\Si'_\nu/{\sim}$ up
to symmetric states. Note first that the statement is trivial for
$\nu=0,1,N-1,N$, since in this case the states of the
form~\eqref{chis} obviously generate the whole space
$\Si_\nu/{\sim}$\,. Suppose, therefore, that $2\leq\nu\leq N-2$, so
that
\[
\Si_\nu/{\sim}=\big\langle\,\ket{{++}\cdots},\,\ket{{+-}\cdots},\,\ket{{--}\cdots}\,\big\rangle\,.
\]
Since the state~\eqref{chi1} and the symmetric state (up to
equivalence)
\[
\nu(\nu-1)\ket{{++}\cdots}+2\nu(N-\nu)\ket{{+-}\cdots}+(N-\nu)(N-\nu-1)\ket{{--}\cdots}
\]
both belong to $\Si'_\nu/{\sim}$, we need only show that (for
instance) $\ket{{+-}\cdots}$ is not in $\Si'_\nu/{\sim}$, i.e., that
$A\ket{{+-}\cdots}$ is not symmetric. But this is certainly the
case, since a state of the form
\[
\ket{\overbrace{\vphantom{|}{+-}\cdots{+-}}^{2k}\ms
\overbrace{\vphantom{|}{-}\cdots-}^{N-\nu-k}\ms
\overbrace{\vphantom{|}{+}\cdots+}^{\nu-k}\:}\,,\qquad
k=1,2,\dots,\min(\nu,N-\nu)\,,
\]
appears in $A\ket{{+-}\cdots}$ with coefficient
$2k(\nu-1)!(N-\nu-1)!$ depending on $k$.
\end{proof}

\section{The algebraic eigenfunctions}\label{sec.eig}

In the previous section we have provided a detailed description of the spaces
$\cH^n_\ep\subset\La\big(\CC[\bz]\otimes\Si\big)$ invariant under the
corresponding gauge Hamiltonian $\BH_\ep$. In this section we shall explicitly
compute all the eigenfunctions of the restrictions of the operators $\BH_\ep$
to their invariant spaces $\cH^{\ms n}_\ep$. This yields several
infinite\footnote{For the model~\eqref{V2} we shall see below that the number
of eigenfunctions with a given total momentum is finite.} families of
eigenfunctions for each of the models~\eqref{Vs}, which is the main result of
this paper. We shall use the term \emph{algebraic} to refer to these
eigenfunctions and their corresponding energies. It is important to observe
that the eigenfunctions of the gauge Hamiltonian $\BH_\ep$ that can be
constructed in this way are necessarily invariant under the whole symmetric
group, in spite of the fact that $\BH_\ep$ is symmetric only under cyclic
permutations. In fact, the explicit solutions of all known CS models with
near-neighbors interactions (both in the scalar and spin cases) can be
factorized as the product of a simple gauge factor analogous to $\mu$ times a
completely symmetric function~\cite{JK99,AJK01,DG01,EGKP05}. This, however,
does not rule out the existence of other eigenfunctions of the gauge
Hamiltonian $\BH_\ep$ invariant only under the subgroup of cyclic permutations,
which is indeed an interesting open problem.

\subsection*{Case a}

We shall begin with the model~\eqref{V0}, which is probably the most
interesting one due to the rich structure of its associated
invariant flag. In order to find the algebraic energies of the
model, note first that one can clearly construct a basis $\cB_0^n$
of $\cH_0^n$ whose elements are homogeneous polynomials in $\bz$
with coefficients in $\Si$. If $f\in\cB_0^n$ has degree $k$, then
$J^0f=kf$ and $T_0f$ has degree at most $k-2$. If $\cB_0^n$ is
ordered according to the degree, the operator $\BH_0$ is represented
in this basis by a triangular matrix with diagonal elements
$E_0+kc_0$, where $k=0,\dots,n$ is the degree. Thus the algebraic
eigenfunctions are the numbers
\[
E_k=E_0+2k\om,\qquad k=0,1,\dots.
\]

We shall next show that the algebraic eigenfunctions of $\BH_0$ can
be expressed in closed form in terms of generalized Laguerre and
Jacobi polynomials. The computation is basically a two-step
procedure. In the first place, one encodes the eigenvalue problem in
the invariant space $\cH_0^n$ as a system of linear partial
differential equations. The second step then consists in finding the
polynomial solutions of this system.

Regarding the first step, we shall need the following preliminary
lemma:
\begin{lemma}
The operator $\BH_0$ preserves the following subspaces of $\cH_0^n$:
\begin{align}
& \BcH^n_{0,\ket\bs}=\langle
f\Phi^{(0)},g\Phi^{(1)},h\Phi^{(2)}\rangle\,,
& &\ket\bs\in\Si\,,\label{BcH}\\
&
\tcH^n_{0,\ket\bs}=\BcH^n_{0,\ket\bs}+\langle\Th\tPhi^{(2)}\rangle\,,
& &\ket\bs\in\Si'\,,\quad S_{12}\ket\bs=\ket\bs\,,\label{tcH}\\
& \hcH^n_{0,\ket\bs}=\BcH^n_{0,\ket\bs}+\langle
w\hPhi^{(3)}\rangle\,, & &\ket\bs\in\Si\,,\quad
S_{12}\ket\bs=-\ket\bs\,,\label{hcH}
\end{align}
where $f$, $g$, $h$, $\Th$, $w$ are as in the definition of
$\cT_0^n$ in Theorem~\ref{thm.1}, and $\Phi^{(k)}$, $\tPhi^{(2)}$,
$\hPhi^{(3)}$ are respectively given by~\eqref{Phik}, \eqref{tPhi2}
and \eqref{hPhi3}.
\end{lemma}
\begin{proof}
The identity $T_0\big(f\Phi^{(0)}\big)=(T_0f)\Phi^{(0)}$ and
Eqs.~\eqref{Phi01}, \eqref{Phi02} and~\eqref{J0Phis} clearly imply
that the subspace $\BcH^n_{0,\ket\bs}$ is invariant under $\BH_0$.
Consider next the action of $\BH_0$ on a function of the form
$\Th\tPhi^{(2)}$. Since $\ket\bs$ is symmetric under $S_{12}$, we
can replace $\La\big[(z_1+z_2)\ket\bs\big]$ by $2\Phi^{(1)}$ in
Eq.~\eqref{tPhi02}. Secondly, any state $\ket\bs\in\Si'$ satisfies
the identity
\begin{equation}\label{AsLas2}
\sum_i\ket{\bs^+_{i,i+1}}=\frac2{N-1}\,\La\ket\bs\,.
\end{equation}
Indeed, we already know that the previous identity holds for
symmetric states (cf.~Eq.~\eqref{AsLas}) and for states of the
form~\eqref{chis} (cf.~Eqs.~\eqref{A} and~\eqref{AchiLachi}). On the
other hand, by Proposition~\ref{prop.span} every state in $\Si'$ is
a linear combination of a symmetric state, states of the
form~\eqref{chis}, and a state $\ket\bs$ such that
\mbox{$\La\big(z_1z_2\ket\bs\big)=0$}. But for the latter ``null''
state $\ket{\bs_{ij}}=0$ for all $i<j$, and hence
$A\ket\bs=\La\ket\bs=0$. Therefore, Eq.~\eqref{tPhi02} can be
written as
\begin{equation}\label{tPhi02mod}
T_0\big(\Th\tPhi^{(2)}\big)
=(T_0\Th+8\Th_2)\tPhi^{(2)}+4\Th_1\Phi^{(1)}-\frac{4a}{N-1}\,\Th\Phi^{(0)}\,.
\end{equation}
{}From the previous equation and Eq.~\eqref{J0Phis} it follows that
$\BH_0\big(\Th\tPhi^{(2)}\big)\in\tcH^n_{0,\ket\bs}$. Finally, if
$S_{12}\ket\bs=-\ket\bs$, Eq.~\eqref{hPhi03} reduces to
\begin{equation}
T_0\big(w\hPhi^{(3)}\big)=(T_0w+12w_2)\hPhi^{(3)}+4w_1\Phi^{(2)}-4(2a+1)w\Phi^{(1)}\,,
\label{hPhi03mod}
\end{equation}
which, together with Eq.~\eqref{J0Phis}, implies that
$\BH_0\big(w\hPhi^{(3)}\big)\in\hcH^n_{0,\ket\bs}$.
\end{proof}
\begin{remark}
The requirement that $\ket\bs$ be symmetric (respectively
antisymmetric) under $S_{12}$ in the definition of the space
$\tcH^n_{0,\ket\bs}$ (respectively $\hcH^n_{0,\ket\bs}$) is no real
restriction, since the antisymmetric (respectively symmetric) part
of $\ket\bs$ does not contribute to the state $\tPhi^{(2)}$
(respectively $\hPhi^{(3)}$).
\end{remark}

By the previous lemma, we can consider without loss of generality
eigenfunctions of $\BH_0$ of the form
\begin{equation}\label{Phi}
\Phi=f\Phi^{(0)}+g\Phi^{(1)}+h\Phi^{(2)}+\Th\tPhi^{(2)}+w\hPhi^{(3)}\,,\qquad\deg\Phi=k\,,
\end{equation}
where the spin functions $\Phi^{(k)}$, $\tPhi^{(2)}$ and
$\hPhi^{(3)}$ are all built from the same spin state $\ket\bs$. Note
that we can assume that $\Th\ms w=0$, and that the spin state
$\ket\bs$ is symmetric under $S_{12}$ and belongs to $\Si'$ if
$\Th\neq 0$, whereas it is antisymmetric under $S_{12}$ if $w\neq
0$.

Using Eqs.~\eqref{Phi01}, \eqref{Phi02}, \eqref{J0Phis},
\eqref{tPhi02mod} and~\eqref{hPhi03mod}, it is straightforward to
show that the eigenvalue equation $\BH_0\Phi=(E_0+2k\om )\Phi$ is
equivalent to the system
\begin{subequations}
\begin{align}
& \big[{-T_0}+2\om(J^0+3-k)\big]w-12w_2=0\,,\label{systemw}\\
& \big[{-T_0}+2\om(J^0+2-k)\big]\Th-8\Th_2=0\,,\label{systemTh}\\
& \big[{-T_0}+2\om(J^0+2-k)\big]h-8h_2=6g_3+4w_1\,,\label{systemh}\\
& \big[{-T_0}+2\om(J^0+1-k)\big]g-4g_2=4(h_1+\Th_1)-4(2a+1)w\,,\label{systemg}\\
&
\big[{-T_0}+2\om(J^0-k)\big]f=2\Big(g_1+(2a+1)h-\frac{2a}{N-1}\,\Th\Big)\,.\label{systemf}
\end{align}
\end{subequations}
Since $f$ and $g$ are linear in $\si_3$, we can write
\begin{equation}\label{fg}
f=p+\si_3 q\,,\qquad g=u+\si_3 v\,,
\end{equation}
where $p$, $q$, $u$ and $v$ are polynomials in $\si_1$ and $\si_2$.
Taking into account that the action of $T_0$ on scalar symmetric
functions is given by the RHS of Eq.~\eqref{Tep-res} with $\ep=0$,
and using Eqs.~\eqref{X0f}, \eqref{T20f} and~\eqref{J0f0}, we
finally obtain the following linear system of PDEs:
\begin{subequations}\label{system2}
\begin{align}
& \big[L_0-2\om(k-3)\big]w-12w_2=0\,,\label{system2w}\\
& \big[L_0-2\om(k-2)\big]\Th-8\Th_2=0\,,\label{system2th}\\
& \big[L_0-2\om(k-2)\big]h-8h_2=6v+4w_1\,,\label{system2h}\\
& \big[L_0-2\om(k-1)\big]u-4u_2=4h_1+4\Th_1+6\si_2v_1\notag\\
& \hphantom{\big[L_0-2\om(k-1)\big]u-4u_2=4h_1}+6(2a+1)\si_1 v-4(2a+1)w\,,\label{system2u}\\
& \big[L_0-2\om(k-4)\big]v-16v_2=0\,,\label{system2v}\\
& \big(L_0-2\om k\big)p=2u_1+2(2a+1)h-\frac{4a}{N-1}\,\Th+6\si_2q_1+6(2a+1)\si_1q\,,\label{system2p}\\
& \big[L_0-2\om(k-3)\big]q-12q_2=2v_1\,,\label{system2q}
\end{align}
\end{subequations}
where
\begin{multline}\label{L0}
L_0=-\big(N\pa_{\si_1}^2+4\si_1\pa_{\si_1}\pa_{\si_2}+4\si_2\pa_{\si_2}^2+2(2a+1)N\pa_{\si_2}\big)
\\+2\om(\si_1\pa_{\si_1}+2\si_2\pa_{\si_2})\,.
\end{multline}
As a consequence of the general discussion of the previous Section,
the latter system is guaranteed to possess polynomials solutions. In
fact, these polynomial solutions can be expressed in closed form in
terms of generalized Laguerre polynomials $L^\la_\nu$ and Jacobi
polynomials $P^{(\ga,\de)}_\nu$.
\begin{theorem}\label{thm.H0}
Let
\[
\al=N\Big(a+\frac12\Big)-\frac32\,,\qquad
\be\equiv\be(m)=1-m-N\Big(a+\frac12\Big)\,,\qquad
t=\frac{2r^2}{N\Bx^2}-1\,,
\]
where $\Bx=\frac 1N\sum_ix_i$ is the center of mass coordinate. The
Hamiltonian $H_0$ possesses the following families of spin
eigenfunctions with eigenvalue \mbox{$E_{lm}=E_0+2\om(2l+m)$}, with
$l\geq 0$ and $m$ as indicated in each case:
\begin{align*}
\Psi^{(0)}_{lm}&=\mu\ms\Bx^mL^{-\be}_l(\omega r^2)
P^{(\al,\be)}_{\lfloor\frac m2\rfloor}(t)\,\Phi^{(0)}\,,\qquad m\geq0\,,\\
\Psi^{(1)}_{lm}&=\mu\ms\Bx^{m-1}L^{-\be}_l(\om
r^2)P^{(\al+1,\be)}_{\lfloor\frac{m-1}2\rfloor}(t)\,\big(\Phi^{(1)}
-\Bx \,\Phi^{(0)}\big)\,,\qquad m\geq1\,,\displaybreak[0]\\
\Psi^{(2)}_{lm}&=\mu\ms\Bx^{m-2}L^{-\be}_l(\om
r^2)\bigg[P^{(\al+2,\be)}_{\lfloor\frac m2\rfloor-1}(t)\big(\Phi^{(2)}-2\Bx\,\Phi^{(1)}\big)\\
&\qquad\qquad\qquad+\Bx^2\bigg(P^{(\al+2,\be)}_{\lfloor\frac
m2\rfloor-1}(t)-\frac{2(\al+1)}{2\lfloor\tfrac{m-1}2\rfloor+1}\,P^{(\al+1,\be)}_{\lfloor\frac
m2\rfloor-1}(t)\bigg)\Phi^{(0)}\bigg]\,,\quad m\geq2\,,\displaybreak[0]\\
\widetilde\Psi^{(2)}_{lm}&=\mu\ms\Bx^{m-2}L^{-\be}_l(\om
r^2)\bigg[P^{(\al+2,\be)}_{\lfloor\frac m2\rfloor-1}(t)\big(\widetilde\Phi^{(2)}-2\Bx\,\Phi^{(1)}\big)\\
&\qquad+\Bx^2\bigg(P^{(\al+2,\be)}_{\lfloor\frac
m2\rfloor-1}(t)+\frac{2(\al+1)}{\big(2\lfloor\tfrac{m-1}2\rfloor+1\big)(N-1)}\,P^{(\al+1,\be)}_{\lfloor\frac
m2\rfloor-1}(t)\bigg)\Phi^{(0)}\bigg]\,,\quad m\geq2\,,\displaybreak[0]\\
\Psi^{(3)}_{lm}&=\mu\ms\Bx^{m-3}L^{-\be}_l(\omega r^2)\bigg[
\frac2{3N}\,P^{(\al+3,\be)}_{\lfloor\frac{m-3}2\rfloor}(t)\sum_i
x_i^3+\Bx^3\vp_m(t)\bigg]\,\Phi^{(0)}\,, \qquad m\geq3\,,\displaybreak[0]\\
\hPsi^{(3)}_{lm}&=\mu\ms\Bx^{m-3}L^{-\be}_l(\omega r^2)\bigg[
P^{(\al+3,\be)}_{\lfloor\frac{m-3}2\rfloor}(t)\big(\hPhi^{(3)}-2\Bx\,\Phi^{(2)}\big)\\
&\qquad\qquad\qquad+2\Bx^2\bigg(P^{(\al+3,\be)}_{\lfloor\frac
{m-3}2\rfloor}(t)+\frac{2(\al+2)}{2\lfloor\frac
m2\rfloor-1}\,P^{(\al+2,\be)}_{\lfloor\frac
{m-3}2\rfloor}(t)\bigg)\Phi^{(1)}\\
&\qquad\qquad\qquad-2\Bx^3\bigg(\frac13\,P^{(\al+3,\be)}_{\lfloor\frac{m-3}2\rfloor}(t)
+\frac1{2\lfloor\frac m2\rfloor-1}\,P^{(\al+2,\be)}_{\lfloor\frac{m-3}2\rfloor}(t)\\
&\qquad\qquad\qquad\qquad
+\big(1-(-1)^m\big)\frac{2\al+3}{2m(m-2)}\,P^{(\al+1,\be)}_{\lfloor\frac{m-3}2\rfloor}(t)\bigg)\Phi^{(0)}\bigg]
\,,\quad m\geq3\,,\displaybreak[0]\\
\Psi^{(4)}_{lm}&=\mu\ms \Bx^{m-4}L^{-\be}_l(\om
r^2)\bigg[\frac3{2(\lfloor\frac{m-3}2\rfloor+\frac12)}\,\Bx^2P^{(\al+3,\be)}_{\lfloor\frac
m2\rfloor-2}(t)\,\Phi^{(2)}\\
&\qquad\qquad\qquad+\Big(\frac32\,\Bx^3\phi_m(t)-\frac{1}N\,P^{(\al+4,\be)}_{\lfloor\frac
m2\rfloor-2}(t)\sum_i x_i^3\Big)\Phi^{(1)}\\
&\qquad\qquad\qquad+\Big(\frac1N\,\Bx P^{(\al+4,\be)}_{\lfloor\frac
m2\rfloor-2}(t)\sum_ix_i^3+\frac32\,\Bx^4\chi_m(t)\Big)\Phi^{(0)}\bigg]\,,\quad
m\geq4\,.
\end{align*}
Here
\[
\Phi^{(k)}=\La(x_1^k\ket\bs),\qquad\tPhi^{(2)}=\La(x_1x_2\ket\bs),\qquad
\hPhi^{(3)}=\La(x_1x_2(x_1-x_2)\ket\bs),
\]
where the spin state $\ket\bs$ is symmetric under $S_{12}$ and
belongs to $\Si'$ for the eigenfunction $\widetilde\Psi^{(2)}_{lm}$,
and is antisymmetric under $S_{12}$ for the eigenfunction
$\hPsi^{(3)}_{lm}$. The functions $\vp_m$, $\phi_m$ and $\chi_m$ are
polynomials given explicitly by
\begin{align*}
\varphi_m&=\frac{m+2\al+2}{m-1}\,P^{(\al+2,\be-2)}_{\frac m2}
-P^{(\al+3,\be-1)}_{\frac
m2-1}-\frac{4\al+7}{m-1}\,P^{(\al+2,\be-1)}_{\frac m2-1}
+\frac13\,P^{(\al+3,\be)}_{\frac m2-2}\,,\displaybreak[0]\\
\phi_m&=P^{(\al+4,\be-1)}_{\frac m2-1}-2P^{(\al+3,\be-1)}_{\frac
m2-1}-\frac{m+2\al+3}{(m-1)(m-3)}\,P^{(\al+2,\be-1)}_{\frac
m2-1}\\
&\hphantom{{}=P^{(\al+4,\be-1)}_{\frac
m2-1}}-\frac13\,P^{(\al+4,\be)}_{\frac
m2-2}+\frac{m+2\al-1}{m-3}\,P^{(\al+3,\be)}_{\frac m2-2}\,,\displaybreak[0]\\
\chi_m&=\frac{3m+2\al}{(m-1)(m-3)}\,P^{(\al+2,\be-1)}_{\frac
m2-1}+\frac{2m-7}{m-3}\,P^{(\al+3,\be-1)}_{\frac
m2-1}-P^{(\al+4,\be-1)}_{\frac m2-1}\\
&\hphantom{{}=P^{(\al+4,\be-1)}_{\frac
m2-1}}-\frac{m+2\al+2}{(m-1)(m-3)}\,P^{(\al+2,\be)}_{\frac
m2-2}-\frac{m+2\al}{m-3}\,P^{(\al+3,\be)}_{\frac
m2-2}+\frac13\,P^{(\al+4,\be)}_{\frac m2-2}\,,
\end{align*}
for even $m$, and
\begin{align*}
\varphi_m&=2P^{(\al+2,\be-1)}_{\frac{m-1}2}-P^{(\al+3,\be-1)}_{\frac{m-1}2}
+\frac13\,P^{(\al+3,\be)}_{\frac{m-3}2}
+\frac{m+2\al+2}{m(m-2)}\,P^{(\al+1,\be)}_{\frac{m-3}2}\\
&\hphantom{{}=2P^{(\al+2,\be-1)}_{\frac{m-1}2}}
-\frac{m+2\al+2}{m-2}\,P^{(\al+2,\be)}_{\frac{m-3}2}\,,\displaybreak[0]\\
\phi_m&=P^{(\al+4,\be-1)}_{\frac{m-3}2}-\frac{2m-5}{m-2}\,P^{(\al+3,\be)}_{\frac{m-3}2}-
\frac13\,P^{(\al+4,\be)}_{\frac{m-5}2}+\frac{m+2\al-1}{m-2}\,P^{(\al+3,\be)}_{\frac{m-5}2}\,,
\displaybreak[0]\\
\chi_m&=\frac{2m-3}{m(m-2)}\,P^{(\al+2,\be-1)}_{\frac{m-3}2}+\frac{2(m-3)}{m-2}
\,P^{(\al+3,\be-1)}_{\frac{m-3}2}-P^{(\al+4,\be-1)}_{\frac{m-3}2}\\
&\hphantom{{}=2P^{(\al+2,\be-1)}_{\frac{m-1}2}}-\frac{m+2\al+1}{m(m-2)}\,P^{
(\al+2,\be)}_{\frac{m-5}2}-\frac{m+2\al}{m-2}\,P^{(\al+3,\be)}_{\frac{m-5}2}
+\frac13\,P^{(\al+4,\be)}_{\frac{m-3}2}\,,
\end{align*}
for odd $m$. For every $n=0,1,\dots$, the above eigenfunctions with
$2l+m\leq n$ span the whole $H_0$-invariant space $\mu\ms\cH_0^n$.
\end{theorem}
\begin{proof}
Recall, to begin with, that the algebraic eigenfunctions of $H_0$
are of the form $\Psi=\mu\Phi$, with $\mu$ given in
Table~\ref{table:params} and $\Phi$ an eigenfunction of $\BH_0$ of
the form~\eqref{Phi}-\eqref{fg}. In order to determine $\Phi$, we
must find the most general polynomial solution of the inhomogeneous
linear system~\eqref{system2}. {}From the structure of this system
it follows that there are seven types of independent solutions,
characterized by the vanishing of certain subsets of the unknown
functions $p,q,u,v,h,\Th,w$. These types are listed in
Table~\ref{table:sols0}, where in the last column we have indicated
the eigenfunction of $H_0$ obtained from each case.
\begin{table}[t]
\begin{center}
\caption{The seven types of solutions of the system~\eqref{system2} and their
corresponding eigenfunctions.\vspace{3mm}}\label{table:sols0}
\begin{tabular}{lc}\hline
\vrule height 15pt depth 9pt width0pt {\hfill
Conditions\hphantom{\qquad}\hfill} & Eigenfunction\\ \hline
\BStrut $q=u=v=h=\Th=w=0,\quad p\neq0$\hphantom{\qquad} &
$\Psi^{(0)}_{lm}$\\ \hline
\BStrut $u=v=h=\Th=w=0,\quad q\neq0$\hphantom{\qquad} &
$\Psi^{(3)}_{lm}$\\ \hline
\BStrut $q=v=h=\Th=w=0,\quad u\neq0$\hphantom{\qquad} &
$\Psi^{(1)}_{lm}$\\ \hline
\BStrut $q=v=\Th=w=0,\quad h\neq0$\hphantom{\qquad} &
$\Psi^{(2)}_{lm}$\\ \hline
\BStrut $q=v=h=w=0,\quad \Th\neq0$\hphantom{\qquad} &
$\widetilde\Psi^{(2)}_{lm}$\\ \hline
\BStrut $q=v=\Th=0,\quad w\neq0$\hphantom{\qquad} &
$\hPsi^{(3)}_{lm}$\\ \hline
\BStrut $\Th=w=0,\quad v\neq0$\hphantom{\qquad} &
$\Psi^{(4)}_{lm}$\\ \hline
\end{tabular}
\end{center}
\end{table}
We shall present here in detail the solution of the
system~\eqref{system2} for the case $q=v=h=\Th=w=0$ and $u\neq0$,
which yields the eigenfunctions of the form $\Psi^{(1)}_{lm}$ (the
procedure for the other cases is essentially the same). In this case
the system~\eqref{system2} reduces to
\begin{equation}
\big[L_0-2\om(k-1)\big]u-4u_2=0\,,\qquad \big(L_0-2\om
k\big)p=2u_1\,.\label{system3}
\end{equation}
Let us begin with the homogeneous equation for $u$. We shall look
for polynomial solutions of this equation of the form
$u=Q(\si_1,\si_2)R(\si_2)$, where $Q$ is a homogeneous polynomial of
degree $m-1$ in $\bz$ and $R$ is a polynomial of degree $l$
in~$\si_2$, so that $k=\deg\Phi=2l+m$ by Eq.~\eqref{Phi}. {}From
Eq.~\eqref{L0} and the homogeneity of $Q$ we have
\begin{align*}
L_0(QR)&=(L_0Q)R+Q(L_0R)-4\si_1Q_1R_2-8\si_2Q_2R_2\\
&=(L_0Q)R+Q(L_0-4(m-1)\pa_{\si_2})R\,.
\end{align*}
Hence the equation for $u$ can be written as
\[
R\big(\widehat
L_0-4\pa_{\si_2}\big)Q=Q\big(-L_0+4m\pa_{\si_2}+4l\om\big)R\,,
\]
where $\widehat L_0=L_0|_{\om=0}$. Since $\big(\widehat
L_0-4\pa_{\si_2}\big)Q$ is a homogeneous polynomial of degree $m-3$
in $\bz$, both sides of the latter equation must vanish separately.
We are thus led to the following decoupled equations for $Q$ and
$R$:
\begin{align}
&\big(\widehat L_0-4\pa_{\si_2}\big)Q=0\label{eqQ}\,,\\
&\big(-L_0+4m\pa_{\si_2}+4l\om\big)R=0\label{eqR}\,.
\end{align}
In terms of the variable $\rho=\om\si_2$, Eq.~\eqref{eqR} can be
written as
\[
4\om\cL^{-\be}_l(R)=0\,,
\]
where
\begin{equation}\label{Lop}
\cL^\la_\nu=\rho\ms\pa_\rho^2+(\la+1-\rho)\pa_\rho+\nu
\end{equation}
is the generalized Laguerre operator. Hence $R$ is proportional to
the generalized Laguerre polynomial $L^{-\be}_l(\om\si_2)$. On the
other hand, we can write $Q=\si_1^{m-1}P(t)$ where $P$ is a
polynomial in the homogeneous variable
$t=\frac{2N\si_2}{\si_1^2}-1$. With this substitution,
Eq.~\eqref{eqQ} becomes
\[
4N\si_1^{m-3}\cJ^{(\al+1,\be)}_{\lfloor\frac{m-1}2\rfloor}(P)=0\,,
\]
where the Jacobi operator $\cJ^{(\ga,\de)}_\nu$ is given by
\[
\cJ^{(\ga,\de)}_\nu=(1-t^2)\ms\pa_t^2+\big[\de-\ga-(\ga+\de+2)\ms
t\big]\ms\pa_t+\nu(\nu+\ga+\de+1)\,.
\]
Thus $P(t)$ is proportional to the Jacobi polynomial
$P^{(\al+1,\be)}_{\lfloor\frac{m-1}2\rfloor}(t)$, so that we can
take
\begin{equation}\label{u}
u=\si_1^{m-1}L^{-\be}_l(\om\si_2)P^{(\al+1,\be)}_{\lfloor\frac{m-1}2\rfloor}(t)\,.
\end{equation}

We must next find a particular solution of the inhomogeneous
equation for $p$ in~\eqref{system3}, since the general solution of
the corresponding homogeneous equation yields an eigenfunction of
the simpler type $\Psi^{(0)}_{lm}$. Since
\[
u_1=\si_1^{m-2}L^{-\be}_l(\om\si_2)\Big[
(m-1)P^{(\al+1,\be)}_{\lfloor\frac{m-1}2\rfloor}(t)-2(t+1)\dot
P^{(\al+1,\be)}_{\lfloor\frac{m-1}2\rfloor}(t)\Big]
\]
(where the dot denotes derivative with respect to $t$), we make the
ansatz $p=\BQ(\si_1,\si_2)\BR(\si_2)$, where $\BQ$ is a homogeneous
polynomial of degree $m$ in $\bz$ and $\BR$ is a polynomial of
degree $l$ in $\si_2$. Substituting this ansatz into the second
equation in~\eqref{system3} and proceeding as before we immediately
obtain
\[
\BR\big(\widehat
L_0\BQ\big)+\BQ\big(L_0-4m\pa_{\si_2}-4l\om\big)\BR=2u_1\,.
\]
If we set $\BR=L^{-\be}_l(\om\si_2)$ the second term of the LHS vanishes, and
canceling the common factor $L^{-\be}_l(\om\si_2)$ we are left with the
following equation for $\BQ$:
\[
\widehat L_0\BQ=2\si_1^{m-2}\Big[
(m-1)P^{(\al+1,\be)}_{\lfloor\frac{m-1}2\rfloor}(t)-2(t+1)\dot
P^{(\al+1,\be)}_{\lfloor\frac{m-1}2\rfloor}(t)\Big]\,.
\]
The form of the RHS of this equation suggests the ansatz
$\BQ=\si_1^m\BP(t)$, with $\BP$ a polynomial in the variable $t$.
The previous equation then yields
\begin{equation}\label{cJBP}
\cJ^{(\al,\be)}_{\lfloor\frac m2\rfloor}(\BP)=\frac1{2N}\,\Big[
(m-1)P^{(\al+1,\be)}_{\lfloor\frac{m-1}2\rfloor}(t)-2(t+1)\dot
P^{(\al+1,\be)}_{\lfloor\frac{m-1}2\rfloor}(t)\Big]\,.
\end{equation}
{}From the definition of the Jacobi operator we easily obtain
\[
\cJ^{(\al,\be)}_{\lfloor\frac
m2\rfloor}=\cJ^{(\al+1,\be)}_{\lfloor\frac{m-1}2\rfloor}+(1+t)\,\pa_t
-\frac12\,(m-1)\,,
\]
which implies that
$\ds\BP=-\frac1N\,P^{(\al+1,\be)}_{\lfloor\frac{m-1}2\rfloor}$ is a
particular solution of Eq.~\eqref{cJBP}. Hence
\begin{equation}\label{p}
p=-\frac1N\,\si_1^mL^{-\be}_l(\om\si_2)P^{(\al+1,\be)}_{\lfloor\frac{m-1}2\rfloor}(t)
\end{equation}
is a particular solution of the inhomogeneous equation
in~\eqref{system3}. We have thus shown that
$\Phi=p\,\Phi^{(0)}+u\,\Phi^{(1)}$, with $u$ and $p$ respectively
given by Eqs.~\eqref{u} and~\eqref{p}, is an eigenfunction of
$\BH_0$ with eigenvalue $E_0+2\om(2l+m)$. Multiplying~$\Phi$ by the
gauge factor $\mu$ we obtain the eigenfunction $\Psi^{(1)}_{lm}$ of
$H_0$ in the statement.

It remains to show that the states $\Psi^{(k)}_{lm}$
($k=0,\dots,4$), $\widetilde\Psi^{(2)}_{lm}$ and $\hPsi^{(3)}_{lm}$
with \mbox{$2l+m\leq n$} generate the
spaces~\eqref{BcH}--\eqref{hcH}. Consider first the ``monomials'' of
the form $\mu\si_1^m\si_2^l\Phi^{(0)}$, which belong to
$\mu\ms\BcH^n_{0,\ket\bs}$ if $2l+m\leq n$. We can order such
monomials as follows: we say that
$\mu\si_1^{m'}\si_2^{l'}\Phi^{(0)}\prec\mu\si_1^m\si_2^l\Phi^{(0)}$
if $2l'+m'<2l+m$, or $2l'+m'=2l+m$ and $m'<m$. {}From the
expansion~\cite[Eq.~8.962.1]{GR00}
\[
P^{(\ga,\de)}_\nu(t)=\frac1{\nu!}\,\sum_{k=0}^\nu\frac1{2^kk!}\,
(-\nu)_k(\ga+\de+\nu+1)_k(\ga+k+1)_{\nu-k}\,(1-t)^k\,,
\]
where $(x)_k$ is the Pochhammer symbol
\[
(x)_k=x(x+1)\cdots(x+k-1)\,,
\]
it follows that $P^{(\ga,\de)}_\nu(0)>0$ provided that $\ga+1>0$ and
$\ga+\de+2\nu<0$. In particular, $P^{(\al,\be)}_{\lfloor\frac
m2\rfloor}(0)>0$ since
\[
\al+1=N\bigg(a+\frac12\bigg)-\frac12>N-\frac12>0\,,\quad
\al+\be+2\bigg\lfloor\frac m2\bigg\rfloor=2\bigg\lfloor\frac
m2\bigg\rfloor-m-\frac12\leq-\frac12<0\,.
\]
Hence we can write
\[
\Psi^{(0)}_{lm}=\mu\Phi^{(0)}\big(c_{lm}\si_1^m\si_2^l+\text{l.o.t.}\big)\,,
\]
where $c_{lm}\neq 0$, so that
\[
\big\langle\Psi^{(0)}_{lm} : 2l+m\leq n\big\rangle
=\big\langle\mu\si_1^m\si_2^l\Phi^{(0)} : 2l+m\leq n\big\rangle\,.
\]
Likewise, a similar argument shows that for $m\geq 1$
\[
\big\langle \mu\ms\Bx^{m-1}L^{-\be}_l(\om
r^2)P^{(\al+1,\be)}_{\lfloor\frac{m-1}2\rfloor}(t)\,\Phi^{(1)}:
2l+m\leq n\big\rangle =\big\langle\mu\si_1^{m-1}\si_2^l\Phi^{(1)} :
2l+m\leq n\big\rangle\,,
\]
and therefore
\[
\big\langle\Psi^{(0)}_{lm},\,\Psi^{(1)}_{lm} : 2l+m\leq n\big\rangle
=\big\langle\mu\si_1^m\si_2^l\Phi^{(0)},\,\mu\si_1^{m-1}\si_2^l\Phi^{(1)}
: 2l+m\leq n\big\rangle\,.
\]
Proceeding in the same way with the remaining spin eigenfunctions we
can finally show that
\[
\big\langle\Psi^{(k)}_{lm}: k=0,\dots,4\,,\,2l+m\leq n\big\rangle
=\mu\ms\BcH^n_{0,\ket\bs}\,,
\]
and that
\begin{align*}
&\mu\ms\BcH^n_{0,\ket\bs}+\big\langle\widetilde\Psi^{(2)}_{lm}:
2l+m\leq n\big\rangle
=\mu\ms\tcH^n_{0,\ket\bs}\,,\qquad\ket\bs\in\Si',\quad S_{12}\ket\bs=\ket\bs,\\
&\mu\ms\BcH^n_{0,\ket\bs}+\big\langle\hPsi^{(3)}_{lm}: 2l+m\leq
n\big\rangle =\mu\ms\hcH^n_{0,\ket\bs}\,,\qquad
S_{12}\ket\bs=-\ket\bs,
\end{align*}
as claimed.
\end{proof}
%\begin{remark}
%By Remark~\ref{rem1}, the coefficients of $\Phi^{(0)}$ in the spin
%eigenfunctions $\Psi^{(0)}_{lm}$ and $\Psi^{(3)}_{lm}$ yield the two
%families of eigenfunctions of the scalar reduction $H^{\text{sc}}_0$
%of the model~\eqref{V0} presented without proof in our previous
%paper~\cite{EFGR05b}. Earlier work on the scalar model
%$H^{\text{sc}}_0$ had established the existence of two families of
%eigenfunctions of the form $\mu L_l^{-\be}(\om r^2)p_\nu(\bx)$, with
%$p_\nu$ a homogeneous polynomial of degree $\nu\geq 3$, only for
%$\nu\leq 6$ and $N\geq\nu$~\cite{AJK01}. More recently, Ezung
%et~al.~\cite{EGKP05} have rederived a very small subset of these
%scalar eigenfunctions by mapping $H^{\text{sc}}_0$ to $N$ decoupled
%oscillators.
%\end{remark}\goodbreak

\subsection*{Case b}

Since $c_0=4\om$ in this case, reasoning as before we conclude that
the algebraic energies are the numbers
\[
E_k=E_0+4k\om,\qquad k=0,1,\dots,
\]
where $k$ is the degree in $\bz$ of the corresponding eigenfunctions
of $\BH_1$. We shall see below that these eigenfunctions can be
written in terms of generalized Laguerre polynomials. To this end,
we begin by identifying certain subspaces of $\cH_1^n$ invariant
under $\BH_1$.

\begin{lemma}
For any given spin state $\ket\bs\in\Si$, the operator $\BH_1$
preserves the subspace
\begin{equation}\label{BcH1n}
\cH_{1,\ket\bs}^n= \big\langle
f(\si_1,\si_2)\,\Phi^{(0)},\,g(\si_1)\,\Phi^{(1)}:{}f_{22}=0\big\rangle
\subset\cH_1^n\,,
\end{equation}
where $f$ and $g$ are polynomials of degrees at most $n$ and $n-1$
in $\bz$, respectively, and $\Phi^{(k)}$ is given by~\eqref{Phik}.
\end{lemma}

\begin{proof}
The statement follows from the obvious identity
$T_1\big(f\Phi^{(0)}\big)=(T_1f)\Phi^{(0)}$ and Eqs.~\eqref{X1f},
\eqref{Tep-res}, \eqref{T21f}, \eqref{Phi11}, \eqref{J0Phis}
and~\eqref{JmPhis}.
\end{proof}

By the previous lemma we can assume that the eigenfunctions of
$\BH_1$ in $\cH_{1,\ket\bs}^n$ are of the form
\begin{equation}\label{Phib}
\Phi=f\Phi^{(0)}+g\Phi^{(1)}\,,\qquad\deg\Phi=k\leq n\,.
\end{equation}
{}From Eqs.~\eqref{Phi11}, \eqref{J0Phis} and \eqref{JmPhis} it
easily follows that the eigenvalue equation
$\BH_1\Phi=(E_0+4k\om)\Phi$ can be cast into the system
\begin{subequations}\label{systemb}
\begin{align}
& \Big[{-T_1}+\om(J^0+1-k)-\Big(b+\frac12\Big)J^-\Big]g-2g_1=0\,,\label{systembg}\\
&
\Big[{-T_1}+\om(J^0-k)-\Big(b+\frac12\Big)J^-\Big]f=\Big(2a+b+\frac12\Big)g\,.\label{systembf}
\end{align}
\end{subequations}
Since $f$ is linear in $\si_2$ (cf.~Eq.~\eqref{BcH1n}), we can write
\begin{equation}\label{fpq}
f=p+\si_2 q,
\end{equation}
where $p$ and $q$ are polynomials in $\si_1$. Using
Eqs.~\eqref{X1f}, \eqref{Tep-res}, \eqref{T21f}, \eqref{J0f1}
and~\eqref{Jmf1} we can easily rewrite the system~\eqref{systemb} as
follows:
\begin{subequations}\label{systemb2}
\begin{align}
& \big[L_1-\om(k-1)\big]g-2g_1=0\,,\label{systemb2g}\\
& \big[L_1-\om(k-2)\big]q-4q_1=0\,,\label{systemb2q}\\
& \big(L_1-\om
k\big)p=\Big(2a+b+\frac12\Big)g+2\Big(4a+b+\frac32\Big)\si_1q\,,\label{systemb2p}
\end{align}
\end{subequations}
where
\begin{equation}\label{L1}
L_1=-\si_1\pa_{\si_1}^2+\Big[\om\si_1-\Big(2a+b+\frac12\Big)N\Big]\pa_{\si_1}\,.
\end{equation}
The last step is to construct the polynomials solutions of the
system~\eqref{systemb2}, which can be expressed in terms of
generalized Laguerre polynomials, according to the following
theorem.
\begin{theorem}\label{thm.H1}
The Hamiltonian $H_1$ possesses the following families of spin
eigenfunctions with eigenvalue $E_k=E_0+4k\om$:
\begin{align*}
\Psi^{(0)}_k&=\mu L^{\al-1}_k(\omega r^2)\,\Phi^{(0)}\,,\qquad k\geq0\,,\\[1mm]
\Psi^{(1)}_k&=\mu L^{\al+1}_{k-1}(\omega
r^2)\big[N\Phi^{(1)}-r^2\Phi^{(0)}\big]
\,,\qquad k\geq1\,,\\[1mm]
\Psi^{(2)}_k&=\mu L^{\al+3}_{k-2}(\omega r^2)\Big[N(\al+1)\sum_i
x_i^4-\be\ms r^4\Big]\,\Phi^{(0)}\,, \qquad k\geq 2\,,
\end{align*}
where $\al=N(2a+b+\frac12)$, $\be=N(4a+b+\frac32)$, and
$\Phi^{(j)}=\La\big(x_1^{2j}\ket\bs\big)$, with $j=0,1$ and
$\ket\bs\in\Si$. For each $\ket\bs\in\Si$ and $n=0,1,\dots$, the
above eigenfunctions with $k\leq n$ span the whole $H_1$-invariant
space $\mu\cH_{1,\ket\bs}^n$.
\end{theorem}

\begin{proof}
As in the previous case, the algebraic eigenfunctions of $H_1$ are
of the form $\Psi=\mu\Phi$, where $\mu$ is given in
Table~\ref{table:params} and $\Phi$ is an eigenfunction of $\BH_1$
of the form~\eqref{Phib}-\eqref{fpq}. The functions $p$, $q$ and $g$
are polynomials in $\si_1$ determined by the
system~\eqref{systemb2}, which in terms of the variable
$t=\om\si_1\equiv\om r^2$ can be written as
\begin{equation}\label{systemb3}
\cL^{\al+1}_{k-1}g=\cL^{\al+3}_{k-2}q=0\,,\qquad
\cL^{\al-1}_kp=-\frac\al{N\om}\,g-\frac{2\be}{N\om^2}\,tq\,,
\end{equation}
where $\cL^{\la}_\nu$ is the generalized Laguerre operator (cf.
Eq.~\eqref{Lop}). The general polynomial solutions of the first two
equations in~\eqref{systemb3} are respectively given by
\begin{equation}\label{gq}
g=c_1L^{\al+1}_{k-1}(t),\qquad q=c_2L^{\al+3}_{k-2}(t).
\end{equation}
On the other hand, from the elementary identity
\[
\cL^{\la}_\nu\Big(t^lL^{\la+2l}_{\nu-l}(t)\Big)=l(l+\la)t^{l-1}L^{\la+2l}_{\nu-l}(t)\,,
\]
it follows that the general polynomial solution of the third
equation in~\eqref{systemb3} is given by
\begin{equation}\label{bp}
p=c_0\ms L^{\al-1}_k(t)-\frac{c_1}{N\om}\,t\ms
L^{\al+1}_{k-1}(t)-\frac{c_2\be}{N\om^2(\al+1)}\,t^2
L^{\al+3}_{k-2}(t)\,.
\end{equation}
Equations~\eqref{gq} and~\eqref{bp} immediately yield the formulas
of the eigenfunctions in this case. The last assertion in the
statement of the theorem follows from the fact that the functions
$p$, $q$ and $g$ in Eqs.~\eqref{gq} and~\eqref{bp} are the most
general polynomial solution of the system~\eqref{systemb3}.
\end{proof}

\begin{remark}
It should be noted that for $\om=0$ the potentials~\eqref{V0} and~\eqref{V1}
scale as $r^{-2}$ under dilations of the coordinates (as is the case for the
original Calogero model). The standard argument used in the solution of the
Calogero model shows that there is a basis of eigenfunctions of these models of
the form $\mu(\bx) L^\la_\nu(\om r^2)F(\bx)$, where $F$ is a homogeneous
spin-valued function. The eigenfunctions presented in Theorems~\ref{thm.H0}
and~\ref{thm.H1} are indeed of this form.
\end{remark}

\subsection*{Case c}

In this case the number of independent algebraic eigenfunctions is
essentially finite. We shall take, for definiteness, the plus sign
in the change of variable listed in Table~\ref{table:params} (it
will be apparent from the discussion that follows that the minus
sign does not yield additional solutions).

Let us first note that the potential for this model is
translationally invariant, so that the total momentum operator
$P=-\iu\sum_k\pa_{x_k}$ commutes with the Hamiltonian $H_2$. Hence
the eigenfunctions of $H_2$ can be assumed to have well-defined
total momentum. Equivalently, since
\begin{equation}\label{PJ0}
{\mu^{-1}\ms P\ms\mu\ms\Big|}_{x_k=-\frac\iu2\ms\log{z_k}}=2J^0,
\end{equation}
the eigenfunctions of $\BH_2$ can be assumed to be homogeneous in
$\bz$. Let $\Phi$ be a homogeneous eigenfunction of $\BH_2$ of
degree $k$ and eigenvalue $E$, so that $\Psi=\mu\Phi$ is an
eigenfunction of $H_2$ with total momentum $2k$
(cf.~Eq.~\eqref{PJ0}) and energy $E$. By Eq.~\eqref{PJ0}, the
function $\tau_N^\la\Psi$ clearly has total momentum $2(k+N\la)$. In
fact, the following lemma implies that $\tau_N^\la\Psi$ is also an
eigenfunction of $H_2$ with a suitably boosted energy:
\begin{lemma}\label{lemma.5}
Let $\Phi$ be a homogeneous eigenfunction of $\BH_2$ of degree $k$
and eigenvalue~$E$. Then $\tau_N^\la\Phi$ is an eigenfunction of
$\BH_2$ with eigenvalue $E+8k\la+4N\la^2$.
\end{lemma}
\begin{proof}
{}From the identity
\[
\sum_i\frac1{z_i-z_{i+1}}\,\big(z_i^2\pa_i-z_{i+1}^2\pa_{i+1}\big)
=J^0+\frac12\sum_i\frac{z_i+z_{i+1}}{z_i-z_{i+1}}\,(D_i-D_{i+1})\,,
\]
where $D_i=z_i\pa_i$, we immediately obtain the following expression
for the gauge Hamiltonian $\BH_2$:
\begin{multline}
\frac14\,(\BH_2-E_0)=\sum_iD_i^2+a\sum_i\frac{z_i+z_{i+1}}{z_i-z_{i+1}}\,(D_i-D_{i+1})\\
-2a\sum_i\frac{z_iz_{i+1}}{(z_i-z_{i+1})^2}\,(1-K_{i,i+1})\,.
\end{multline}
Since $\tau_N^{-\la}D_i\tau_N^\la=D_i+\la$ for any real $\la$, it
follows that
\[
\tau_N^{-\la}\BH_2\tau_N^\la=\BH_2+8\la J^0+4N\la^2\,.
\]
Taking into account that $J^0\Phi=k\Phi$, we conclude that
\[
\BH_2\big(\tau_N^\la\Phi\big)=\big(E+8k\la+4N\la^2\big)(\tau_N^\la\Phi)\,,
\]
as claimed.
\end{proof}
By the previous discussion, in what follows any two eigenfunctions
of $\BH_2$ that differ by a power of $\tau_N$ shall be considered
equivalent. {}From Theorem~\ref{thm.1} and Corollary~\ref{cor.1} it
easily follows that in this case the number of independent algebraic
eigenfunctions is finite, up to equivalence. More precisely:
\begin{lemma}\label{lemma.6}
Up to equivalence, the algebraic eigenfunctions of $\BH_2$ can be
assumed to belong to a space of the form
\begin{equation}\label{BcH2kets}
\cH_{2,\ket\bs}=\langle\si_1,\,\tau_{N-1},\,\si_1\tau_{N-1},\,\tau_N\rangle\,\Phi^{(0)}
+\langle 1,\,\tau_{N-1}\rangle\,\Phi^{(1)} +\langle
1,\,\si_1\rangle\,\tau_N\Phi^{(-1)}
\end{equation}
for some spin state $\ket\bs$, where $\Phi^{(k)}$ is given
by~\eqref{Phik}.
\end{lemma}
\begin{proof}
Given a spin state $\ket\bs$, the obvious identity
\begin{equation}\label{T2fPhi0}
T_2\big(f\Phi^{(0)}\big)=(T_2f)\Phi^{(0)}
\end{equation}
and Eqs.~\eqref{BHep}, \eqref{Phi21}, \eqref{Phi2-1},
and~\eqref{J0Phis} imply that the gauge Hamiltonian $\BH_2$
preserves the space
\begin{equation}\label{BcH2n}
\cH_{2,\ket\bs}^n=\big\langle f\Phi^{(0)},g\ms\Phi^{(1)},
\tau_Nq\ms\Phi^{(-1)}\big\rangle\,,
\end{equation}
where $f$, $g$ and $q$ are as in the definition of $\cT_2^n$ in
Theorem~\ref{thm.1}. Let $\Phi\in\cH_{2,\ket\bs}^n$ be an
eigenfunction of $\BH_2$, which as explained above can be taken as a
homogeneous function of $\bz$. {}From the conditions satisfied by
the functions $f$, $g$ and $q$ in~\eqref{BcH2n} and the homogeneity
of $\Phi$, it readily follows that $\Phi\in\tau_N^l\cH_{2,\ket\bs}$
for some $l$, as claimed.
\end{proof}
\begin{theorem}\label{thm.H2}
The Hamiltonian $H_2$ possesses the following spin eigenfunctions
with zero momentum
\begin{gather*}
\Psi_0=\mu\,\Phi^{(0)},\qquad
\Psi_{1,2}=\mu\sum_i\bigg\{\begin{matrix}\cos\\\sin\end{matrix}\bigg\}
\big(2(x_i-\Bx)\big)\ket{\bs_i},\\
\Psi_3=\mu\bigg[\frac{2a}{2a+1}\,\Phi^{(0)}+\sum_{i\neq
j}\cos\!\big(2(x_i-x_j)\big)\ket{\bs_j}\bigg],
\quad\Psi_4=\mu\sum_{i\neq j}\sin\!\big(2(x_i-x_j)\big)\ket{\bs_j},
\end{gather*}
where $\Bx$ is the center of mass coordinate. Their energies are
respectively given by
\[
E_0\,,\qquad E_{1,2}=E_0+4\Big(2a+1-\frac1N\Big)\,,\qquad
E_{3,4}=E_0+8(2a+1)\,.
\]
Any algebraic eigenfunction with well-defined total momentum is
equivalent to a linear combination of the above eigenfunctions.
\end{theorem}
\begin{proof} By Lemma~\ref{lemma.6}, in order to compute the algebraic eigenfunctions of $\BH_2$
it suffices to diagonalize $\BH_2$ in the spaces $\cH_{2,\ket\bs}$
given by~\eqref{BcH2kets}. Equations~\eqref{Phi21}, \eqref{Phi2-1},
\eqref{J0Phis} and~\eqref{T2fPhi0}, and the fact that $\BH_2$
preserves the degree of homogeneity, imply that the following
subspaces of $\cH_{2,\ket\bs}$ are invariant under $\BH_2$:
%\begin{subequations}\label{invsubs2}
%\begin{align}
%& \langle\si_1\tau_{N-1}\,,\tau_N\rangle\,\Phi^{(0)}\,,\hspace*{-2cm}\label{invsubs20}\\
%& \langle\si_1\ms\Phi^{(0)}\rangle\,, & & \langle
%\si_1\ms\Phi^{(0)},\Phi^{(1)}\rangle\,, & &
%\langle\si_1\tau_{N-1}\ms\Phi^{(0)},\tau_N\ms\Phi^{(0)},\tau_{N-1}\ms\Phi^{(1)}\rangle\,\,,\label{invsubs21}\\
%& \langle\tau_{N-1}\ms\Phi^{(0)}\rangle\,, & & \langle
%\tau_{N-1}\ms\Phi^{(0)},\tau_N\Phi^{(-1)}\rangle\,, & &
%\langle\si_1\tau_{N-1}\ms\Phi^{(0)},\tau_N\ms\Phi^{(0)},\si_1\tau_N\ms\Phi^{(-1)}\rangle\,.\label{invsubs2-1}
%\end{align}
%\end{subequations}
\begin{subequations}\label{invsubs2}
\begin{align}
&\langle\si_1\tau_{N-1}\,,\tau_N\rangle\,\Phi^{(0)}\,,\hspace*{-2cm}\label{invsubs20}\\
&\langle\si_1\ms\Phi^{(0)}\rangle\,,\;\langle
\si_1\ms\Phi^{(0)},\Phi^{(1)}\rangle\,,\;
\langle\si_1\tau_{N-1}\ms\Phi^{(0)},\tau_N\ms\Phi^{(0)},\tau_{N-1}\ms\Phi^{(1)}\rangle\,\,,\label{invsubs21}\\
&\langle\tau_{N-1}\ms\Phi^{(0)}\rangle\,,\; \langle
\tau_{N-1}\ms\Phi^{(0)},\tau_N\Phi^{(-1)}\rangle\,,\;
\langle\si_1\tau_{N-1}\ms\Phi^{(0)},\tau_N\ms\Phi^{(0)},\si_1\tau_N\ms\Phi^{(-1)}\rangle\,.\label{invsubs2-1}
\end{align}
\end{subequations}
{}From Eqs.~\eqref{invsubs2} it follows that the alternative change
of variables $z_k=\e^{-2\iu x_k}$ does not yield additional
eigenfunctions of $H_2$. Indeed, the latter change corresponds to
the mapping $z_k\mapsto 1/z_k$, which up to equivalence leaves the
subspace~\eqref{invsubs20} invariant and exchanges each subspace
in~\eqref{invsubs21} with the corresponding one
in~\eqref{invsubs2-1}. For this reason, we can safely ignore the
subspaces~\eqref{invsubs2-1} in the computation that follows,
provided that we add to the eigenfunctions of $\BH_2$ obtained from
the subspaces~\eqref{invsubs21} their images under the mapping
$z_k\mapsto 1/z_k$.

For the subspaces~\eqref{invsubs20}-\eqref{invsubs21}, using
Eqs.~\eqref{BHep}, \eqref{Tep-res}, \eqref{T22f}, \eqref{Phi21},
\eqref{J0Phis}, and \eqref{J0f2} we easily obtain the following
eigenfunctions of $\BH_2$:
\begin{subequations}\label{eigenf2}
\begin{align}
& \tau_N\ms\Phi^{(0)}, & & \quad E=E_0+4N,\label{eigenf2a}\\
& \Phi^{(1)}, & & \quad E=E_0+4(2a+1),\label{eigenf2b}\\
& \tau_{N-1}\ms\Phi^{(1)}-\frac{\tau_N}{2a+1}\,\Phi^{(0)}, & &\quad
E=E_0+4(N+4a+2).\label{eigenf2c}
\end{align}
\end{subequations}
We have omitted the two additional eigenfunctions
\[
\si_1\ms\Phi^{(0)}\,,\qquad
\Big(\si_1\tau_{N-1}-\frac{N\tau_N}{2a+1}\Big)\ms\Phi^{(0)}
\]
from the above list, since they are respectively obtained
from~\eqref{eigenf2b} and~\eqref{eigenf2c} when the spin state
$\ket\bs$ is symmetric. The eigenfunctions~\eqref{eigenf2} are
equivalent to the following ``zero momentum'' eigenfunctions:
\begin{subequations}\label{eigenf2zero}
\begin{align}
& \Phi^{(0)}, & & \quad E=E_0,\label{eigenf2zeroa}\\
& \tau_N^{-1/N}\Phi^{(1)}, & & \quad E=E_0+4\Big(2a+1-\frac1N\Big),\label{eigenf2zerob}\\
& \frac{\tau_{N-1}}{\tau_N}\,\Phi^{(1)}-\frac1{2a+1}\,\Phi^{(0)}, &
&\quad E=E_0+8(2a+1),\label{eigenf2zeroc}
\end{align}
\end{subequations}
where the energies have been computed from those in
Eqs.~\eqref{eigenf2} using Lemma~\ref{lemma.5}. The eigenfunctions
of $H_2$ listed in the statement are readily obtained from these
spin functions together with the transforms of~\eqref{eigenf2zerob}
and~\eqref{eigenf2zeroc} under the mapping $z_k\mapsto 1/z_k$.
\end{proof}
\begin{remark}
If the spin state $\ket\bs$ is symmetric, then
$\ket{\bs_i}=\Phi^{(0)}/N$ for all $i$, and one easily obtains from
Theorem~\ref{thm.H2} the following eigenfunctions of the scalar
Hamiltonian $H^{\mathrm{sc}}_2$:
\[
\psi_0=\mu,\;\;
\psi_{1,2}=\mu\sum_i\bigg\{\begin{matrix}\cos\\\sin\end{matrix}\bigg\}
\big(2(x_i-\Bx)\big),\;\; \psi_3=\mu\bigg[\frac
{aN}{2a+1}+\sum_{i<j}\cos\!\big(2(x_i-x_j)\big)\bigg].
\]
\end{remark}

\chapter{NN spin chains of HS type}
\label{Ch:NNchain}

\section{Introduction}

As we saw in Chapter~\ref{Ch:BCN}, one can successfully solve spin chains of HS
type by means of Polychronakos's freezing trick, which essentially consists in
computing the large coupling constant limit of the quotient of the partition
functions of a spin CS model and its scalar counterpart.

In the present chapter we shall extend this technique to the case of QES
models of CS type with spin. This extension will prove crucial in the analysis
of the spectrum of the novel $N$-site chain
\begin{align}\label{ssH.NN}
\ssH&=\sum_i(\xi_i-\xi_{i+1})^{-2}\,(1-S_{i,i+1})\,,
\end{align}
with sites $\bxi\in C_0\subset\RR^N$ defined by the algebraic system
\begin{equation}\label{sites.NN}
\xi_i=\frac1{\xi_i-\xi_{i-1}}+\frac1{\xi_i-\xi_{i+1}}\,,\qquad 1\leq i\leq N\,.
\end{equation}
Here and in what follows, we identify $\xi_0\equiv\xi_N$,
$\xi_{N+1}\equiv\xi_1$ (and similarly for $x_i$ and $S_{i,i+1}$),
and keep denoting by $C_0$ the domain~\eqref{C0} and by $M$ the
particles' spin. It is not difficult to check that the spin
chain~\eqref{ssH.NN}, which we shall simply call the \Emph{NN
chain}, is related to the NN Hamiltonian $H_0$ in Eq.~\eqref{V0}
along the lines of the freezing trick.

We shall see in this chapter that this spin chain is in some sense
intermediate between the Heisenberg chain (short-range, position-independent
interactions) and the PF chain (long-range, position-dependent interactions),
from which one can derive the Hamiltonian~\eqref{ssH.NN} by retaining only
nearest-neighbors interactions. Moreover, $\ssH$ turns out to be QES in the
sense that a nontrivial proper subset of the spectrum can be explicitly
computed for any value of $N$ and $M$. While QES models have long played a
relevant role in the theory of CS systems~\cite{Sh89,ST89}, this is to our
best knowledge the first QES phenomenon spotted in the (as has been made clear
in this dissertation) closely related field of HS spin chains.

The results in this chapter can be easily modified to deal with the NN chains
\begin{align*}
\ssH_1&=\sum_i\frac{\xi_i^2+\xi_{i+1}^2}{(\xi_i^2-\xi_{i+1}^2)^2}(1-S_{i,i+1})\,,\\
\ssH_2&=\sum_i\sin^{-2}(\xi_i-\xi_{i+1})\,(1-S_{i,i+1})\,,
\end{align*}
where the chain sites are respectively defined by
\begin{align*}
\xi_i&=\frac1{\xi_i}+\frac{\xi_i}{\xi_i^2-\xi_{i+1}^2}+
\frac{\xi_i}{\xi_i^2-\xi_{i-1}^2}\,,\\
0&=\cot(\xi_i-\xi_{i+1})+\cot(\xi_i-\xi_{i-1})\,.
\end{align*}
These spin chains are obtained from the NN models~\eqref{V1} and~\eqref{V2}
respectively by a freezing trick argument. Note, however, that the sites of
the second chain turn out to be equally spaced, so $\ssH_2$ is essentially the
Heisenberg chain.

This chapter is organized as follows. In Section~\ref{S:sitesNN} we analyze the
distribution of the chain sites using both exact results and numerical
calculations. In Section~\ref{S:freeze.NN} we show the QES nature of the chain
by providing several families of eigenvectors, valid for any choice of $N$ and
$M$. The main technical tool is a nontrivial refinement of the freezing trick
developed in Chapter~\ref{Ch:BCN} which also applies to QES systems. Finally,
we discuss some qualitative properties of the spectrum and formulate an
intriguing conjecture claiming that an eigenvalue of the chain is algebraic if
and only if it is an integer.

The material presented in this chapter is taken from Ref.~\cite{EFGR07b}.

\section{The chain sites}
\label{S:sitesNN}

We shall begin with a detailed analysis of the distribution of the chain sites
on the line in which we shall combine exact results and numerical computations
to extract as much information as possible. An easy observation, which stems
from the spin chain's connection with the dynamical model~\eqref{V0} that will
be developed in the next section, is that the sites are given by the
coordinates of a critical point of the function
\begin{equation}\label{la.NN}
\la(\bx)=\sum_i\log|x_i-x_{i+1}|-\frac{r^2}2\,,
\end{equation}
which is the scaled logarithm of the ground state function of the NN
Hamiltonian~\eqref{V0}.

The first questions to ascertain refer to the existence and uniqueness of the
critical points of $\la$. In this direction we have the following result, which
is roughly analogous to Proposition~\eqref{crit.BC}:

\begin{proposition}\label{crit.NN}
  The function $\la$ has a unique critical point $\bxi$ in $C_0$, which is a
  hyperbolic maximum. Moreover, $\bxi$ satisfies
\begin{equation}\label{symm.xi}
\xi_i=-\xi_{N-i+1}\,,\qquad i=1,\dots,N.
\end{equation}
\end{proposition}
\begin{proof}
  The existence of a maximum of $\la$ in $C_0$ is clear, since it is
  continuous in $C_0$ and tends to $-\infty$ both on its boundary and as $r\to\infty$.
  Uniqueness and hyperbolicity follow from the fact that the Hessian of $\la$
  is negative definite in $C_0$. Indeed, by Gerschgorin's
  theorem~\cite[15.814]{GR00}, the eigenvalues of the Hessian of $\la$ at
  $\bx$ lie in the union of the intervals
\[
\Big[\,\frac{\pd^2 \la}{\pd x_i^2}-\gamma_i\,,\frac{\pd^2 \la}{\pd
x_i^2}+\gamma_i\,\Big]\,, \quad\text{where } \gamma_i=\sum_{j\neq
i}\Big|\,\frac{\pd^2 \la}{\pd x_i\pd x_j}\,\Big|\,.
\]
Since
\begin{gather*}
\frac{\pd^2\la}{\pd x_i^2}=-1-(x_i-x_{i+1})^{-2}-(x_i-x_{i-1})^{-2},\\[1mm]
\frac{\pd^2\la}{\pd x_i\pd x_{i±1}}=(x_i-x_{i±1})^{-2},\qquad
\frac{\pd^2\la}{\pd x_i\pd x_{j}}=0\;\text{ if }\;j\neq i,i±1,
\end{gather*}
we have
\[
\frac{\pd^2 \la}{\pa x_i^2}+\ga_i=-1\,,
\]
and thus all the eigenvalues of the Hessian of $\la$ are strictly
negative.

Furthermore, the mapping
\[
x_i\mapsto -x_{N-i+1}\,,\qquad i=1,\dots, N\,,
\]
is a symmetry of $\la$ and maps the domain $C_0$ into itself. Since
$\bxi$ is the unique critical point of $\la$ in $C_0$, it must be a
fixed point of the above transformation, and thus
Eqs.~\eqref{symm.xi} follow.
\end{proof}
From the previous proposition (and also directly from Eq.~\eqref{sites.NN}),
it immediately follows that the center of mass of the spins vanishes, i.e.,
\begin{equation}
  \label{xibar}
  \sum_i\xi_i=0\,.
\end{equation}

\begin{proposition}\label{xi2}
$\|\bxi\|^2=N$.
\end{proposition}
\begin{proof}
Let us write $\la$ in terms of the variables $r=\|\bx\|\in\RR^+$ and $\mathbf
y=\bx/r\in\mathbb S^{N-1}$, i.e.,
\[
\la=\sum_i\log|y_i-y_{i+1}|+N\log r-\frac{r^2}2\,.
\]
Obviously
\[
0=\frac{\pd\la}{\pd r}(\bxi)=\frac N{\|\bxi\|}-\|\bxi\|\,,
\]
as we wanted to show.
\end{proof}

\begin{figure}[t]
\begin{center}
\psfrag{x}{\footnotesize $\xi_i$}
\includegraphics[height=6cm]{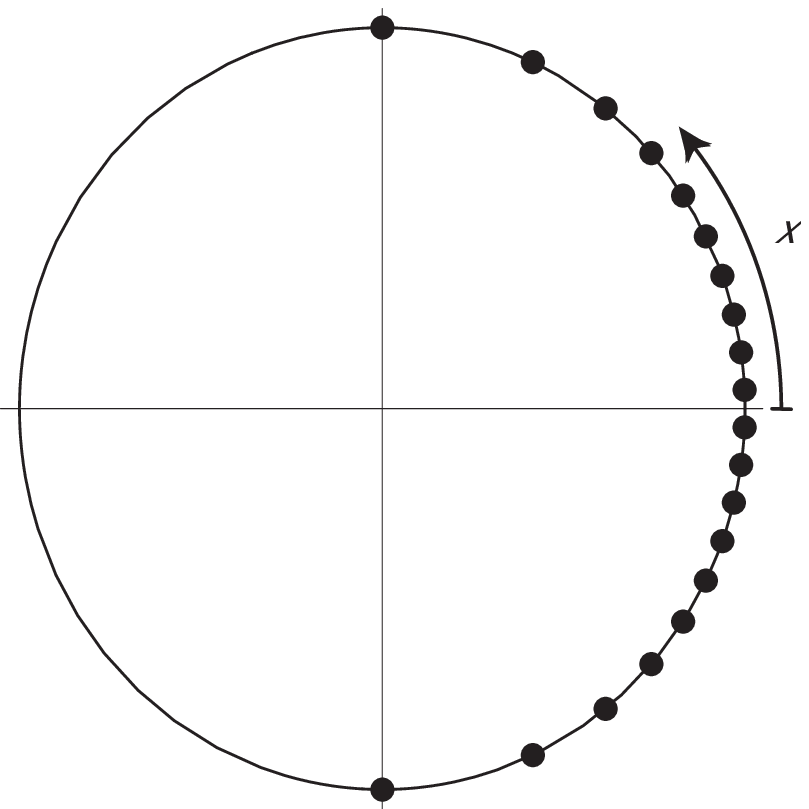}
\begin{quote}
\caption{Sites of the chain $\ssH$ for $N=20$ spins.\label{fig:sites20}}
\end{quote}
\end{center}
\end{figure}

It is convenient to think of the spins as if located on a circle. In this way,
the chain~\eqref{ssH.NN} truly presents only near-neighbors interactions in
spite of the fact that the first spin interacts with the last one. Indeed, if
the site coordinate $\xi_i$ is understood as an arc length in a circle of
radius
\begin{equation}\label{rN}
r_N=\frac{2\ms\xi_N}\pi\,,
\end{equation}
as shown in Fig.~\ref{fig:sites20}, the spins at the sites $\xi_1$ and $\xi_N$
become nearest-neighbors and, moreover, the strength of the interaction is
inversely proportional to the squared distance between the spins, measured
along the arc.

We saw in Section~\ref{S:HS} that the site coordinates of the PF chain
define the $N$-th Hermite polynomial. It is natural to wonder whether a similar
result holds for the family of monic polynomials
\begin{equation}
  p_N(t)=\prod_i(t-\xi_i)\,,\qquad \bxi\in C_0\subset\RR^N\,,
  \label{pN}
\end{equation}
defined by $\ssH$, which can be regarded as the closest NN analog of the PF
chain. In the following proposition we answer this question negatively:

\begin{proposition}
The polynomials \eqref{pN} do not form an orthogonal family.
\end{proposition}
\begin{proof}
  In order to show that the family $\big\{p_N\big\}$ is not orthogonal, it
  suffices to prove that the polynomials~\eqref{pN} do not satisfy a
  three-term recursion relation of the form
\[
p_{N+1}(t)=t\ms p_N(t)+c_N\ms p_{N-1}(t)\,,
\]
with $c_N$ constant. In fact, solving the system~\eqref{sites.NN} for
$N=2,3,4$ one immediately obtains
\[
p_2(t)=t^2-1\,,\qquad p_3(t)=t^3-\frac{3t}2\,,\qquad p_4(t)=t^4-2
t^2+\frac14\,.
\]
Thus the previous recursion relation already fails for $N=3$.
\end{proof}

The sites are certainly not equally spaced; in fact, our numerical computations
lead to the following

\begin{claim}\label{Gauss}
For $N\gg1$, the site distribution follows the Gaussian law with
zero mean and unit variance.
\end{claim}

\begin{figure}
\begin{center}
\psfrag{F}{\footnotesize\,$\cF(x),F(x)$} \psfrag{x}{\footnotesize
$x$}
\includegraphics[height=6cm]{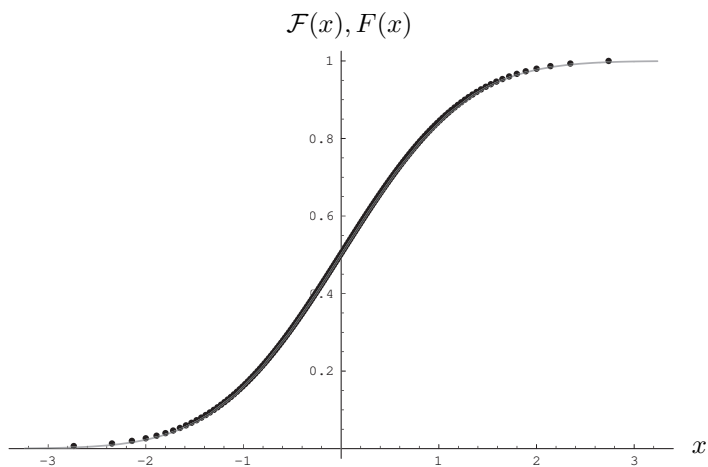}
\begin{quote}
\caption{Cumulative distribution functions $\cF(x)$ (at its
discontinuity points) and $F(x)$ (continuous grey line) for $N=150$
spins.\label{fig:sitesdist150}}
\end{quote}
\end{center}
\end{figure}

Note that, should the site distribution follow a Gaussian law, its mean would
be zero on account of Corollary~\ref{xibar}. We have solved numerically
Eq.~\eqref{sites.NN} for the positions of the chain sites for up to $N=200$
spins, and the results completely support the validity of the above
conjecture. More precisely, we claim that the cumulative density of sites
\begin{equation}\label{cF}
\cF(x)=N^{-1}\sum_i\theta(x-\xi_i)\,,
\end{equation}
where $\theta$ is Heaviside's step function, is asymptotically
given by
\begin{equation}\label{F.NN}
F(x)=\frac12\,\Big[1+\erf\big(x/\sqrt2\ms\big)\Big]\,.
\end{equation}
The agreement between the functions $\cF$ and $F$ is remarkably good
for $N\gtrsim100$ (see~Fig.~\ref{fig:sitesdist150} for the case
$N=150$) and increases steadily with $N$. For example, the mean square
error of the fit for $100$, $150$ and 200 spins is respectively
$2.6×10^{-5}$, $1.1×10^{-5}$ and $7.9× 10^{-6}$.

\begin{proof}[Heuristic proof of Claim~\ref{Gauss}]
The fact that for large $N$ the cumulative density of sites is well
approximated by the Gaussian law~\eqref{F.NN} can be justified by the following
semi-rigorous argument. Let $x(t,N)$ be a smooth function such that
$x(i,N)=\xi_i$ for $i=1,\dots,N$, and define the rescaled function
$y(s,\ep)=x(s/\ep,1/\ep)$. By Eq.~\eqref{sites.NN}, the latter function must
satisfy the relation
\begin{equation}\label{sitesy}
\frac1{y(s,\ep)-y(s-\ep,\ep)}+\frac1{y(s,\ep)-y(s+\ep,\ep)}=y(s,\ep)
\end{equation}
for $\ep=1/N\ll 1$ and $s=\frac1N,\frac2N,\dots,1$. Let us now assume that
Eq.~\eqref{sitesy} holds for all $s\in\RR$ and all $\ep\ll 1$. Writing
\[
y(s,\ep)=\sum_{k=0}^\infty y_k(s)\ep^k\,,
\]
and using the expansion
\[
y(s,\ep)-y(s±\ep,\ep)=\mp\,y'_0(s)\ep-\Big(\frac{y''_0(s)}2±
y'_1(s)\Big)\ep^2+O(\ep^3)
\]
the leading term in Eq.~\eqref{sitesy} yields the differential equation
\[
y_0''=y_0\,{y_0'}^2\,.
\]
The general solution of this equation is implicitly given by
\begin{equation}\label{sy0}
s=c_0+c_1\erf\!\big(y_0(s)/\sqrt2\ms\big)\,.
\end{equation}
Hence, up to terms of order $\ep=1/N$, the cumulative distribution function of
the chain sites (normalized to unity) is approximated by the continuous
function
\[
F(x)=c_0+c_1\erf\!\big(x/\sqrt2\ms\big)\,.
\]
The normalization conditions $F(-\infty)=0$ and $F(\infty)=1$ imply that
$c_0=c_1=1/2$, and thus the empiric law~\eqref{F.NN} is recovered.

{}From Eq.~\eqref{sy0} (with $c_0=c_1=1/2$) it follows that the site $\xi_k$
can be determined up to terms of order $1/N$ by the formula
\begin{equation}\label{xibad}
\xi_k\simeq\sqrt2\,\erf^{-1}\Big(\frac{2k-N}N\Big)\,.
\end{equation}
Note that, were the sites $\xi_k$ exactly given by the previous formula, they
would satisfy the identity
\[
\erf\!\big(\xi_k/\sqrt 2\big)+\erf\!\big(\xi_{N-k+1}/\sqrt 2\big)=\frac1N\,,
\]
which is clearly inconsistent with the exact relation~\eqref{symm.xi}. However,
the slightly modified formula
\begin{equation}\label{xigood}
\xi_k\simeq\sqrt2\,\erf^{-1}\Big(\frac{2k-N-1}N\Big)
\end{equation}
differs from~\eqref{xibad} by a term of order $1/N$ and is fully consistent
with the relation~\eqref{symm.xi}. Although both~\eqref{xibad}
and~\eqref{xigood} provide an excellent approximation to the chain sites for
large $N$, the latter equation is always more accurate than the former, and can
be used to estimate $\xi_k$ with remarkable precision even for relatively low
values of $N$, cf.~Fig.~\ref{fig:xis20}.
\begin{figure}[t]
\begin{center}
\psfrag{x}{\footnotesize$\xi_k$} \psfrag{k}{\footnotesize $k$}
\includegraphics[height=6cm]{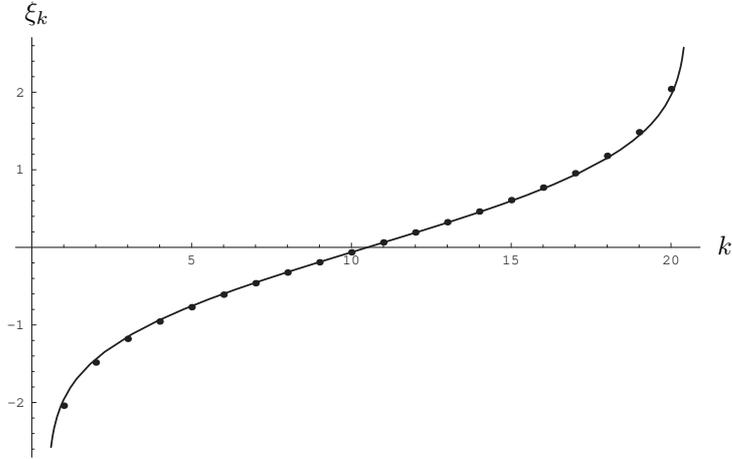}
\begin{quote}
\caption{Sites coordinates $\xi_k$ and their continuous
approximation~\eqref{xigood} for $N=20$ spins.\label{fig:xis20}}
\end{quote}
\end{center}
\end{figure}
\end{proof}

%Interestingly, not only the sites seem to be distributed according to the
%Gaussian law. Indeed, for large $N$ the strength of the interactions as a
%function of the sites' coordinate also follows the Gaussian distribution with
%zero mean and variance $1/2$, as shown in Fig.~\ref{fig:couplings100}.
%
%\begin{figure}[h]
%\begin{center}
%\psfrag{G}{\footnotesize $(\xi_i-\xi_{i+1})^{-2},G(x)$}
%\psfrag{x}{\footnotesize $x$}
%\includegraphics[height=6cm]{couplings100.eps}
%\begin{quote}
%\caption{Plot of the coupling between the spins $i$ and $i+1$ versus
%their mean position $(\xi_i+\xi_{i+1})/2$, fitted by the Gaussian
%$G(x)=(\xi_{50}-\xi_{51})^{-2}\e^{-x^2}$, for $N=100$
%spins.\label{fig:couplings100}}
%\end{quote}
%\end{center}
%\end{figure}

\begin{figure}[t]
\begin{center}
\psfrag{x}[Bc][Bc][1][0]{\footnotesize $\xi_N$} \psfrag{N}{\footnotesize $N$}
\includegraphics[height=6cm]{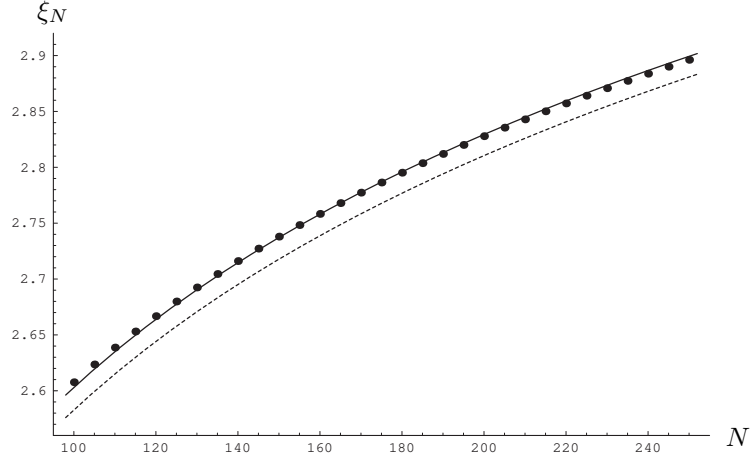}
\begin{quote}
\caption{Position of the last spin $\xi_N$ for $N=100,105,\dots,250$ and its
continuous approximation~\eqref{xiNsimpler}-\eqref{etaalfa} for $\al=0.94$
(solid line) and $\al=1$ (dashed line).\label{fig:xmax100}}
\end{quote}
\end{center}
\end{figure}

In the rest of this section we shall analyze several relevant features of the
site distribution using the above heuristic approach. It is certainly of
interest to determine whether the position of the last spin tends to infinity
as $N\to\infty$, since according to our interpretation of the chain's geometry
the number $2\xi_N/\pi$ is the radius of the circle on which the spins lie.
According to Eq.~\eqref{xigood}, for large $N$ the last spin's coordinate
$\xi_N$ is approximately given by
\begin{equation}\label{xiN}
\xi_N\simeq\sqrt2\,\erf^{-1}\Big(1-\frac1N\Big)\,,
\end{equation}
so that $\xi_N$ should diverge as $N\to\infty$. Of course, this assertion
should be taken with some caution, since in Eq.~\eqref{xigood} the argument of
the inverse error function is correct only up to terms of order $1/N$. In order
to check the correctness of the approximation~\eqref{xiN}, we use the
asymptotic expansion of $\erf^{-1}(u)$ for $u\to 1$ in Ref.~\cite{BEJ76} to
replace~\eqref{xiN} by the simpler formula
\begin{equation}\label{xiNsimpler}
\xi_N\simeq\sqrt{2\eta-\log\eta}\,,
\end{equation}
where
\[
\eta=\log\Big(\frac N{\sqrt\pi}\Big)\,.
\]
As can be seen in Fig.~\ref{fig:xmax100}, the approximate
formula~\eqref{xiNsimpler} qualitatively reproduces the growth of $\xi_N$ when
$N$ ranges from $100$ to $250$. A greater accuracy can be achieved by
introducing an adjustable parameter in Eq.~\eqref{xiN} through the replacement
$1/N\to\al/N$, so that in Eq.~\eqref{xiNsimpler} $\eta$ becomes
\begin{equation}\label{etaalfa}
\eta=\log\Big(\frac N{\al\sqrt\pi}\Big)\,.
\end{equation}

In Fig.~\ref{fig:xmax100}, we have also plotted the
law~\eqref{xiNsimpler}-\eqref{etaalfa} with the optimal value $\al=0.94$, which
is in excellent agreement with the numerical values of $\xi_N$ for
$N=100,105,\dots,250$.

\begin{figure}[t]
\begin{center}
\psfrag{X}[Bc][Bc][1][0]{\footnotesize $h_k$} \psfrag{x}{\footnotesize
$\overline\xi_k$}
\includegraphics[height=6cm]{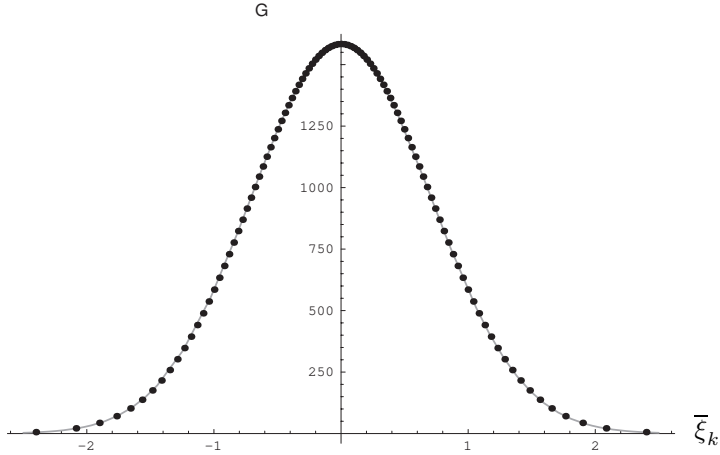}
\begin{quote}
\caption{Plot of the coupling between the spins $k$ and $k+1$ versus their mean
position, fitted by the Gaussian~\eqref{couplings}, for $N=100$
spins.\label{fig:couplings100}}
\end{quote}
\end{center}
\end{figure}

The last property of the spin chain~\eqref{ssH.NN} that we shall analyze in
this section is the dependence of the coupling between neighboring spins on
their mean coordinate. Calling
\[
h_k=(\xi_k-\xi_{k+1})^{-2}\,,\qquad\overline\xi_k=\frac{\xi_k+\xi_{k+1}}2\,,
\]
we shall now see that when $N\gtrsim100$ the Gaussian law
\begin{equation}\label{couplings}
h_k\simeq\frac{N^2}{2\pi}\,\e^{-{\overline\xi_k}^{\ms 2}}
\end{equation}
holds with remarkable precision, cf.~Fig.~\ref{fig:couplings100}. Indeed, if
$x=x(k)$ denotes the RHS of Eq.~\eqref{xigood} we have
\[
2k=N\erf\Big(\frac x{\sqrt2}\Big)+N+1\,,
\]
so that
\[
\frac{\dd x}{\dd k}=\frac{\sqrt{2\pi}}N\,\e^{\frac12x^2}
\]
is of order $1/N$. Hence, up to terms of order $1/N$ we have
\begin{equation}\label{precouplings}
\Big[x\Big(k-\frac12\Big)-x\Big(k+\frac12\Big)\Big]^{-2}\simeq \Big(\frac{\dd
x}{\dd k}\Big)^{-2}=\frac{N^2}{2\pi}\,\e^{-x^2(k)}\,.
\end{equation}
Since (up to terms of order $1/N^2$)
\[
x(k)\simeq\frac12\,\Big[x\Big(k-\frac12\Big)+x\Big(k+\frac12\Big)\Big]\,,
\]
Eq.~\eqref{couplings} follows from~\eqref{precouplings} replacing $k$ by
$k+\frac12$.

A brief comparison of the previous properties with those of the PF
chain \eqref{Ch2.PF} is in order. For large $N$, the density of sites of the PF
chain (normalized to unity), that is the density of zeros of the $N$-th Hermite
polynomial, is asymptotically given by~\cite{CP78b}
\begin{equation}\label{rhoPF}
\rho_N(x)=\frac1{\pi N}\,\sqrt{2N-x^2}\,.
\end{equation}
The last site $\ze_N$ of the PF chain grows with $N$ much faster than the
corresponding site $\xi_N$ of the chain~\eqref{ssH.NN}, for the largest zero of
the $N$-th Hermite polynomial behaves as $\sqrt{2N}+\cO(N^{-1/6})$; see, for
instance, the recent paper~\cite{Do06}.

% A greater accuracy can be achieved by
%introducing an adjustable parameter in Eq.~\eqref{xiN} through the replacement
%$1/N\to\al/N$, so that in Eq.~\eqref{xiNsimpler} $\eta$ becomes
%\begin{equation}\label{etaalfa}
%\eta=\log\Big(\frac N{\al\sqrt\pi}\Big)\,.
%\end{equation}

%When $N$ ranges from $5$ to $200$, we
%have numerically observed that the site coordinate $\xi_N$ can be
%accurately fitted by the logarithmic law
%\begin{equation}\label{loglaw}
%\Xi(x)=\al\log(x+\be)+\ga
%\end{equation}
%with parameters $\al=0.33$, $\be=-1.97$, $\ga=1.10$; see Fig.~\ref{fig:xmax}.
%This is fully consistent with Gloss~\ref{limit.MA}.
%
%
%Its growth in $N$ is certainly small, and the fit is good but perhaps not as
%accurate as in Conjecture~\ref{Gauss}. It is therefore natural to raise the
%following (so far unanswered)
%
%\begin{question}
%Does the log law~\eqref{loglaw} hold asymptotically? If not, is
%$\xi_N$ bounded?
%\end{question}
%
%In any case, the growth of $\xi_N$ in~\eqref{ssH.NN} and in the PF
%chains are clearly distinct, since in the latter case it is given by
%the $N$-th zero of the Hermite polynomial of degree $N$, which
%behaves as $(2N)^{1/2}+\cO(N^{-1/6})$ (see, e.g., \cite{Do06}).
%
%\begin{figure}[b]
%\begin{center}
%\psfrag{x}{\footnotesize $\xi_N$,$\,\Xi(N)$}
%\psfrag{N}{\footnotesize $N$}
%\includegraphics[height=6cm]{xmax.eps}
%\begin{quote}
%\caption{Position of the last spin $\xi_N$ for $N=5,10,\dots,200$
%and its continuous approximation $\Xi(N)$.\label{fig:xmax}}
%\end{quote}
%\end{center}
%\end{figure}

\section{Quasi-exact solvability of the NN chain}
\label{S:freeze.NN}

In this section we shall explore the solvability properties of the
spin chain~\eqref{ssH.NN} by exploiting its connection with the
QES model $H_0$ defined in~\eqref{V0}. As in
Section~\ref{S:chainBCN}, we shall rescale the constant $\om$ in the
latter Schrödinger operator and its scalar counterpart $\Hsc$ and
remove the ground state energy, defining
\[
H=(H_0-E_0)|_{\om=a}\,,\qquad \Hsc=(\Hsc_0-E_0)|_{\om=a}\,.
\]
In terms of the matrix multiplication operator
\[
\ssh(\bx)=\sum_i(x_i-x_{i+1})^{-2}(1-S_{i,i+1})\,,
\]
the connection between $H$ and $\Hsc$ can be simply written as
\begin{equation}\label{HHsc.NN}
H=\Hsc+2a\ms\ssh\,,
\end{equation}
whereas the spin chain Hamiltonian~\eqref{ssH.NN} is recovered as
\begin{equation}\label{ssHssh.NN}
\ssH=\ssh(\bxi)\,.
\end{equation}

Recall that the ground state function of the scalar Hamiltonian $\Hsc$ can be
written as $\mu(\bx)=\e^{a\ms\la(\bx)}$, cf.~Table~\ref{table:params} and
Eq.~\eqref{la.NN}. We shall also denote by
\[
\vp_0=\|\mu\|^{-2}\mu^2\in L^1(C_0)\cap C^\infty(C_0)
\]
its normalized square. The following result is easily established by
repeating the proof of Lemma~\ref{delta.BC}.

\begin{lemma}\label{delta.NN}
$\vp_0$ converges to the delta distribution supported at $\bxi$ and
decays exponentially fast away from $\bxi$.
\end{lemma}

One would be tempted to believe that, mutatis mutandis, the
discussion of Section~\eqref{S:chainBCN} would provide an
appropriate bridge to link the algebraic states of $H$ and $\ssH$.
However, the fact that the dynamical models $H$ and $\Hsc$ are only
quasi-exactly solvable, and thus only a proper subset of their
spectra is known, makes it impossible to compute their partition
functions from their algebraic energies. Hence Lemma~\ref{freezing}, which
constitutes the core of the application of CS models to the study of
spin chains, cannot be invoked in this case.

In the rest of this chapter we shall develop a finer version of the
freezing trick that will enable us to circumvent this limitation. We
shall obtain a number of eigenstates of the chain (which we call
\emph{algebraic}\/), valid for any choice of $N$ and $M$. As
mentioned in the introduction, these algebraic states do not span
the whole spin space, the chain Hamiltonian~\eqref{ssH.NN} thus
inheriting the QES structure of the dynamical model.

In the following lemma we prove the key convergence result (``freezing trick'')
that shall enable us to compute the algebraic eigenvectors of $\ssH$ in closed
form. This result is most easily stated in terms of the function
$\Bmu=\|\mu\|^{-1}\mu$, that is, the normalized ground state of $\Hsc$.
Moreover, given a scalar function $\vp\in L^2(C_0)$ and a spin state $\Phi\in
L^2(C_0)\otimes\Si$, we shall use the notation
\[
(\vp,\Phi)=\int_{C_0}\overline{\vp}\ms\Phi
\]
to denote their $\Si$-valued scalar product.

\begin{lemma}\label{freeze.NN}
  Let $\Psi$ be an algebraic eigenfunction of $H$ with energy $E$ (cf.\
  Theorem~\ref{thm.H0}). Assume that there exists an eigenfunction $\psi$ of
  $\Hsc$ and a $\Si$-valued polynomial $F\in\CC[\bx,a^{-1}]\otimes\Si$ such that
\begin{equation}\label{property}
\Bmu^{-1}(\Psi-\psi F)=\cO(a^{-1})\,,
\end{equation}
and suppose, moreover, that these eigenfunctions can be normalized so that
\begin{equation}\label{norm.Psi}
\Bmu^{-1}\Psi\in\CC[\bx,a^{-1}]\otimes\Si\,,\qquad\Bmu^{-1}\psi\in\CC[\bx,a^{-1}]\,.
\end{equation}
Then the limits
\begin{align*}
c=\lim_{a\to\infty}\frac{\psi(\bxi)}{\Bmu(\bxi)}\,,\qquad
\chi=\lim_{a\to\infty} F(\bxi)\,,\qquad \ssE=\lim_{a\to\infty}\frac{E}{2a}
\end{align*}
exist, and
\begin{equation}\label{ssH-ssE}
c\ms\ms(\ssH-\ssE) \chi=0\,.
\end{equation}
\end{lemma}
\begin{proof}
The existence of the above limits trivially follows from the polynomial
dependence of $\Bmu\psi$ and $F$ on $a^{-1}$ and the fact that $E$ is linear in
$a$. Furthermore, the self-adjointness of $\Hsc$,
Eqs.~\eqref{ssHssh.NN}-\eqref{property} and Lemma~\ref{delta.NN} readily imply
that
\begin{align*}
(\Bmu,\Hsc\Psi)&=(\Hsc\Bmu,\Psi)=0\,,\\
(\Bmu,\ssh\Psi)&=\int_{C_0}\vp_0\, \big[\Bmu^{-1}\psi\ms \ssh
F+\cO(a^{-1})\big]=c\ssH\chi+\cO(a^{-1})\,.
\end{align*}
If we now use Eq.~\eqref{HHsc.NN} and Lemma~\ref{delta.NN} to write
\[
(\Bmu,\Hsc\Psi)+2a(\Bmu,\ssh\Psi)=(\Bmu,H\Psi)=Ec\chi+\cO(1)\,,
\]
substitute the former equations into this identity and take the limit $a\to\infty$,
we readily derive Eq.~\eqref{ssH-ssE}.
\end{proof}

\begin{corollary}\label{coro.NN}
If $c\chi\neq0$, then $\chi$ is an eigenvector of $\ssH$ with energy $\ssE$.
\end{corollary}

\begin{remark}
  Since, by Lemma~\ref{delta.NN},
  \[
  c = \lim_{a\to\infty}(\Bmu,\psi)\,,
  \]
  the previous corollary is of no practical use unless $\psi$ can be taken as
  the normalized ground state $\Bmu$, in which case $c=1$.
\end{remark}

We are now ready to prove the main result of this chapter.

\begin{theorem}\label{main.NN}
If $\ket\bs\in\Si$ and $\ket{\bs'}\in\Si'$, the states
\begin{subequations}\label{s0}
\begin{align}\label{s00}
\chi_0&\equiv\chi_0(\ket\bs)=\La\ket\bs\,,\\[1ex] \label{s01}
\chi_1&\equiv\chi_1(\ket\bs)=\sum_i\xi_i\,\ket{\bs_i}\,,\\\label{s02}
\chi_2&\equiv\chi_2(\ket{\bs'})=\sum_i\xi_i^2\ket{\bs'_i}+(N-1)\sum_{i<j}\xi_i\xi_j\,\ket{\bs_{ij}'^+}\,,\quad
S_{12}\ket{\bs'}=\ket{\bs'}\,,
\\\label{s03}
\chi_3&\equiv\chi_3(\ket\bs)=\sum_{i<j}\xi_i\xi_j(\xi_i-\xi_j)\,\ket{\bs_{ij}^-}+2\sum_i\xi_i\ket{\bs_i}\,,
\quad S_{12}\ket\bs=-\ket\bs\,,
\end{align}
\end{subequations}
satisfy the equations
\begin{equation*}
(\ssH-i)\chi_i=0\,,\qquad i=0,1,2,3.
\end{equation*}
\end{theorem}
\begin{proof}
  The proof consists in the application of Lemma~\ref{freeze.NN} to
  appropriate triples $(\Psi,\psi,F)$, where in fact $\psi=\Bmu$ (and hence $c=1$)
  in view of the previous remark.

  In the first place, it is easy to show that $\chi_0\in\ker\ssH$ directly.
  However, we prefer to prove it using the same freezing trick argument as in
  the other cases. To this end, consider the functions
\begin{align*}
\Psi=\frac{\Psi^{(0)}_{00}}{\|\Bmu\|}=\Bmu\ms\La\ket\bs\,,\qquad
F=\La\ket\bs\,,
\end{align*}
which trivially satisfy the hypotheses in Lemma~\ref{freeze.NN} with $E=0$ and
$\psi=\Bmu$. Since in this case
\[
\ssE=0\,,\qquad \chi=\La\ket\bs\,,
\]
from Corollary~\ref{coro.NN} it follows that $\chi_0$, if non-vanishing, is an
eigenvector of $\ssH$ with eigenvalue $0$.

The next case is slightly more complicated. Let us set
\begin{align*}
\Psi=\frac{\Psi^{(1)}_{01}}{\|\mu\|}=\Bmu(\Phi^{(1)}-\Bx\ms\Phi^{(0)})\,,\qquad F=\Phi^{(1)}-\Bx\ms\Phi^{(0)}\,.
\end{align*}
It is not difficult to show that $(\Psi,\Bmu,F)$ satisfies the hypotheses in
Lemma~\ref{freeze.NN} with $E=2a$. Since $\ssE=1$ in this case,
Corollary~\ref{coro.NN} shows that the state
\begin{align*}
\chi&=\lim_{a\to\infty}F(\bxi)=\sum_i\xi_i\ket{\bs_i}
\end{align*}
is either zero or an eigenvector of $\ssH$ with energy 1. Here we have used
the identity $\sum_i\xi_i=0$, cf.~Corollary~\ref{xibar}.

The other two states are obtained in a similar manner starting with $\psi=\Bmu$
and the functions
\begin{gather*}
  \Psi=\frac{\Psi^{(2)}_{02}+(N-1)\ms\tPsi^{(2)}_{02}}{\|\mu\|}=\Bmu\big(\Phi^{(2)}+(N-1)\ms
  \tPhi^{(2)}-2N\Bx\ms\Phi^{(1)}+(N+2)\Bx^2\Phi^{(0)}\big)\,,\\
  F=\frac{\Psi}{\Bmu}\,,\qquad E=4a\,,\qquad \ket{\bs'}\in\Si'\,,\qquad
  S_{12}\ket{\bs'}=\ket{\bs'}\,,
%F=\Phi^{(2)}+(N-1)\ms \tPhi^{(2)}-2N\Bx\ms\Phi^{(1)}+(N+2)\Bx^2\Phi^{(0)}\\
\end{gather*}
and
\begin{gather*}
\Psi=\frac{\hPsi^{(3)}_{03}-4\Psi^{(1)}_{03}-\frac83\Psi^{(0)}_{03}}{\|\mu\|}=\Bmu\bigg[\hPhi^{(3)}-2\Bx\ms\Phi^{(2)}+\frac{2r^2}N\Phi^{(1)}+\Bx\bigg(
\frac{r^2}{2N}-\frac{4\Bx^2}3\bigg)\Phi^{(0)}
\bigg]\,,\\
\qquad F=\frac{\Psi}{\Bmu}\,,\qquad E=6a\,,
\qquad S_{12}\ket{\bs}=-\ket{\bs}
\end{gather*}
in each case. Indeed, with the former set of functions we immediately arrive at
\[
\chi=\sum_i\xi_i^2\ket{\bs_i'}+(N-1)\sum_{i<j}\xi_i\xi_j\ket{\bs_{ij}'^+}\,,
\]
while in the latter case we obtain
\[
\chi=\sum_{i<j}\xi_i\xi_j(\xi_i-\xi_j)\ket{\bs_{ij}^-}+\frac2N\|\bxi\|^2\sum_i\xi_i\ket{\bs_i}\,.
\]
Using Proposition~\ref{xi2}, the result follows.
\end{proof}
\begin{remark}\label{integerenergies}
All the algebraic energies of $H$ being integral multiples of $a$, it follows
that any eigenvalue of the chain constructed from them via a freezing argument
must be an integer.
\end{remark}
\begin{remark}
The linear combinations of eigenfunctions with energy 2 and 3 used in
Theorem~\ref{main.NN} are not casual. Should one start, e.g., with
\[
\Psi=\frac{\Psi^{(2)}_{02}}{a\|\mu\|}\,,\qquad\psi=\Bmu\,,\qquad
F=\Bmu^{-1}\Psi\,,
\]
for which $F=\cO(1)$ and $c=1$, one would be lead to
\[
\chi=-\frac{2N}{N-1}\,\Big(\sum_i\xi_i\Big)^2\Phi^{(0)}=0\,.
\]
The above linear combinations have been chosen so as to suppress the dominant
parts in $\tPsi^{(2)}_{02}$ and $\hPsi^{(3)}_{03}$, which respectively are
proportional to $\Bx^2$ and $\Bx^3$.

It is also clear that the state $\chi_2$ (resp.\ $\chi_3$) is zero whenever the
spin vector $\ket\bs$ (resp.\ $\ket{\bs'}$) is not symmetric (resp.\
antisymmetric) under $S_{12}$, which accounts for the condition on $\ket\bs$
and $\ket{\bs'}$ imposed in Eqs.~\eqref{s01} and~\eqref{s03}.
\end{remark}

It is natural to wonder whether additional eigenvectors of the chain can be
computed using some different linear combinations of the algebraic
eigenfunctions of $H$ listed in Theorem~\ref{thm.H0}. Nevertheless, since
\[
c\chi=\lim_{a\to\infty}F(\bxi)=\lim_{a\to\infty}\frac{\Psi(\bxi)}{\Bmu(\bxi)}\,,
\]
it is clear that an algebraic eigenfunction in the linear span of
$\Psi^{(j)}_{lm}$, $\tPsi^{(j)}_{lm}$ and $\hPsi^{(j)}_{lm}$ can only give rise
to a nonvanishing vector $c\chi$ when $l=0$ and $m=j$. Indeed, $\phi=a^{-l}\Bmu
L_l^{\be(0)}(ar^2)\in\Bmu\CC[\bx,a^{-1}]$, with $\be(0)=1-N(a+\frac12)$, is an
eigenfunction of $\Hsc$ with energy $4al$, which implies that
\[
\lim_{a\to\infty}a^{-l}L_l^{-\be-k}(aN^2)=0\qquad\forall l>0
\]
as $\|\bxi\|^2=N$, proving the former statement. The latter condition follows
directly from the factor $\Bx^{m-j}$ present at the above eigenfunctions of
$H$. This discussion strongly suggests (albeit does not rigorously prove) that
one cannot obtain any additional eigenvectors of $\ssH$ using the technique
developed in Lemma~\ref{freeze.NN} and Theorem~\ref{main.NN}.

\section{Qualitative properties of the algebraic spectrum}
\label{S:qual.NN}

In this section we shall analyze which portion of the spectrum
of~\eqref{ssH.NN} is algebraic and how the algebraic eigenvalues are embedded
into the whole spectrum. Regarding the first point, it is convenient to start
with the following states $\Phi_i\in L^2(C_0)\otimes\Si$, whose evaluation at $\bxi$
yields the chain eigenvectors~\eqref{s0}:
\begin{align*}
\Phi_0(\ket\bs)&=\Phi^{(0)}(\ket\bs)\,,\\
\Phi_1(\ket\bs)&=\Phi^{(1)}(\ket\bs)\,,\\
\Phi_2(\ket\bs)&=\Phi^{(2)}(\ket\bs)+(N-1)\ms\tPhi^{(2)}(\ket\bs)\,,\\
\Phi_3(\ket\bs)&=\hPhi^{(3)}(\ket\bs)-2\Bx\Phi^{(2)}(\bs)+\frac{2r^2}N\Phi^{(1)}(\ket\bs)\,.
\end{align*}
It may be easily shown that if $\ket\bs$ is symmetric then $\Phi_1(\ket\bs)$ and
$\Phi_2(\ket\bs)$ vanish at $\bx=\bxi$. Likewise, if $\ket\bs\sim0$ then
$\Phi_2(\ket\bs)$ is also zero when $\bx=\bxi$. For this reason, we shall
consider the following spaces:
\begin{align*}
  \cV_0&=\La(\Si)\,,\qquad
  \cV_1=\Big\lan\Phi_1(\ket\bs):\ket\bs\in\Si/\La(\Si)\Big\ran\,,\\[1mm]
  \cV_2&=\Big\lan\Phi_2(\ket\bs):\ket\bs\in\Big(\Si'/\La(\Si)\Big)/{\sim}
  \Big\ran\,,\\[1mm]
  \cV_3&=\Big\lan\Phi_3(\ket\bs):\ket\bs\in\Si\,,\
  S_{12}\ket\bs=-\ket\bs\Big\ran\,.
\end{align*}
\begin{proposition}\label{num.dyn}
The dimensions of the spaces $\cV_i$ are given by
\begin{align}
  \label{dimcV0}
\dim\cV_0&=\binom{N+2M}N\,,\\[1mm]
\label{dimcV12}
\dim\cV_1&=\dim\cV_2=(N-1)\binom{N+2M-1}N\,,\\[1mm]
\label{dimcV3}
\dim\cV_3&=\binom{N-1}2\binom{N+2M-2}N\,.
\end{align}
\end{proposition}
\begin{proof}
  First of all, the dimension of $\cV_0$ is simply the number of permutations
  with repetitions of $N$ elements from $2M+1$.

  Consider next the space $\cV_1$. If two basic states $\ket\bs$ and
  $\ket{\bs'}$ differ by a permutation of the last $N-1$ spins, then
  $\Phi_1(\ket\bs)=\Phi_1(\ket{\bs'})$. Hence the dimension of $\cV_1$ is given by
\[
(2M+1)\binom{N+2M-1}{2M}-\binom{N+2M}{2M}\,,
\]
where the first quantity is the number of possible choices for $s_1$, the
second one is that of symmetric states of $N-1$ particles, and the last one,
that of symmetric states of $N$ particles.

The dimension of $\cV_2$ coincides with that of the space
$\big(\Si'/\La(\Si)\big)/{\sim}\,$, a basis of which was computed in
Propositions~\ref{P:basic} and~\ref{P:basic2}. Using the previous notation we
can say that, for a given spin content $\{s^1,\dots,s^n\}\subset\{-M,\dots,M\}$
with multiplicities $(\nu_1,\dots,\nu_n)\in\NN^n$ (with $\sum_{j=1}^n\nu_j=N$),
there are $n-1$ independent states in $\big(\Si'/\La(\Si)\big)/{\sim}\,$. Hence
\[
\dim\cV_2=\sum_{n=1}^{\min\{2M+1,N\}}(n-1)\ms\card(\fP_N\cap
\NN^n)\binom{2M+1}n\,,
\]
where the binomial term accounts for the different choices of
$\{s^1,\dots,s^n\}$, and the cardinal for all the possible values of the
multiplicities. Noting $\card(\fP_N\cap \NN^n)=\binom{N-1}{n-1}$ and
using the identity
\[
\sum_{k=0}^\infty\binom nk\binom m{j+k}=\binom{m+n}{m-j}\,,
\]
we arrive at the desired formula.

It only remains to prove Eq.~\eqref{dimcV3}. To this end, let us
concentrate on the spin vectors $\Si_0$ with a fixed spin content
$\{s^1,\dots,s^n\}$, $(\nu_1,\dots,\nu_n)$. It is clear that
$\cV_3(\Si_0)=\lan\Phi_3(\ket\bs):\ket\bs\in\Si_0\ran$ is spanned by the
$\binom n2$ states
\[
\bv(s^i,s^j)=\nu_i\nu_j\,\Phi_3\big(\ket{s^is^j\cdots}-\ket{s^js^i\cdots}\big)\,.
\]
We shall shortly show that, for all $i$,
\begin{equation}\label{uves}
\sum_j\bv(s^i,s^j)=0\,.
\end{equation}
By antisymmetry, the double sum $\sum_{i,j}\bv(s^i,s^j)$ vanishes, so there
are at most $n-1$ independent relations of the above form. It can be verified
that $\{\bv(s^i,s^j):1\leq i<j<n\}$ is in fact a basis of $\cV_3(\Si_0)$, so
the dimension of $\cV_3$ is obtained from (ii) by substituting $n-1$ for
$\binom {n-1}2$.

Finally, let us prove~\eqref{uves}. A first observation is that
\[
r^2\Phi^{(1)}(\ket{s^is^j\cdots})-N\Bx
\Phi^{(2)}(\ket{s^is^j\cdots})=-\sum_{k}\nu_k\hPhi^{(3)}(\ket{s^is^k\cdots})\,.
\]
Hence
\begin{multline}\label{RHS}
  \frac N{2\nu_i}\sum_{i,j}\bv(s^i,s^j)=N\sum_j\nu_j\hPhi^{(3)}(\ket{s^is^j\cdots})\\
  -\sum_{j,k}\nu_j\nu_k\hPhi^{(3)}(\ket{s^is^k\cdots})
  +\sum_{j,k}\nu_j\nu_k\hPhi^{(3)}(\ket{s^js^k\cdots})\,.
\end{multline}
The first two terms of the RHS obviously cancel, and the last one vanishes by
antisymmetry.
\end{proof}

This proposition, which is interesting in its own right because of
its connection with the dynamical model~\eqref{V0}, has the
following consequences for the spin chain~\eqref{ssH.NN}:

\begin{corollary}
For each $i=0,1,2,3$, the number of algebraic eigenvectors with energy $i$ is
bounded by $\dim\cV_i$.
\end{corollary}

\begin{corollary}
The ratio
\[
\frac{\text{number of algebraic states}}{\text{total number of
states}}
\]
tends to zero as $N\to\infty$. The whole kernel of $\ssH$ is
composed of algebraic eigenvectors.
\end{corollary}
\begin{proof}
The first part of the statement follows by direct comparison with the total
number of states $(2M+1)^N$. The second part is due to the fact that $\ssH$ is
a linear combination with positive coefficients of the nonnegative operators
$1-S_{i,i+1}$, which implies that $\ker\ssH=\La(\Si)$.
\end{proof}

These observations are corroborated by our numerical simulations: the number
of algebraic levels increase as $M$ and $N$ become larger, but nevertheless
they span a lesser and lesser part of the spin space. The kernel of the chain,
e.g., shows clearly this behavior. But most remarkable is the following
conjecture regarding the embedding of the algebraic eigenvalues in the
spectrum.

\begin{conjecture}\label{mainconj.NN}
  The algebraic eigenvectors~\eqref{s00}--\eqref{s02} span the three lowest
  levels of the spin chain~\eqref{ssH.NN}. When $M\geq1$ the fourth algebraic
  energy, $\ssE=3$, is however the fifth lowest level of the chain, and its
  whole eigenspace is spanned by the states of the form~\eqref{s03}. The
  algebraic energies are singled out in the spectrum of $\ssH$ as being the
  only integer ones, i.e.,
\[
\spec(\ssH)\cap\ZZ=\{0,1,2,3\}\,.
\]
\end{conjecture}
We have numerically diagonalized the matrix representing the Hamiltonian
$\ssH$ for $M\leq 2$ and up to $12$ particles, and our results fully support the
above conjecture. Another important observation is that the standard freezing
trick relation between the energies of the spin chain $\ssH$ and of the
dynamical models $H$ and $\Hsc$, namely
\[
\ssE = \lim_{a\to\infty}\frac{E-E\scal}{2a}\,,
\]
and the fact that $\ssH$ has noninteger eigenvalues imply that the spin
dynamical model~\eqref{V0} has noninteger energies, and hence that some of its
eigenfunctions are not among those listed in Theorem~\ref{thm.H0}.

\chapter{Conclusions and perspectives}
\label{Ch:Final}

In this thesis we have studied three QES models of CS type presenting
short-range interactions and developed a rigorous treatment of the CS/HS
correspondence, valid for both ES and QES Hamiltonians. In a nutshell, the
main conclusions and results that one should extract from this dissertation
are the following:

\begin{enumerate}
\item\emph{On the $BC_N$ HS spin chain:}\/ We have computed the spectrum of
  the trigonometric $BC_N$ Sutherland model by expressing its Hamiltonian as
  the sum of squares of an appropriate set of commuting Dunkl operators. The
  $BC_N$ HS chain is obtained from this system by ``freezing'' the particles
  at their classical equilibrium positions. We have proved a rigorous version
  of the freezing trick which has enabled us to compute the partition function
  of the chain in terms of those of the corresponding scalar and spin models.
  We have resorted to numerical computations to explore the level distribution
  of the chain, finding out that the energies accurately follow the Gaussian
  law. A constant magnetic field was added to the picture in a rather
  straightforward way.

\item\emph{On NN spin models:}\/ We defined three QES spin models with
  near-neighbors interactions that are related to NN versions of the sum of
  the squares of appropriate sets of $A_N$ Dunkl operators. Furthermore, we
  have obtained infinite flags of finite-dimensional spaces invariant under
  these models, and computed several families of eigenfunctions in closed form
  by diagonalizing the restriction of the Hamiltonians to these spaces. All
  the algebraic solutions, whose energies are integers, can be expressed as
  the ground state times a symmetric function. We conjecture that some of
  their eigenfunctions are not of this form, and that their corresponding
  eigenvalues are not natural numbers.

\item\emph{On NN spin chains:}\/ We have defined a spin chain presenting
  nearest-neighbors interactions that is related to one of the previous NN
  models by a freezing trick argument. We have discussed its site distribution
  in detail, both analytically and numerically, and obtained a number of
  algebraic eigenvectors and energies using a refinement of the techniques
  developed in Chapter~\ref{Ch:BCN} that also applies to QES systems. We have
  raised the conjecture that the only integer energies of this chain are
  $0,1,2,$ and $3$, and that the eigenstates corresponding to these energies
  are all algebraic.
\end{enumerate}

The conjectures scattered throughout the text do not exhaust all the open
questions that we find worth exploring in connection with our work. In what
follows, we shall briefly mention some wide research lines that in our opinion
would be of interest, and that we intend to follow in the near future:
\begin{enumerate}
\item\emph{The structure of the (non-algebraic) eigenfunctions of $H_0$ (and
    other QES operators):\/} From the previous discussion, it looks unlikely
  that all the eigenfunctions of $H_0$ admit the ground state as a Jastrow
  factor. Nevertheless, this does happen in the theory of CS models, and the
  functions that are explicitly constructed in the usual QES models are
  generally of this form. A thorough study of this phenomenon is, to the best
  of our knowledge, still lacking, and would probably shed some light on the
  connections between Schrödinger operators and orthogonal polynomials. The
  particular case of $H_0$ would be of special interest.

\item\emph{HS/CS and symmetric spaces:\/} A powerful procedure for
  constructing CS models with spin starting out with a symmetric space of
  nonpositive curvature has been recently developed by Fehér and Pusztai, as
  described in Section~\ref{S:CM}. It should not be difficult to extend the
  techniques described in this thesis to solve the associated spin chains.

\item\emph{Statistical properties of the spectrum of CS models:\/}
The remarkable statistical properties of the $BC_N$ HS chain occupied a
significant portion of Section~\ref{S:stat.BCN}. It is difficult to obtain pen
and paper results on these aspects, but the rigorous literature studying these
topics is always increasing (see~\cite{CM04,De00} and references therein).
Further study of the topics mentioned in this section and extensions to the
case of NN models would make a valuable addition to the existing literature.

\item\emph{Integrability vs.\ solvability in spin chains of HS
type:\/} A satisfactory analysis of the integrability of the chains of HS type
has only been accomplished in a few instances. A natural approach to this
problem would be to define a set of integrals of the chain using the freezing
trick, but in general it is not easy to grant the linear independence of these
conserved operators. If one managed to overcome these technical problems, this
method (which has already been exploited in the study of some particular
models) would provide a clean proof of the integrability of the chains
associated with integrable spin models.
\end{enumerate}

\backmatter

%\bibliography{cmprefs}\frenchspacing

\begin{thebibliography}{100}
\providecommand{\url}[1]{\texttt{#1}}
\providecommand{\urlprefix}{URL }
\providecommand{\eprint}[2][]{\url{#2}}

\bibitem{AW05}
Abanov, A.~G. and Wiegmann, P.~A.: {Q}uantum hydrodynamics, the quantum
  {B}enjamin--{O}no equation, and the {C}alogero model.
\newblock Phys. Rev. Lett. \textbf{95}, 076402(5) (2005)

\bibitem{AM78}
Abraham, R. and Marsden, J.~E.: \emph{{F}oundations of {M}echanics}.
\newblock New York: Benjamin/Cummings, 1978

\bibitem{AKLT87}
Affleck, I., Kennedy, T., Lieb, E.~H., and Tasaki, H.: {R}igorous results on
  valence-bond ground states in antiferromagnets.
\newblock Phys. Rev. Lett. \textbf{59}, 799--802 (1987)

\bibitem{Ag82}
Agmon, S.: \emph{{L}ectures on Exponential Decay of Solutions of Second-Order
  Elliptic Equations}.
\newblock Princeton: Princeton University Press, 1982

\bibitem{ADS06}
Agrotis, M.~A., Damianou, P.~A., and Sophocleous, C.: {T}he {T}oda lattice is
  super-integrable.
\newblock Physica A \textbf{365}, 235--243 (2006)

\bibitem{AKLM03}
Aleekseevsky, D., Kriegl, A., Losik, M., and Michor, P.~W.: {T}he {R}iemannian
  geometry of orbit spaces. {T}he metric, geodesics, and integrable systems.
\newblock Pub. Math. Debrecen \textbf{62}, 247--276 (2003)

\bibitem{ABZH87}
Anderson, P.~W., Baskaran, G., Zou, Z., and Hsu, T.: {R}esonating-valence-bond
  theory of phase transitions and superconductivity in {$\rm La_2CuO_4$}-based
  compounds.
\newblock Phys. Rev. Lett. \textbf{58}, 2790--2793 (1987)

\bibitem{AJ06}
Aniceto, A. and Jevicki, A.: {N}otes on collective field theory of matrix and
  spin {C}alogero models.
\newblock J. Phys. A: Math. Gen. \textbf{39}, 12765(91) (2006)

\bibitem{AF98}
Arutyunov, G.~E. and Frolov, S.~A.: {O}n the {H}amiltonian structure of the
  spin {R}uijsenaars--{S}chneider model.
\newblock J. Phys. A: Math. Gen. \textbf{31}, 4203--4216 (1998)

\bibitem{AJK01}
Auberson, G., Jain, S.~R., and Khare, A.: {A} class of {$N$}-body problems with
  nearest- and next-to-nearest-neighbour interactions.
\newblock J. Phys. A: Math. Gen. \textbf{34}, 695--724 (2001)

\bibitem{AI94}
Azuma, H. and Iso, S.: {E}xplicit relation of the quantum {H}all effect and the
  {C}alogero--{S}utherland model.
\newblock Phys. Lett. B \textbf{331}, 107--113 (1994)

\bibitem{Ba82}
Babujian, H.: {E}xact solution of the one-dimensional isotropic {H}eisenberg
  chain with arbitrary spin {$S$}.
\newblock Phys. Lett. A \textbf{90}, 479--482 (1982)

\bibitem{BF97b}
Baker, T.~H. and Forrester, P.~J.: {A} {$q$}-analogue of the type {A} {D}unkl
  operator and integral kernel.
\newblock Int. Math. Res. Not. \textbf{14}, 667--686 (1997)

\bibitem{BF97}
Baker, T.~H. and Forrester, P.~J.: {T}he {C}alogero--{S}utherland model and
  generalized classical polynomials.
\newblock Comm. Math. Phys. \textbf{188}, 175--216 (1997)

\bibitem{BEHR07}
Ballesteros, A., Enciso, A., Herranz, F.~J., and Ragnisco, O.: {A} maximally
  superintegrable system on an {$n$}-dimensional space of nonconstant
  curvature.
\newblock Submitted to Physica D; {\texttt{math-ph/0612080}}

\bibitem{BHSS03}
Ballesteros, A., Herranz, F.~J., Santander, M., and Sanz-Gil, T.: {M}aximal
  superintegrability on {$N$}-dimensional curved spaces.
\newblock J. Phys. A: Math. Gen. \textbf{36}, L93--L99 (2003)

\bibitem{BGG03}
Basu-Mallick, B., Ghosh, P., and Gupta, K.: {I}nequivalent quantizations of the
  rational {C}alogero model.
\newblock Phys. Lett. A \textbf{311}, 87--92 (2003)

\bibitem{BG01}
Basu-Mallick, B. and Gupta, K.: {B}ound states in one-dimensional quantum
  {$N$}-body systems with inverse square interaction.
\newblock Phys. Lett. A \textbf{292}, 36--42 (2001)

\bibitem{Be91}
Beerends, R.~J.: {C}hebyshev polynomials in several variables and the radial
  part of the {L}aplace--{B}eltrami operator.
\newblock Trans. Amer. Math. Soc. \textbf{328}, 779--814 (1991)

\bibitem{BS05}
Beisert, N. and Staudacher, M.: {L}ong-range {$\mathfrak{psu}(2,2|4)$} {B}ethe
  ans{{ä}}tze for gauge theory and strings.
\newblock Nucl. Phys. B \textbf{727}, 1--62 (2005)

\bibitem{BC04}
Berenstein, D. and Cherkis, S.: {D}eformations of {$N=4$} {SYM} and integrable
  spin chain models.
\newblock Nucl. Phys. B \textbf{702}, 49--85 (2004)

\bibitem{BEG03}
Berest, Y., Etingof, P., and Ginzburg, V.: {C}herednik algebras and
  differential operators on quasi-invariants.
\newblock Duke Math. J. \textbf{118}, 279--337 (2003)

\bibitem{BGHP93}
Bernard, D., Gaudin, M., Haldane, F. D.~M., and Pasquier, V.: {Y}ang--{B}axter
  equation in long-range interacting systems.
\newblock J. Phys. A: Math. Gen. \textbf{26}, 5219--5236 (1993)

\bibitem{BPS95}
Bernard, D., Pasquier, V., and Serban, D.: {E}xact solution of long-range
  interacting spin chains with boundaries.
\newblock Europhys. Lett. \textbf{30}, 301--306 (1995)

\bibitem{Be31}
Bethe, H.: {Z}ur {T}heorie der {M}etalle {I}. {E}igenwerte und
  {E}igenfunktionen der linearen {A}tomkette.
\newblock Z. Physik \textbf{71}, 205--226 (1931)

\bibitem{BEJ76}
Blair, J. M., Edwards, C. A., and Johnson, J. H.: Rational {C}hebyshev
approximations for the inverse of the error function.
\newblock Math. Comp. \textbf{30}, 827--830 (1976)

\bibitem{Bl33}
Bloch, F.: {B}remsverm{ö}gen von {A}tomen mit mehreren {E}lektronen.
\newblock Z. Physik \textbf{81}, 363--376 (1933)

\bibitem{BGS99}
Bogomolny, E.~B., Gerland, U., and Schmit, C.: {M}odels of intermediate
  spectral statistics.
\newblock Phys. Rev. E \textbf{59}, R1315--R1318 (1999)

\bibitem{BCS98}
Bordner, A., Corrigan, E., and Sasaki, R.: {C}alogero--{M}oser models. {I}. {A}
  new formulation.
\newblock Prog. Theor. Phys. \textbf{100}, 1107--1129 (1998)

\bibitem{BCS99}
Bordner, A., Corrigan, E., and Sasaki, R.: {G}eneralised {C}alogero--{M}oser
  models and universal {L}ax pair operators.
\newblock Prog. Theor. Phys. \textbf{102}, 499--529 (1999)

\bibitem{BMS00}
Bordner, A., Manton, N., and Sasaki, R.: {C}alogero--{M}oser models. {V}.
  {S}upersymmetry and quantum {L}ax pair.
\newblock Prog. Theor. Phys. \textbf{103}, 463--487 (2000)

\bibitem{BTV05}
Boreskov, K.~G., Turbiner, A.~V., and Vieyra, J. C.~L.: {S}olvability of the
  {H}amiltonians related to exceptional root spaces: {R}ational case.
\newblock Comm. Math. Phys. \textbf{260}, 17--44 (2005)

\bibitem{BH96}
Braden, H.~W. and Hone, A. N.~W.: {A}ffine {T}oda solitons and systems of
  {C}alogero--{M}oser type.
\newblock Phys. Lett. B \textbf{380}, 296--302 (1996)

\bibitem{BHV92}
Brink, L., Hansson, T.~H., and Vasiliev, M.~A.: {E}xplicit solution to the
  {$N$}-body {C}alogero problem.
\newblock Phys. Lett. B \textbf{286}, 109--111 (1992)

\bibitem{Ca71}
Calogero, F.: {S}olution of the one-dimensional {$N$}-body problems with
  quadratic and/or inversely quadratic pair potentials.
\newblock J. Math. Phys. \textbf{12}, 419--436 (1971)

\bibitem{CP78b}
Calogero, F. and Perelomov, A.: {A}symptotic density of the zeros of {H}ermite
  polynomials of diverging order, and related properties of certain singular
  integral operators.
\newblock Lett. Nuovo Cimento \textbf{23}, 650--652 (1978)

\bibitem{CL99}
Carey, A.~L. and Langmann, E.: {L}oop groups, anyons and the
  {C}alogero--{S}utherland model.
\newblock Comm. Math. Phys. \textbf{201}, 1--34 (1999)

\bibitem{CM04}
Caselle, M. and Magnea, U.: {R}andom matrix theory and symmetric spaces.
\newblock Phys. Rep. \textbf{394}, 41--156 (2004)

\bibitem{CC04}
Caudrelier, V. and Crampé, N.: {I}ntegrable {$N$}-particle hamiltonians with
  {Y}angian or reflection algebra symmetry.
\newblock J. Phys. A: Math. Gen. \textbf{37}, 6285--6298 (2004)

\bibitem{CV93}
Chalykh, O.~A. and Veselov, A.~P.: {I}ntegrability in the theory of
  {S}chr{ö}dinger operators and harmonic analysis.
\newblock Comm. Math. Phys. \textbf{152}, 29--40 (1993)

\bibitem{Ch91}
Cherednik, I.: {A} unification of {K}nizhnik--{Z}amolodchikov and {D}unkl
  operators via affine {H}ecke algebras.
\newblock Invent. Math. \textbf{106}, 411--431 (1991)

\bibitem{Ch94}
Cherednik, I.: {I}ntegration of quantum many-body problems by affine
  {K}nizhnik--{Z}amolodchi­kov equations.
\newblock Adv. Math. \textbf{106}, 65--95 (1994)

\bibitem{Ch95}
Cherednik, I.: {D}ouble affine {H}ecke algebras and {M}acdonald's conjectures.
\newblock Ann. of Math. \textbf{141}, 191--216 (1995)

\bibitem{CU96}
Cicuta, G.~M. and Ushveridze, A.~G.: {Q}uasi-exactly solvable problems and
  random matrix theory.
\newblock Phys. Lett. A \textbf{215}, 167--175 (1996)

\bibitem{CP90}
Cirelli, R. and Pizzocchero, L.: {O}n the integrability of quantum mechanics as
  an infinite-dimensional hamiltonian system.
\newblock Nonlinearity \textbf{3}, 1057--1080 (1990)

\bibitem{Je93}
de~Jeu, M. F.~E.: {T}he {D}unkl {T}ransform.
\newblock Invent. Math. \textbf{113}, 147--162 (1993)

\bibitem{Je06}
de~Jeu, M. F.~E.: {P}aley--{W}iener theorems for the {D}unkl {T}ransform.
\newblock Trans. Amer. Math. Soc. \textbf{358}, 4225--4250 (2006)

\bibitem{DG01}
Deguchi, T. and Ghosh, P.~K.: {S}pin chains from super-models.
\newblock J. Phys. Soc. Japan \textbf{70}, 3225--3237 (2001)

\bibitem{De00}
Deift, P.: \emph{{O}rthogonal polynomials and random matrices: {A}
  {R}iemann--{H}ilbert approach}.
\newblock Providence: AMS, 2000

\bibitem{CP62}
des Cloizeaux, J. and Pearson, J.: {S}pin-wave spectrum of the
  antiferromagnetic linear chain.
\newblock Phys. Rev. \textbf{128}, 2131--2135 (1962)

\bibitem{DLM01}
Desrosiers, P., Lapointe, L., and Mathieu, P.: {S}upersymmetric
  {C}alogero--{M}oser--{S}utherland models and {J}ack superpolynomials.
\newblock Nucl. Phys. B \textbf{606}, 547--582 (2001)

\bibitem{DLM03}
Desrosiers, P., Lapointe, L., and Mathieu, P.: {J}ack polynomials in
  superspace.
\newblock Comm. Math. Phys. \textbf{242}, 331--360 (2003)

\bibitem{DLM04}
Desrosiers, P., Lapointe, L., and Mathieu, P.: {E}xplicit formulae for the
  generalized {H}ermite polynomials in superspace.
\newblock J. Phys. A: Math. Gen. \textbf{37}, 1251--1268 (2004)

\bibitem{DP98}
D'Hoker, E. and Phong, D.~H.: {C}alogero--{M}oser systems in {SU($N$)}
  {S}eiberg--{W}itten theory.
\newblock Nucl. Phys. B \textbf{513}, 405--444 (1998)

\bibitem{FMS99}
di~Francesco, P., Mathieu, P., and Senechal, D.: \emph{{C}onformal Field
  Theory}.
\newblock New York: Springer-Verlag, 1999

\bibitem{Do06}
Dominici, D.: \texttt{math.CA/0601078}

\bibitem{Dr85}
Drinfeld, V.~G.: {H}opf algebras and the quantum {Y}ang--{B}axter equation.
\newblock Soviet Math. Dokl. \textbf{32}, 254--258 (1985)

\bibitem{Dr86}
Drinfeld, V.~G.: {D}egenerate affne Hecke algebras and Yangians.
\newblock Funct. Anal. Appl. \textbf{20}, 58--60 (1986)

\bibitem{Du89}
Dunkl, C.~F.: {D}ifferential-difference operators associated to reflection
  groups.
\newblock Trans. Amer. Math. Soc. \textbf{311}, 167--183 (1989)

\bibitem{Du98}
Dunkl, C.~F.: {O}rthogonal polynomials of types ${A}$ and ${B}$ and related
  {C}alogero models.
\newblock Comm. Math. Phys. \textbf{197}, 451--487 (1998)

\bibitem{DX01}
Dunkl, C.~F. and Xu, Y.: \emph{{O}rthogonal {P}olynomials of {S}everal
  {V}ariables}.
\newblock Cambridge: Cambridge University Press, 2001

\bibitem{Dy70}
Dyson, F.~J.: {C}orrelations between eigenvalues of a random matrix.
\newblock Comm. Math. Phys. \textbf{19}, 235--250 (1970)

\bibitem{EFGR07b}
Enciso, A., Finkel, F., González-López, A., and Rodr{í}guez, M.~A.: {A}
  novel quasi-exactly solvable spin chain with nearest-neighbors interactions.
\newblock To be submitted to Nucl. Phys. B

\bibitem{EFGR05c}
Enciso, A., Finkel, F., González-López, A., and Rodr{í}guez, M.~A.: {A}
  {H}aldane--{S}hastry spin chain of {$BC_N$} type in a constant magnetic
  field.
\newblock J. Nonlin. Math. Phys. \textbf{12}, 253--265 (2005)

\bibitem{EFGR05}
Enciso, A., Finkel, F., González-López, A., and Rodr{í}guez, M.~A.:
  {H}aldane--{S}hastry spin chains of {$BC_N$} type.
\newblock Nucl. Phys. B \textbf{707}, 553--576 (2005)

\bibitem{EFGR05b}
Enciso, A., Finkel, F., González-López, A., and Rodr{í}guez, M.~A.:
  {S}olvable scalar and spin models with near-neighbors interactions.
\newblock Phys. Lett. B \textbf{605}, 214--222 (2005)

\bibitem{EFGR06}
Enciso, A., Finkel, F., González-López, A., and Rodr{í}guez, M.~A.:
  {Q}uasi-exactly solvable {$N$}-body spin {H}amiltonians with short-range
  interaction potentials.
\newblock SIGMA \textbf{2}, 73(11) (2006)

\bibitem{EFGR07}
Enciso, A., Finkel, F., González-López, A., and Rodr{í}guez, M.~A.:
  {E}xchange operator formalism for {$N$}-body spin models with near-neighbors
  interactions.
\newblock J. Phys. A: Math. Theor. \textbf{40}, 1857--1883 (2007)

\bibitem{EP06}
Enciso, A. and Peralta-Salas, D.: {C}lassical and quantum integrability of
  {H}amiltonians without scattering states.
\newblock Theor. Math. Phys. \textbf{148}, 1086--1099 (2006)

\bibitem{EP06b}
Enciso, A. and Polychronakos, A.~P.: {T}he fermion density operator in the
  droplet bosonization picture.
\newblock Nucl. Phys. B \textbf{751}, 376--389 (2006)

\bibitem{Et06}
Etingof, P.: {L}ectures on {C}alogero--{M}oser systems.
\newblock \texttt{math.QA/0606233}

\bibitem{EG02}
Etingof, P. and Ginzburg, V.: {S}ymplectic reflection algebras,
  {C}alogero--{M}oser space, and deformed {H}arish--{C}handra homomorphism.
\newblock Invent. Math. \textbf{147}, 243--348 (2002)

\bibitem{Ev90}
Evans, N.~W.: {S}uperintegrability in classical mechanics.
\newblock Phys. Rev. A \textbf{41}, 5666--5576 (1990)

\bibitem{Ev90b}
Evans, N.~W.: {S}uperintegrability of the {W}internitz system.
\newblock Phys. Lett. A \textbf{147}, 483--486 (1990)

\bibitem{EGKP05}
Ezung, M., Gurappa, N., Khare, A., and Panigrahi, P.~K.: {Q}uantum many-body
  systems with nearest and next-to-nearest neighbor long-range interactions.
\newblock Phys. Rev. B \textbf{71}, 125121(8) (2005)

\bibitem{FP06c}
Fehér, L. and Pusztai, B.~G.: {A} class of {C}alogero type reductions of free
  motion on a simple {L}ie group.
\newblock \texttt{math-ph/0609085}

\bibitem{FP06b}
Fehér, L. and Pusztai, B.~G.: {S}pin {C}alogero models associated with
  {R}iemannian symmetric spaces of negative curvature.
\newblock Nucl. Phys. B \textbf{751}, 436--458 (2006)

\bibitem{FP06}
Fehér, L. and Pusztai, B.~G.: {S}pin {C}alogero models obtained from
  dynamical {$r$}-matrices and geodesic motion.
\newblock Nucl. Phys. B \textbf{734}, 304--325 (2006)

\bibitem{FTF05}
Feh{é}r, L., Tsutsui, I., and F{ü}l{ö}p, T.: {I}nequivalent quantizations
  of the three-particle {C}alogero model constructed by separation of
  variables.
\newblock Nucl. Phys. B \textbf{715}, 713--757 (2005)

\bibitem{FGP05}
{Fernández Nú{ñ}ez}, J., {Garc{í}a Fuertes}, W., and Perelomov, A.~M.:
  {Q}uantum trigonometric {C}alogero--{S}utherland model, irreducible
  characters and {C}lebsch--{G}ordan series for the exceptional algebra
  ${E}_{7}$.
\newblock J. Math. Phys. \textbf{46}, 103505(27) (2005)

\bibitem{FGGRZ01}
Finkel, F., Gómez-Ullate, D., González-López, A., Rodr{í}guez, M.~A.,
  and Zhdanov, R.: ${A}_{N}$-type {D}unkl operators and new spin
  {C}alogero--{S}utherland models.
\newblock Comm. Math. Phys. \textbf{221}, 477--497 (2001)

\bibitem{FGGRZ01b}
Finkel, F., Gómez-Ullate, D., González-López, A., Rodr{í}guez, M.~A.,
  and Zhdanov, R.: {N}ew spin {C}alogero--{S}utherland models related to
  ${B}_{N}$-type {D}unkl operators.
\newblock Nucl. Phys. B \textbf{613}, 472--496 (2001)

\bibitem{FGGRZ03}
Finkel, F., Gómez-Ullate, D., González-López, A., Rodr{í}guez, M.~A.,
  and Zhdanov, R.: {O}n the {S}utherland model of ${B}_{{N}}$ type and its
  associated spin chain.
\newblock Comm. Math. Phys. \textbf{233}, 191--209 (2003)

\bibitem{FG05}
Finkel, F. and González-López, A.: {G}lobal properties of the spectrum of
  the {H}aldane--{S}hastry spin chain.
\newblock Phys. Rev. B \textbf{72}, 174411(6) (2005)

\bibitem{FGNR00}
Fock, V., Gorsky, A., Nekrasov, N., and Rubtsov, V.: {D}uality in integrable
  systems and gauge theories.
\newblock J. High Energy Phys. \textbf{2000}, 28(40)

\bibitem{Fo92}
Forrester, P.: {E}xact eigenstates of some spin-{$\frac12$} {H}eisenberg chains
  with {$1/r^2$} exchange.
\newblock J. Phys. A: Math. Gen. \textbf{21}, 5447--5462 (1992)

\bibitem{FM93}
Fowler, M. and Minahan, A.: {I}nvariants of the {H}aldane--{S}hastry
  {$\mathrm{SU}(N)$} chain.
\newblock Phys. Rev. Lett. \textbf{70}, 2325--2328 (1993)

\bibitem{Fr93}
Frahm, H.: {S}pectrum of a spin chain with inverse-square exchange.
\newblock J. Phys. A: Math. Gen. \textbf{26}, L473--L479 (1993)

\bibitem{FKM05}
Freyhult, L., Kristjansen, C., and M{å}nsson, T.: {I}ntegrable spin chains
  with {$\mathrm{U}(1)^3$} symmetry and generalized {L}unin--{M}aldacena
  backgrounds.
\newblock J. High Energy Phys. \textbf{2005}, 8(17)

\bibitem{FMSUW65}
Fris, I., Mandrosov, V., Smorodinsky, J., Uhlir, M., and Winternitz, P.: {O}n
  higher- order symmetries in quantum mechanics.
\newblock Phys. Lett. \textbf{16}, 354--356 (1965)

\bibitem{GV87}
Gebhard, F. and Vollhardt, D.: {C}orrelation functions for {H}ubbard-type
  models: the exact results for the {G}utzwiller wave function in one
  dimension.
\newblock Phys. Rev. Lett. \textbf{59}, 1472--1475 (1987)

\bibitem{Go05}
Golubeva, V.: {H}amiltonians of the {C}alogero--{S}utherland type models
  associated to the root systems and corresponding {F}ock spaces.
\newblock In Fainsilber, L. and Hobbs, C., eds., \emph{European Women in
  Mathematics}. New York: Hindawi Publishing, 2005, pp. 217--229

\bibitem{GT04}
Gonz\'{a}lez-L\'{o}pez, A. and Tanaka, T.: A new family of {$\mathcal N$}-fold
  supersymmetry: type {B}.
\newblock Phys. Lett. B \textbf{586}, 117--124 (2004)

\bibitem{Go05b}
Gorsky, A.: {S}pin chains and gauge-string duality.
\newblock Theor. Math. Phys. \textbf{142}, 153--165 (2005)

\bibitem{GN94}
Gorsky, A. and Nekrasov, N.: {H}amiltonian systems of {C}alogero type, and
  2-dimensional {Y}ang--{M}ills theory.
\newblock Nucl. Phys. B \textbf{414}, 213--238 (1994)

\bibitem{GR00}
Gradshteyn, I.~S. and Ryzhik, I.~M.: \emph{{T}able of {I}ntegrals, {S}eries,
  and {P}roducts}.
\newblock San Diego: Academic Press, 2000, sixth ed.

\bibitem{GJR87}
Gros, C., Joynt, R., and Rice, T.~M.: {A}ntiferromagnetic correlations in
  almost-localized {F}ermi liquids.
\newblock Phys. Rev. B \textbf{36}, 381--393 (1987)

\bibitem{Gu63}
Gutzwiller, M.~C.: {E}ffect of correlation on the ferromagnetism of transition
  metals.
\newblock Phys. Rev. Lett. \textbf{10}, 159--162 (1963)

\bibitem{Ha94}
Ha, Z. N.~C.: {E}xact dynamical correlation functions of
  {C}alogero--{S}utherland model and one-dimensional fractional statistics.
\newblock Phys. Rev. Lett. \textbf{73}, 1574--1577 (1994)

\bibitem{Ha88}
Haldane, F. D.~M.: {E}xact {J}astrow--{G}utzwiller resonating-valence-bond
  ground state of the spin-$1/2$ antiferromagnetic {H}eisenberg chain with
  $1/r^2$ exchange.
\newblock Phys. Rev. Lett. \textbf{60}, 635--638 (1988)

\bibitem{Ha91}
Haldane, F. D.~M.: ``{S}pinon gas'' description of the $s=\frac12$ {H}eisenberg
  chain with inverse-square exchange: exact spectrum and thermodynamics.
\newblock Phys. Rev. Lett. \textbf{66}, 1529--1532 (1991)

\bibitem{HHTBP92}
Haldane, F. D.~M., Ha, Z. N.~C., Talstra, J.~C., Bernard, D., and Pasquier, V.:
  {Y}angian symmetry of integrable quantum chains with long-range interactions
  and a new description of states in conformal field theory.
\newblock Phys. Rev. Lett. \textbf{69}, 2021--2025 (1992)

\bibitem{HL06}
Halln{\"{a}}s, M. and Langmann, E.: {E}xplicit formulae for the eigenfunctions
  of the {$N$}-body {C}alogero model.
\newblock J. Phys. A: Math. Gen. \textbf{39}, 3511--3533 (2006)

\bibitem{He87}
Heckman, G.~J.: {R}oot systems and hypergeometric functions {II}.
\newblock Compositio Math. \textbf{64}, 353--373 (1987)

\bibitem{He91}
Heckman, G.~J.: {A}n elementary approach to the hypergeometric shift operators
  of {O}pdam.
\newblock Invent. Math. \textbf{103}, 341--350 (1991)

\bibitem{He94}
Heckman, G.~J.: {H}ypergeometric and spherical functions.
\newblock In Heckman, G.~J. and Schlichtkrull, H., eds., \emph{{H}armonic
  analysis and special functions on symmetric spaces}. San Diego: Academic
  Press, 1994, pp. 1--89

\bibitem{He97}
Heckman, G.~J.: {D}unkl operators.
\newblock Astérisque \textbf{245}, 223--246 (1997)

\bibitem{HO97}
Heckman, G.~J. and Opdam, E.~M.: {H}armonic analysis for affine {H}ecke
  algebras.
\newblock In Mazur, B., ed., \emph{{C}urrent developments in mathematics}.
  Boston: International Press, 1997, pp. 37--60

\bibitem{HO87}
Heckman, G.~J. and Opdam, E.~M.: {R}oot systems and hypergeometric functions
  {I}.
\newblock Compositio Math. \textbf{64}, 329--352 (1987)

\bibitem{He28}
Heisenberg, W.: {Z}ur {T}heorie des {F}erromagnetismus.
\newblock Z. Physik \textbf{49}, 619--636 (1928)

\bibitem{HS84}
Helffer, B. and Sj{ö}strand, J.: {M}ultiple wells in the semiclassical limit.
  {I}.
\newblock Comm. Partial Diff. Equations \textbf{9}, 337--408 (1984)

\bibitem{He00}
Helgason, S.: \emph{{G}roups and geometric analysis}.
\newblock Providence: AMS, 2000

\bibitem{Hi95}
Hikami, K.: {R}epresentation of motifs: new aspect of the
  {R}ogers--{S}zeg{{ö}} polynomial.
\newblock J. Phys. Soc. Japan \textbf{64}, 1047--1050 (1995)

\bibitem{Hi87}
Hitchin, N.: {S}table bundles and integrable systems.
\newblock Duke Math. J. \textbf{54}, 91--114 (1987)

\bibitem{Ho04}
Hochgerner, S.: {S}ingular cotangent bundle reduction and spin
  {C}alogero--{M}oser systems.
\newblock {\texttt{math.SG/0411068}}

\bibitem{HK06}
Holcman, D. and Kupka, I.: {S}ingular perturbation for the first eigenfunction
  and blow-up analysis.
\newblock Forum Math. \textbf{18}, 445--518 (2006)

\bibitem{Hu38}
Hulth{é}n, L.: {Ü}ber das {A}ustauchsproblem eines {K}ristalles.
\newblock Arkiv. Mat. Astron. Fysik \textbf{26A}, 1--106 (1938)

\bibitem{Hu90}
Humphreys, J.~E.: \emph{{R}eflection groups and {C}oxeter groups}.
\newblock Cambridge: Cambridge University Press, 1990

\bibitem{HN05}
Hurtubise, J. and Nevins, T.: {T}he geometry of {C}alogero--{M}oser systems.
\newblock Ann. Inst. Fourier (Grenoble) \textbf{55}, 2091--2116 (2005)

\bibitem{In90}
Inozemtsev, V.~I.: {O}n the connection between the one-dimensional {$S=1/2$}
  {H}eisenberg chain and {H}aldane--{S}hastry model.
\newblock J. Stat. Phys. \textbf{59}, 1143--1155 (1990)

\bibitem{IS01b}
Inozemtsev, V.~I. and Sasaki, R.: {H}ierarchies of spin models related to
  {C}alogero--{M}oser models.
\newblock Nucl. Phys. B \textbf{618}, 689--698 (2001)

\bibitem{IS01}
Inozemtsev, V.~I. and Sasaki, R.: {U}niversal {L}ax pairs for spin
  {C}alogero--{M}oser models and spin exchange models.
\newblock J. Phys. A: Math. Gen. \textbf{34}, 7621--7632 (2001)

\bibitem{Ja70}
Jack, H.: {A} class of symmetric polynomials with a parameter.
\newblock Proc. Roy. Soc. Edinburgh Sect. A \textbf{69}, 1--18 (1970/1971)

\bibitem{JK99}
Jain, S.~R. and Khare, A.: {A}n exactly solvable many-body problem in one
  dimension and the short-range {D}yson model.
\newblock Phys. Lett. A \textbf{262}, 35--39 (1999)

\bibitem{Ka96}
Kakei, S.: {C}ommon algebraic structure for the {C}alogero--{S}utherland
  models.
\newblock J. Phys. A: Math. Gen. \textbf{29}, L619--L624 (1996)

\bibitem{KL87}
Kalmeyer, V. and Laughlin, R.~B.: {E}quivalence of the resonating valence bond
  and fractional quantum {H}all states.
\newblock Phys. Rev. Lett. \textbf{59}, 2095--2098 (1987)

\bibitem{KN06}
Karabali, D. and Nair, V.~P.: {Q}uantum {H}all effect in higher dimensions,
  matrix models and fuzzy geometry.
\newblock J. Phys. A: Math. Gen. \textbf{39}, 12735(63) (2006)

\bibitem{Ka95}
Kasman, A.: {B}ispectral {K}{P} solutions and linearization of
  {C}alogero--{M}oser particle systems.
\newblock Comm. Math. Phys. \textbf{172}, 427--448 (1995)

\bibitem{KA92}
Kawakami, N.: {A}symptotic {B}ethe-ansatz solution of multicomponent quantum
  systems with $1/r^2$ long-range interaction.
\newblock Phys. Rev. B \textbf{46}, 1005--1014 (1992)

\bibitem{KKS78}
Kazhdan, D., Kostant, B., and Sternberg, S.: {H}amiltonian group actions and
  dynamical systems of {C}alogero type.
\newblock Comm. Pure Appl. Math. \textbf{31}, 481--507 (1978)

\bibitem{Ko72}
Koenigs, G.: {S}ur les g{é}od{é}siques à intégrales quadratiques.
\newblock In Darboux, G., ed., \emph{{L}e{\c{c}}ons sur la théorie
  générale des surfaces, Vol. 4}. New York: Chelsea, 1972, pp. 368--404

\bibitem{KT02}
Komori, Y. and Takemura, K.: {E}mbedded random matrix ensembles for complexity
  and chaos in finite interacting particle systems.
\newblock Phys. Rep. \textbf{227}, 93--118 (2002)

\bibitem{KBBT95}
Krichever, I.~M., Babelon, O., Billey, E., and Talon, M.: {S}pin generalization
  of the {C}alogero--{M}oser system and the matrix {KP} equation.
\newblock In Novikov, S., ed., \emph{Topics in topology and mathematical
  physics}. Providence: AMS, 1995, pp. 83--119

\bibitem{Ku85}
Kupershmidt, B.~A.: {Q}uantum mechanics as an integrable system.
\newblock Phys. Lett. A \textbf{109}, 136--138 (1985)

\bibitem{La00}
Langmann, E.: {A}nyons and the elliptic {C}alogero--{S}utherland model.
\newblock Lett. Math. Phys. \textbf{54}, 279--289 (2000)

\bibitem{La04}
Langmann, E.: {S}econd quantization of the elliptic {C}alogero--{S}utherland
  model.
\newblock Comm. Math. Phys. \textbf{247}, 321--351 (2004)

\bibitem{La91}
Lassalle, M.: {P}olyn{\^{o}}mes de {H}ermite généralisés.
\newblock C. R. Acad. Sci. Paris \textbf{313}, 579--582 (1991)

\bibitem{La91c}
Lassalle, M.: {P}olyn{\^{o}}mes de {J}acobi généralisés.
\newblock C. R. Acad. Sci. Paris \textbf{312}, 425--428 (1991)

\bibitem{La91b}
Lassalle, M.: {P}olyn{\^{o}}mes de {L}aguerre généralisés.
\newblock C. R. Acad. Sci. Paris \textbf{312}, 725--728 (1991)

\bibitem{Li04}
Li, L.~C.: {A} family of hyperbolic spin {C}alogero--{M}oser systems and the
  spin {T}oda lattices.
\newblock Comm. Pure Appl. Math. \textbf{57}, 791--832 (2004)

\bibitem{Li06}
Li, L.~C.: {P}oisson involutions, spin {C}alogero--{M}oser systems associated
  with symmetric {L}ie subalgebras and the symmetric space spin
  {R}uijsenaars--{S}chneider models.
\newblock Comm. Math. Phys. \textbf{265}, 333--372 (2006)

\bibitem{LX00}
Li, L.~C. and Xu, P.: {S}pin {C}alogero--{M}oser systems associated with simple
  {L}ie algebras.
\newblock C. R. Acad. Sci. Paris \textbf{331}, 55--60 (2000)

\bibitem{LL01}
Lieb, E.~H. and Loss, M.: \emph{{A}nalysis}.
\newblock Providence: AMS, 2001

\bibitem{Lu89}
Lusztig, G.: Affine Hecke algebras and their graded version.
\newblock J. Amer. Math. Soc. \textbf{2}, 599--635 (1989)

\bibitem{Ma95}
Macdonald, I.~G.: \emph{{S}ymmetric functions and {H}all polynomials}.
\newblock New York: Oxford University Press, 1995

\bibitem{Ma03}
Macdonald, I.~G.: \emph{{A}ffine {H}ecke Algebras and orthogonal polynomials}.
\newblock Cambridge: Cambridge University Press, 2003

\bibitem{MG69}
Majumdar, C. and Ghosh, D.~K.: {O}n next-nearest-neighbor interaction in linear
  chain. {I}.
\newblock J. Math. Phys. \textbf{10}, 1388--1398 (1969)

\bibitem{MG69b}
Majumdar, C. and Ghosh, D.~K.: {O}n next-nearest-neighbor interaction in linear
  chain. {II}.
\newblock J. Math. Phys. \textbf{10}, 1399--1402 (1969)

\bibitem{Ma99}
Marshakov, A.: \emph{{S}eiberg--{W}itten theory and integrable systems}.
\newblock Singapore: World Scientific, 1999

\bibitem{MZ03}
Minahan, J. and Zarembo, K.: {T}he {B}ethe-ansatz for {$\mathcal N=4$} super
  {Y}ang--{M}ills.
\newblock J. High Energy Phys. \textbf{2003}, 13(29)

\bibitem{MP93}
Minahan, J.~A. and Polychronakos, A.~P.: {I}ntegrable systems for particles
  with internal degrees of freedom.
\newblock Phys. Lett. B \textbf{302}, 265--270 (1993)

\bibitem{MF75}
Mon, K.~K. and French, J.~B.: {S}tatistical properties of many-particle
  spectra.
\newblock Ann. Phys. \textbf{95}, 90--111 (1975)

\bibitem{Mo75}
Moser, J.: {T}hree integrable {H}amiltonian systems connected with isospectral
  deformations.
\newblock Adv. Math. \textbf{16}, 197--220 (1975)

\bibitem{Ne96}
Nekrasov, N.: {H}olomorphic bundles and many-body systems.
\newblock Comm. Math. Phys. \textbf{180}, 587--603 (1996)

\bibitem{OP76}
Olshanetsky, M. A. and Perelomov, A. M.: Explicit solution of the Calogero
model in the classical case and geodesic flows on symmetric spaces of zero
curvature.
\newblock Lett. Nuovo Cimento \textbf{16}, 333--339 (1976)

\bibitem{OP76b}
Olshanetsky, M. A. and Perelomov, A. M.: Explicit solutions of some completely
integrable systems.
\newblock Lett. Nuovo Cimento \textbf{17}, 97--101 (1976)

\bibitem{OP81}
Olshanetsky, M.~A. and Perelomov, A.~M.: {C}lassical integrable
  finite-dimensional systems related to {L}ie algebras.
\newblock Phys. Rep. \textbf{71}, 313--400 (1981)

\bibitem{OP83}
Olshanetsky, M.~A. and Perelomov, A.~M.: {Q}uantum integrable systems related
  to {L}ie algebras.
\newblock Phys. Rep. \textbf{94}, 313--404 (1983)

\bibitem{Op88}
Opdam, E.~M.: {R}oot systems and hypergeometric functions {III}.
\newblock Compositio Math. \textbf{67}, 21--49 (1988)

\bibitem{Op88b}
Opdam, E.~M.: {R}oot systems and hypergeometric functions {IV}.
\newblock Compositio Math. \textbf{67}, 191--209 (1988)

\bibitem{OR04}
Ortega, J.~P. and Ratiu, T.: \emph{{M}omentum maps and {H}amiltonian
  reduction}.
\newblock Berlin: Birkh{ä}user, 2004

\bibitem{OS95}
Oshima, T. and Sekiguchi, H.: {C}ommuting families of differential operators
  invariant under the action of a {W}eyl group.
\newblock J. Math. Sci. Univ. Tokyo \textbf{2}, 1--75 (1995)

\bibitem{Po89}
Polychronakos, A.~P.: {N}on-relativistic bosonization and fractional
  statistics.
\newblock Nucl. Phys. B \textbf{324}, 597--622 (1989)

\bibitem{Po92}
Polychronakos, A.~P.: {E}xchange operator formalism for integrable systems of
  particles.
\newblock Phys. Rev. Lett. \textbf{69}, 703--705 (1992)

\bibitem{Po93}
Polychronakos, A.~P.: {L}attice integrable systems of {H}aldane--{S}hastry
  type.
\newblock Phys. Rev. Lett. \textbf{70}, 2329--2331 (1993)

\bibitem{Po94}
Polychronakos, A.~P.: {E}xact spectrum of {$\mathrm{SU}(n)$} spin chain with
  inverse-square exchange.
\newblock Nucl. Phys. B \textbf{419}, 553--566 (1994)

\bibitem{Po99}
Polychronakos, A.~P.: {G}eneralized {C}alogero models through reductions by
  discrete symmetries.
\newblock Nucl. Phys. B \textbf{543}, 485--498 (1999)

\bibitem{Po99b}
Polychronakos, A.~P.: {G}eneralized statistics in one dimension.
\newblock In Comtet, A., Jolic{\oe}ur, T., Ouvry, S., and David, F., eds.,
  \emph{Topological aspects of low dimensional systems}. Berlin:
  Springer-Verlag, 1999, pp. 415--471

\bibitem{Po01}
Polychronakos, A.~P.: {Q}uantum {H}all states on the cylinder as unitary matrix
  {C}hern--{S}imons theory.
\newblock J. High Energy Phys. \textbf{2001}, 11(20)

\bibitem{Po02}
Polychronakos, A.~P.: {C}alogero--{M}oser models with noncommutative spin
  interactions.
\newblock Phys. Rev. Lett. \textbf{89}, 126403(6) (2002)

\bibitem{Po06}
Polychronakos, A.~P.: {P}hysics and mathematics of {C}alogero particles.
\newblock J. Phys. A: Math. Gen. \textbf{39}, 12793--12845 (2006)

\bibitem{RS75}
Reed, M. and Simon, B.: \emph{{M}ethods of modern {M}athematical {P}hysics},
  vol. {II}: {F}ourier analysis, {S}elf-adjointness.
\newblock San Diego: Academic Press, 1975

\bibitem{RV04}
Roiban, R. and Volovich, A.: {Y}ang--{M}ills correlation functions from
  integrable spin chains.
\newblock J. High Energy Phys. \textbf{2004}, 32(24)

\bibitem{Ro98}
R{ö}sler, M.: {G}eneralized {H}ermite polynomials and the heat equation for
  {D}unkl operators.
\newblock Comm. Math. Phys. \textbf{192}, 519--542 (1998)

\bibitem{Ro03}
R{ö}sler, M.: {D}unkl operators: theory and applications.
\newblock In Koelink, E. and van Assche, W., eds., \emph{{O}rthogonal
  polynomials and special functions}. Berlin: Springer-Verlag, 2003

\bibitem{RS86}
Ruijsenaars, S. N.~M. and Schneider, H.: {A} new class of integrable systems
  and its relation to solitons.
\newblock Ann. Phys. \textbf{170}, 370--405 (1986)

\bibitem{SV04}
Sergeev, A.~N. and Veselov, A.~P.: {D}eformed quantum {C}alogero--{M}oser
  problems and {L}ie superalgebras.
\newblock Comm. Math. Phys. \textbf{245}, 249--278 (2004)

\bibitem{Sh88}
Shastry, B.~S.: {E}xact solution of an ${S}=1/2$ {H}eisenberg antiferromagnetic
  chain with long-ranged interactions.
\newblock Phys. Rev. Lett. \textbf{60}, 639--642 (1988)

\bibitem{Sh89}
Shifman, M.~A.: {N}ew findings in quantum mechanics (partial algebraization of
  the spectral problem).
\newblock Int. J. Mod. Phys. A \textbf{4}, 2897--2952 (1989)

\bibitem{ST89}
Shifman, M.~A. and Turbiner, A.~V.: {Q}uantal problems with partial
  algebraization of the spectrum.
\newblock Comm. Math. Phys. \textbf{126}, 347--365 (1989)

\bibitem{Sh01}
Shukla, P.: {N}on-{H}ermitian random matrices and the {C}alogero--{S}utherland
  model.
\newblock Phys. Rev. Lett. \textbf{87}, 194102(4) (2001)

\bibitem{Si83}
Simon, B.: {S}emiclassical analysis of low lying eigenvalues {I}.
  {N}on-degenerate minima: asymptotic expansions.
\newblock Ann. Inst. Henri Poincar{é} Sect. A \textbf{38}, 295--307 (1983)

\bibitem{Si95}
Simon, B.: \emph{{R}epresentations of finite and compact groups}.
\newblock Providence: AMS, 1995

\bibitem{Si00}
Simon, B.: {S}chr{ö}dinger operators in the twentieth century.
\newblock J. Math. Phys. \textbf{41}, 3523--3555 (2000)

\bibitem{St89b}
Stanley, R.~P.: {S}ome combinatorial properties of {J}ack symmetric functions.
\newblock Adv. Math. \textbf{77}, 76--115 (1989)

\bibitem{Su71}
Sutherland, B.: {E}xact results for a quantum many-body problem in one
  dimension.
\newblock Phys. Rev. A \textbf{4}, 2019--2021 (1971)

\bibitem{Su72}
Sutherland, B.: {E}xact results for a quantum many-body problem in one
  dimension. {I}{I}.
\newblock Phys. Rev. A \textbf{5}, 1372--1376 (1972)

\bibitem{SS93}
Sutherland, B. and Shastry, B.~S.: {S}olution of some integrable
  one-dimensional quantum systems.
\newblock Phys. Rev. Lett. \textbf{71}, 5--8 (1993)

\bibitem{Sz75}
Szeg{ö}, G.: \emph{{O}rthogonal polynomials}.
\newblock Providence: AMS, 1975

\bibitem{Ta03}
Takemura, K.: {T}he {H}eun equation and the {C}alogero--{M}oser--{S}utherland
  system {I}: the {B}ethe {A}nsatz method.
\newblock Comm. Math. Phys. \textbf{235}, 467--494 (2003)

\bibitem{Ta05}
Takemura, K.: {T}he {H}eun equation and the {C}alogero--{M}oser--{S}utherland
  system {IV}: the {H}ermite--{K}richever {A}nsatz.
\newblock Comm. Math. Phys. \textbf{258}, 367--403 (2005)

\bibitem{Ta82}
Takhtajan, L.: {T}he picture of low-lying excitations in the isotropic
  {H}eisenberg chain of arbitrary spins.
\newblock Phys. Lett. A \textbf{87}, 479--482 (1982)

\bibitem{Th92}
Thaller, B.: \emph{{T}he {D}irac Equation}.
\newblock Berlin: Springer-Verlag, 1992

\bibitem{To67}
Toda, M.: {O}ne-dimensional dual transformation.
\newblock J. Phys. Soc. Japan \textbf{22}, 431--436 (1967)

\bibitem{To50}
Tomonaga, S.: {R}emarks on {B}loch's method of sound waves applied to
  many-{F}ermion problems.
\newblock Prog. Theor. Phys. \textbf{5}, 544--569 (1950)

\bibitem{Tu88}
Turbiner, A.~V.: {Q}uasi-exactly solvable problems and $\mathfrak{sl}(2)$
  algebra.
\newblock Comm. Math. Phys. \textbf{118}, 467--474 (1988)

\bibitem{Us94}
Ushveridze, A.~G.: \emph{{Q}uasi-{E}xactly {S}olvable {M}odels in {Q}uantum
  {M}echanics}.
\newblock Bristol: Institute of Physics Publishing, 1994

\bibitem{Di95}
van Diejen, J.~F.: {C}ommuting difference operators with polynomial
  eigenfunctions.
\newblock Compositio Math. \textbf{95}, 183--233 (1995)

\bibitem{Di99}
van Diejen, J.~F.: {P}roperties of some families of hypergeometric orthogonal
  polynomials in several variables.
\newblock Trans. Amer. Math. Soc. \textbf{351}, 233--270 (1999)

\bibitem{UW96}
Wadati, H. U.~M.: {R}odrigues formula for {H}i-{J}ack symmetric polynomials
  associated with the quantum {C}alogero model.
\newblock J. Phys. Soc. Japan \textbf{65}, 2423--2439 (1996)

\bibitem{We92}
Weigert, S.: {T}he problem of quantum integrability.
\newblock Physica D \textbf{56}, 107--119 (1992)

\bibitem{Wi98}
Wilson, G.: {C}ollisions of {C}alogero--{M}oser particles and an adelic
  {G}rassmannian.
\newblock Invent. Math. \textbf{133}, 1--41 (1998)

\bibitem{Wi84}
Witten, E.: {N}onabelian bosonization in two dimensions.
\newblock Comm. Math. Phys. \textbf{92}, 455--472 (1984)

\bibitem{Wo85}
Wojciechowski, S.: {A}n integrable marriage of the {E}uler equations with the
  {C}alogero--{M}oser system.
\newblock Phys. Lett. A \textbf{111}, 101--103 (1985)

\bibitem{Ya95}
Yamamoto, T.: {M}ulticomponent {C}alogero model of ${B}_{N}$-type confined in a
  harmonic potential.
\newblock Phys. Lett. A \textbf{208}, 293--302 (1995)

\bibitem{YT96}
Yamamoto, T. and Tsuchiya, O.: {I}ntegrable $1/r^2$ spin chain with reflecting
  end.
\newblock J. Phys. A: Math. Gen. \textbf{29}, 3977--3984 (1996)

\end{thebibliography}
%\bibliographystyle{cmp}
%\end{document}

\end{document}